\documentclass{aa}
\usepackage{graphicx}
\usepackage{xcolor}
\usepackage[position=top]{subfig}
\usepackage[fleqn]{amsmath}
\usepackage[utf8]{inputenc}
\usepackage{epsfig,amsmath,amssymb}
\usepackage[varg]{txfonts}
\usepackage{multirow}

\usepackage{tcolorbox}

\definecolor{myblue}{rgb}{0.00, 0.0, 0.9}
\definecolor{myred}{rgb}{0.90, 0.0, 0.0}
\definecolor{mygreen}{rgb}{0.0, 0.7, 0.0}
\usepackage[breaklinks,  citecolor=myblue, linkcolor=myred, urlcolor=purple, colorlinks=true, debug, baseurl=' ']{hyperref}
\usepackage{orcidlink}

\definecolor{vincent}{rgb}{0.1, 0.0, 0.7}
\definecolor{raphael}{rgb}{0.7, 0.1, 0.5}

\titlerunning{Magnetic field in thick bubble shells}

\authorrunning{Pelgrims et al.}

\begin{document}

%

%
\title{
An analytical model for the magnetic field in the thick shell of Galactic bubbles with uniform initial conditions
}

\author{V. Pelgrims
      \inst{1}\fnmsep\thanks{vincent.pelgrims@ulb.be}\orcidlink{0000-0002-5053-3847},
      M.~Unger\inst{2}\fnmsep\inst{3}\orcidlink{0000-0002-7651-0272},
      I.~C.~Mari{\c s}\inst{1}\orcidlink{0000-0002-5771-1124}
}          
\institute{
Universit{\'e} Libre de Bruxelles, Science Faculty CP230, B-1050 Brussels, Belgium
\and
Institut f\"ur Astroteilchenphysik, Karlsruher Institut f\"ur Technologie, Karlsruhe 76344, Germany
\and
Institutt for fysikk, Norwegian University of Science and Technology (NTNU), Trondheim, Norway
}

\date{Received -- / Accepted --}

\abstract{
Bubbles and super-bubbles are ubiquitous in the interstellar medium and influence their local magnetic field.
Starting from the assumption that bubbles result from violent explosions that sweep matter away in a thick shell, we derive the analytical equations for the divergence-free, regular magnetic field in the shell. The explosion velocity field is assumed to be radial but not necessarily spherical, making it possible to model various-shaped bubbles.
Assuming an explosion center, the magnetic field at the present time is fully determined by the initial uniform magnetic field, the present-time geometry of the bubble shell, and a radial vector field that encodes the explosion-induced displacement of matter, from its original location to its present-time location.
We present the main characteristics of our magnetic-field model using a simple linear model for the radial displacements.
Next, we use our analytical prescription, informed by a three-dimensional dust density map, to estimate the expected contribution of the shell of the Local Bubble, the super-bubbles in which the Sun resides, to the integrated Faraday rotation measures and synchrotron emission and compare these to full-sky observational data.
We find that, while the contribution to the former is minimal, the contribution to the latter is very significant at Galactic latitudes $|b|>45^\circ$.
Our results underline the need to take the Local Bubble into account in large-scale Galactic magnetic field studies.
}


\maketitle


\section{Introduction}
\label{sec:intro}
Like external galaxies, the Milky Way is permeated by magnetic fields. Magnetic fields are relevant in various processes and scales, from regulating star formation to shaping large-scale structures in the disk and the halo of the Galaxy.
Magnetic fields in the Galaxy also affect our ability to study the Universe’s structure and history and to identify the sources of ultra-high energy cosmic rays (e.g., see \citealt{Beck2015}; \citealt{Haverkorn2015} for recent reviews).

To gain insight into the general structure and amplitude of the Galactic magnetic field (GMF) on the kiloparsec scale, authors have primarily relied on observations of Faraday rotation measures (RM) toward extragalactic sources, Galactic synchrotron emission and polarized emission from Galactic dust.
Using forward modeling frameworks, where model predictions are compared to observations, authors have deduced constraints on a variety of three-dimensional (3D) models for the GMF (e.g., see \citealt{Jaffe2019} for a recent review on model tuning).
One of the main obstacles to such an approach is the line-of-sight-integrated nature of the observables, which leads to degeneracy between models and model parameters and, consequently, uncertainties in the reconstructed picture of the GMF.
In the future, this degeneracy will probably be overcome, at least for the first few kiloparsecs around the Sun, by the advent of tomography data derived from both starlight polarization and Faraday tomography (e.g., \citealt{Pelgrims2023,Pelgrims2024}; \citealt{Erceg2024a,Erceg2024b}) which are made possible by major surveys (\citealt{Magalhaes2005}; \citealt{Tassis2018}; \citealt{Shimwell2022}).
In the meantime, it is difficult to disentangle the respective contributions of the Galaxy's near and far regions to the line-of-sight-integrated observables.

In this context, it should be stressed that most efforts to model the large-scale GMF have overlooked the fact that the Solar system resides in the Local Bubble.
The Local Bubble is a cavity with unusually low density of gas (e.g., \citealt{Cox1987}); \citealt{Lallement2003}), filled with an X-ray emitting plasma (e.g., \citealt{Liu2017}), and surrounded by a shell of cold (ionized) gas and dust (e.g., \citealt{Pelgrims2020}; \citealt{ONeill2024}). This local cavity has most likely been created by successive supernovae explosions that occurred in the past 10-15~Myr (e.g., \citealt{MaizApell2001}; \citealt{Breitschwerdt2016}; \citealt{Schulreich2023}), and has currently an irregular shape, with a radius of around 100~pc in the Galactic plane and around 300~pc perpendicular to it.
As for other bubbles in the interstellar medium (ISM), it is expected that supernovae explosions and the resulting motion of the swept-up matter have deformed the initial magnetic field and amplified its strength in compressed regions.
This is in agreement with the picture obtained from starlight polarization that the magnetic field in the local ISM does not follow the large-scale GMF (e.g., \citealt{Heiles1998}; \citealt{Leroy1999}; \citealt{Santos2011}; \citealt{Frisch2012}; \citealt{Berdyugin2014}; \citealt{Gontcharov2019}) and that the amplitude of the magnetic field is likely enhanced in the shell of the Local Bubble (\citealt{Andersson2005,Andersson2006}; \citealt{Medan2019}).

Although imprints of the magnetized shell of the Local Bubble have recently been detected in tomographic data, from measurements on Faraday depth and starlight polarization, in combination or not with dust polarized emission (\citealt{Erceg2024a,Erceg2024b}; \citealt{Skalidis2019}; \citealt{Pelgrims2024}; \citealt{Uppal2024}), its contribution to line-of-sight integrated observables which are sensitive to the GMF is still unknown, or at least little constrained.
The chance for its contribution to be dominant naturally increases toward high-Galactic latitudes as less matter is encountered beyond the Local Bubble wall. This is even more likely to be the case the smaller the scale height of the density distribution of the relevant ISM matter.
This, for example, has motivated \cite{Alves2018} to use the dust polarized emission at high-Galactic latitudes to constrain a physical model for the magnetic field in the shell of the Local Bubble whose shape was assumed to be spheroidal, and with no thickness. This motivation was later validated by \cite{Skalidis2019} (but see \citealt{Halal2024}) and encouraged \cite{Pelgrims2020} to have the analysis of \cite{Alves2018} reworked with a more realistic shape for the Local Bubble shell that they deduced directly from 3D dust density maps available at the time.
However, it is difficult to predict how the relative contributions from the Local Bubble shell and from the background change with sky regions, toward lower Galactic latitudes. This is primarily due to depolarization effects which occur both along the sightline and within the beam of observation. Surprisingly, perhaps, \cite{ONeill2024b} suggested recently that the wall of the Local Bubble may be a substantial contributor to the dust-polarized emission for large portions of the sky.

Recently, \cite{Korochkin2024} were the first to include a prescription for the magnetic field in the shell of the Local Bubble into the modeling of the regular component of the GMF using full-sky data on Galactic synchrotron polarized emission and Faraday RM from extragalactic sources.
They find that the contribution from the shell of the Local Bubble is substantial.
However, in their implementation, they use a spherical model for the shape of the Local Bubble and a geometrical prescription to derive the magnetic field in its shell, which does not lead to a divergence-free magnetic field.
With this paper, we aim to go beyond this and quantify the effects of a more realistic shape for the shell of the Local Bubble.

\smallskip

In their paper, \cite{Alves2018} derive an analytical expression for a divergence-free magnetic field in the shell of the Local Bubble which they assume to be very thin.
While in principle this formalism makes it possible to describe the magnetic field in non-spherical shells, it has the important drawback that the shell must be infinitely thin and, as a consequence, that the amplitude of the magnetic field becomes very large (infinite) in the shell. Consequently, this analytical prescription is inadequate for characterizing the magnetic field in a realistic shell of finite thickness, and for predicting the shell's contribution to Faraday RM and synchrotron, both of which depend on the amplitude of the magnetic field.
In this work, we move past the assumption that the bubble shells must be infinitely thin. We propose a general analytical expression for the divergence-free magnetic field in the thick shell of bubbles which, additionally might be non-spherical.
The paper is structured as follows.

We derive our analytical model in Sect.~\ref{sec:AnalyticModel} where we also present its main features and some illustrative examples.
In Sect.~\ref{sec:LBcase}, we particularize our model to the case of the Local Bubble and estimate its contribution to the full-sky data on synchrotron polarized emission and Faraday RM, adopting realistic shapes for the shell of the Local Bubble as inferred from a 3D dust map. We discuss the limitations of our model and the possible implications of our findings for the modeling of the large-scale GMF in Sect.~\ref{sec:conclusion}.

\section{An analytical model for the magnetic field in thick bubble shells}
\label{sec:AnalyticModel}
In this paper, we build upon the pioneering work of \cite{Parker1970} and \cite{Alves2018} to describe the divergence-free magnetic field in thick shells of ISM bubbles.
We only consider the coherent (regular) component of the magnetic field.
We first derive the equations in a general case without specifying any functional form for the displacement vector field which relates the present-time position of test particles to their initial locations.
Then, we particularize our solution to the case where the displacement vector field is linear in the radial coordinate (with respect to the center of explosion) and explore our model predictions for various possible geometries for the bubble shell.

\subsection{General expression for the magnetic field}
\label{sec:Bmodel}
To derive the equations for the magnetic field in the thick shell of a bubble, we closely follow the derivation in \cite{Alves2018} which is based on the work of \cite{Parker1970}. Our base assumptions are that an ISM bubble results from a single explosion that has swept up matter into a shell with finite thickness and that the explosion-induced motions are purely radial with respect to the explosion center, but do not necessarily have a spherical symmetry. That is, the explosion-induced velocity flow is not necessarily isotropic. We further adopt the frozen-in approximation which implies that magnetic field lines follow the matter in its motion. As a result of this coupling, the vector field of the magnetic field, which we assume to be initially uniform in strength and direction, is modified by the explosion. The magnetic field lines are deformed to closely follow the geometry of the shell and are confined within it, thus increasing the amplitude of the magnetic field. The amplitude of the magnetic field vanishes inside the bubble and remains unchanged outside it.

To proceed, we consider that the explosion center is at the origin of a Cartesian coordinate system $(x,\,y,\,z)$ to which is associated a spherical coordinate system $(r,\,\theta,\,\phi)$ with its corresponding unit basis vectors $(\mathbf{e}_r,\,\mathbf{e}_\theta,\,\mathbf{e}_\phi)$.
In this reference frame, the velocity field at any time $t$ takes the form $\mathbf{V} = V(r,\theta,\phi) \, \mathbf{e}_r$, and the initial position of a test particle presently at position $\mathbf{r} = r \, \mathbf{e}_r$ can be written as
\begin{align}
    \mathbf{r}^0 &= r^0(r,\theta,\phi)\,\mathbf{e}_r  \nonumber \\
            &= \mathbf{r} - \mathbf{\Lambda} \;,
\end{align} where we introduced the displacement vector field $\mathbf{\Lambda}$ which results from the time integration of $\mathbf{V}$. As it is the case for $\mathbf{V}$, the displacement vector field is radial but not necessarily with spherical symmetry: $\mathbf{\Lambda}= \Lambda(r,\theta,\phi)\,\mathbf{e}_r$.

Generally, the frozen-in approximation is given by the induction equation in the limit of zero magnetic diffusion. Therefore, the magnetic field $\mathbf{B}$ must be given by:
\begin{align}
    \frac{\partial \mathbf{B}}{\partial t} = \mathbf{\nabla} \times \left( \mathbf{V} \times \mathbf{B}\right) \;.
    \label{eq:inducEqB}
\end{align}
Introducing the vector potential $\mathbf{A}$, such that $\mathbf{B} = \mathbf{\nabla} \times \mathbf{A}$, in Eq.~\eqref{eq:inducEqB}, swapping the time and spatial derivatives and de-curling, we obtain
\begin{align}
    \frac{\partial \mathbf{A}}{\partial t} = \mathbf{V} \times \left(\mathbf{\nabla} \times \mathbf{A}\right) + \mathbf{\nabla}f \;,
    \label{eq:inducEqA}
\end{align}
where $f$ is an arbitrary scalar field. Adopting the gauge $f = - \mathbf{V}\cdot\mathbf{A}$, we then write the equation for the Cartesian components of $\mathbf{A}$ as:
\begin{align}
    &\left[ \frac{\partial }{\partial t} + \sum_{j=1}^3 V_j \, \frac{\partial}{\partial x_j} \right] \,  A_i = - \sum_{j=1}^3 A_j \, \frac{\partial V_j}{\partial x_i} \nonumber \\
    & \Leftrightarrow \; \frac{{\rm{d}}}{{\rm{d}}t} \,  A_i = - \sum_{j=1}^3 A_j \, \frac{\partial V_j}{\partial x_i},
\end{align}
where the operator on the left-hand side is known as the advective derivative.
Integrating in time the above equation leads to the expression of the Cartesian components of the potential vector in the present time and at location ($\mathbf{x}$) as a function of the initial potential vector at the initial location ($\mathbf{x}^0$):
\begin{align}
    A_i(\mathbf{x}) = \sum_{j=1}^{3} A^0_{~j}(\mathbf{x}^0) \, \frac{\partial x^0_{~j}}{\partial{x_i}} \;.
    \label{eq:A_gen}
\end{align}
This is Eq.~(10) in \cite{Parker1970}. If $\mathbf{A}^0(\mathbf{x}^0)$ is known, determining $\mathbf{A}(\mathbf{x})$ amounts to determining how the moving coordinate system evolves in space with respect to the fix one.
Since we assume a radial velocity field, the displacement vector field is also radial, and we can write its Cartesian components as
\begin{align}
    \left( \mathbf{\Lambda}(\mathbf{x})\right)_i = \Lambda \, \frac{\partial x_i}{\partial r}\;,
\end{align}
and therefore
\begin{align}
    x^0_{~i} = x_i - \Lambda \, \frac{\partial x_i}{\partial r} \;.
\end{align}
Inserting this into Eq.~\eqref{eq:A_gen} leads to
\begin{align}
A_i = \sum_{j=1}^3 A^0_{~j} \, \delta_{ij} \left(1 - \frac{\Lambda}{r} \right) - \sum_{j=1}^3 A^0_{~j} \, \frac{\partial x_j}{\partial r} \left(\frac{\partial \Lambda}{\partial x_i} - \frac{\Lambda}{r}\,\frac{\partial x_i}{\partial r} \right) \;,
\end{align}
where $\delta_{ij}$ is the Kronecker symbol.
Then, using the fact that the spherical component of $\mathbf{A}$ can be written as
\begin{align}
    A_r &= \sum_{i=1}^3 A_i \, \frac{\partial x_i}{\partial r} = \sum_{i=1}^3 A_i \, \frac{x_i}{r} \nonumber \\
    A_\theta &= \sum_{i=1}^3 A_i \, \frac{1}{r}\,\frac{\partial x_i}{\partial \theta} \\
    A_\phi &= \sum_{i=1}^3 A_i \, \frac{1}{r\sin\theta}\,\frac{\partial x_i}{\partial \phi} \; , \nonumber
\end{align}
and that the same holds true for $\mathbf{A}^0$ (using the same spherical coordinates), we finally write the spherical components of $\mathbf{A}(\mathbf{r})$ as a function of $\mathbf{A}^0(\mathbf{r}^0)$ and the displacement vector and its partial derivatives:
\begin{align}
    A_r &= A^0_{~r} \, \frac{\partial r^0}{\partial r} = A^0_{~r} \, \left( 1 - \frac{\partial \Lambda}{\partial r} \right) \nonumber \\
    A_\theta &= A^0_{~\theta} \, \frac{r^0}{r}
        - \frac{1}{r} \, \frac{\partial \Lambda}{\partial \theta} \, A^0_{~r} \\
    A_\phi &= A^0_{~\phi} \, \frac{r^0}{r}
        - \frac{1}{r\,\sin \theta} \, \frac{\partial \Lambda}{\partial \phi} \, A^0_{~r} \;. \nonumber 
\end{align}
These equations are equivalent to Eqs.~(A4) and~(A5) derived in (\citealt{Alves2018}).

\begin{figure*}
    \centering
    \includegraphics[trim={1.1cm 0.6cm 3.0cm 1.2cm},clip,height=.31\linewidth]{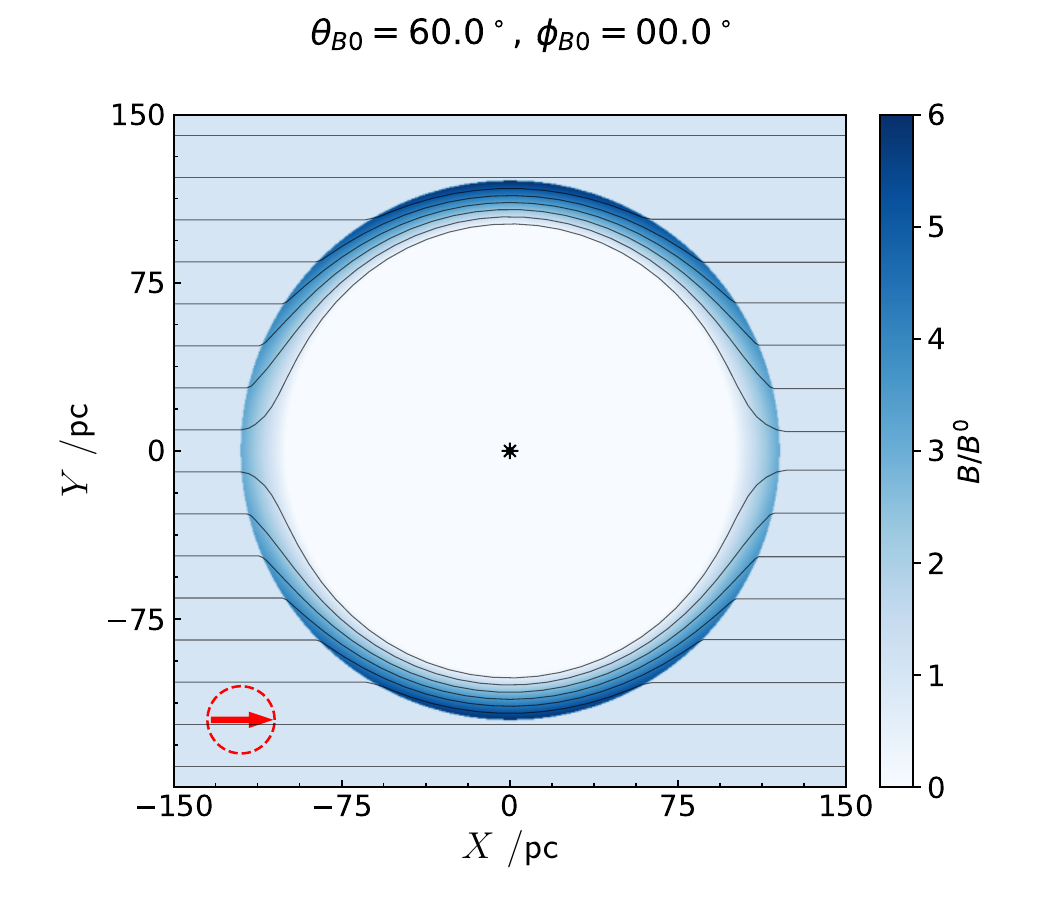}
    \includegraphics[trim={1.1cm 0.6cm 3.0cm 1.2cm},clip,height=.31\linewidth]{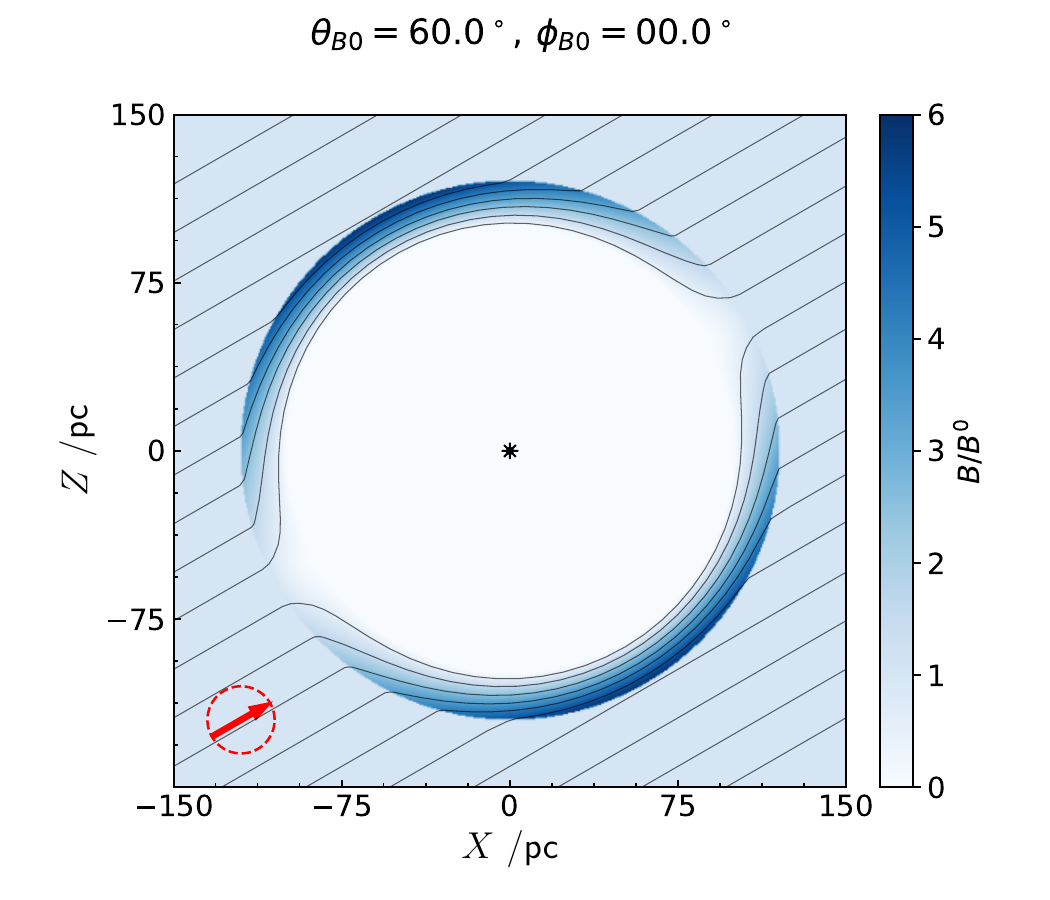}
    \includegraphics[trim={1.1cm 0.6cm 1.1cm 1.2cm},clip,height=.31\linewidth]{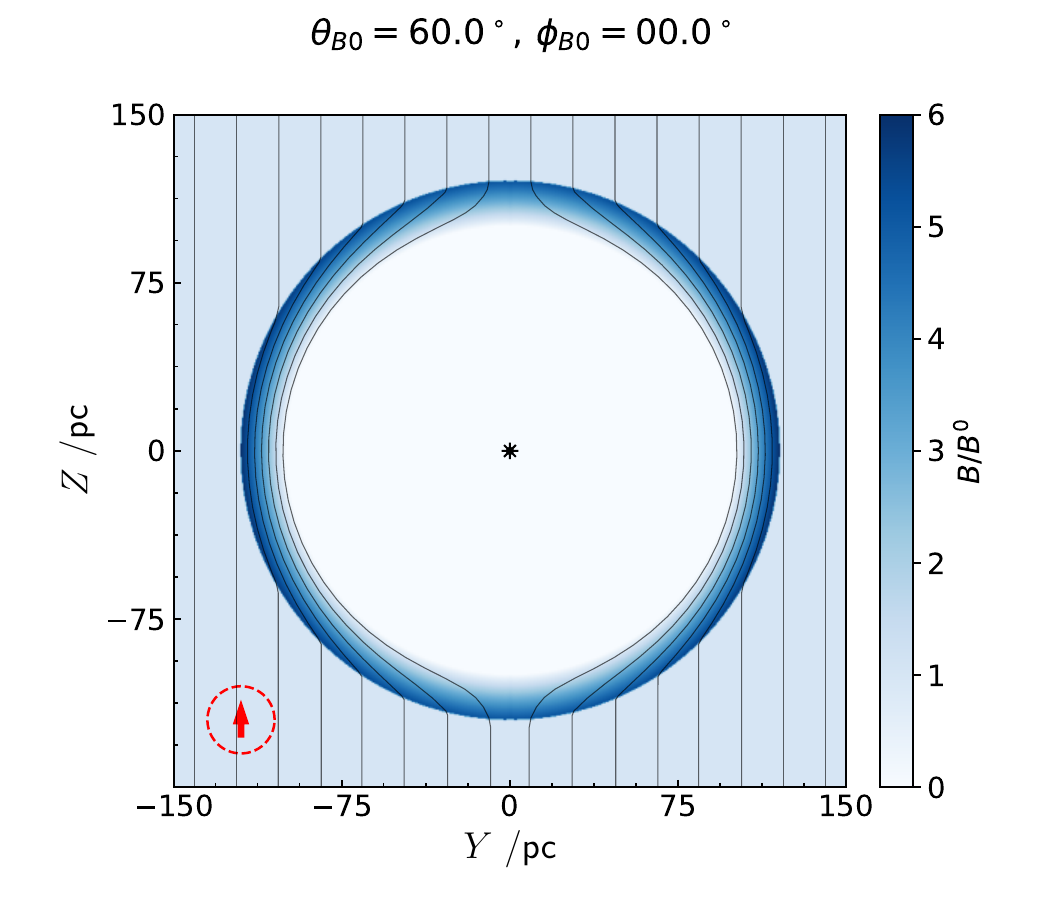}\\[-0.5ex]
    \caption{Crosscuts along the planes $XY$, $XZ$, and $YZ$ through a spherical bubble with inner radius of 100~pc and thickness of 20~pc. The explosion center (black asterisk) is at the origin of the axes and at the center of the bubble. The direction of $\mathbf{B}^0$ is set by $(\theta_{B^0},\,\phi_{B^0}) = (60^\circ,\,0^\circ)$ in a spherical coordinate where $\theta=0$ points to positive $Z$ and $(\theta,\,\phi)=(0,\,0)$ points to positive $X$. $\mathbf{B}^0$ belongs to the $XZ$ plane.
    The background colors indicate the amplification of the total magnetic field in the plane ($|\mathbf{B}|/|\mathbf{B}^0|$) and streamline the orientation of the projected $\mathbf{B}$ in the crosscut planes.
    The red arrow in the lower left corner of each panel indicates the projection of $\mathbf{B}^0$ in the planes. The red circle has a radius of $B^0$ and indicates the maximum length of the arrow.
    }
    \label{fig:CC_Sph00}
\end{figure*}
\begin{figure}
    \centering
    \includegraphics[trim={0.4cm 0.3cm 0.3cm 1.2cm},clip,width=.98\linewidth]{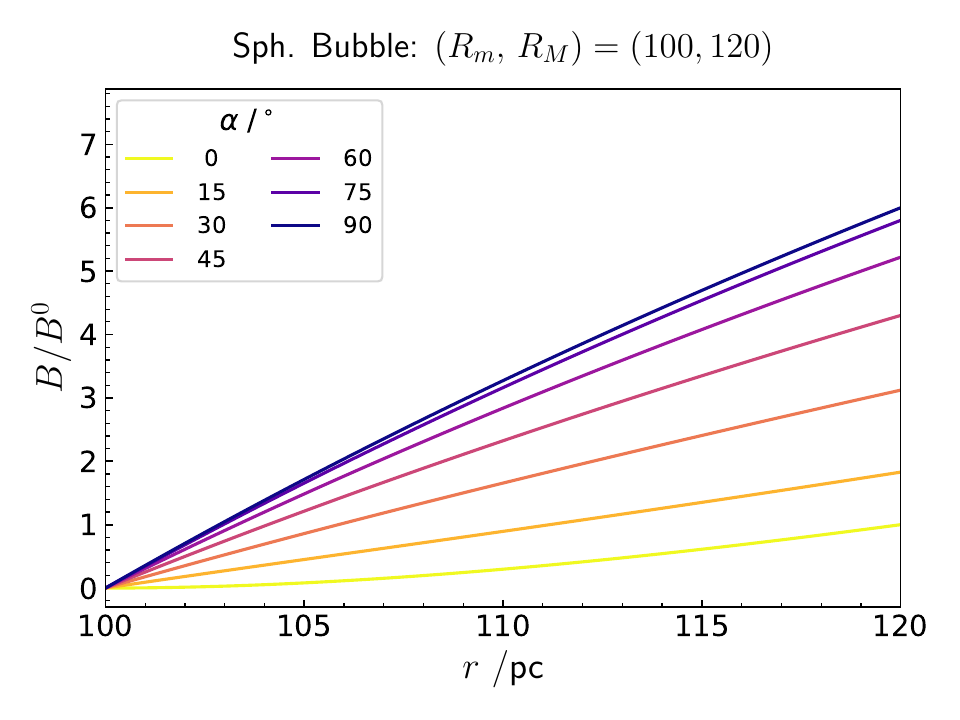}\\
    \includegraphics[trim={0.4cm 0.3cm 0.3cm 1.2cm},clip,width=.98\linewidth]{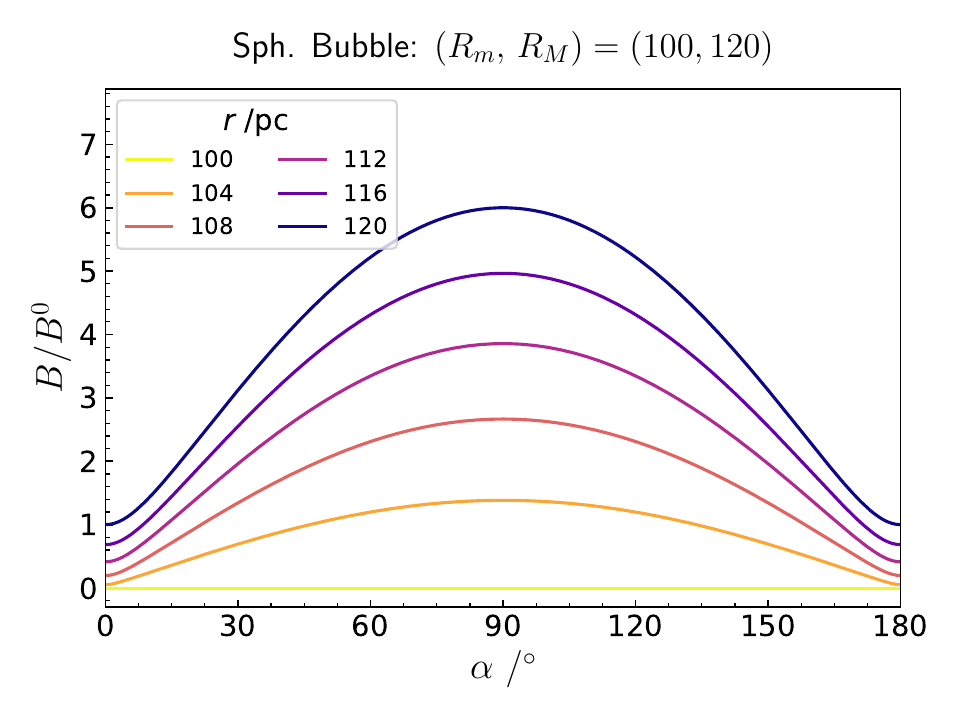}\\[-2.ex]
    \caption{Dependence of the amplification factor of the magnetic field in the shell of a spherical bubble as a function of the radial distance (top) and as a function of the angle $\alpha$ between the radial unit vector and the direction of the initial magnetic field (bottom), for several values of $\alpha$ and $r$, respectively. In this example, $R_{\rm{min}} = 100$ pc and $R_{\rm{max}} = 120$ pc.}
    \label{fig:Bampl}
\end{figure}

Now we can derive the spherical-coordinate components of $\mathbf{B}(\mathbf{r})$ simply by taking the curl of $\mathbf{A}(\mathbf{r})$. In spherical coordinates, we have:
\begin{align}
    B_r &= \frac{1}{r \sin\theta}\left( \frac{\partial(\sin\theta \, A_\phi)}{\partial \theta} - \frac{\partial A_\theta}{\partial \phi} \right) \nonumber \\
    B_\theta &= \frac{1}{r}\left(\frac{1}{\sin\theta} \frac{\partial A_r}{\partial \phi} - \frac{\partial (r\,A_\phi)}{\partial r} \right)\\
    B_\phi &= \frac{1}{r}\left( \frac{\partial(r \, A_\theta)}{\partial r} - \frac{\partial A_r}{\partial \theta} \right) \; . \nonumber
\end{align}
The same holds true for $\mathbf{B}^0(\mathbf{x}^0)$ in the vector basis associated to the moving coordinate system $(r^0,\,\theta^0,\,\phi^0)$.
Because the displacement vector field is a function of $(r,\,\theta,\,\phi)$, the change of coordinates $\mathbf{r} \leftrightarrow \mathbf{r}^0$ (or equivalently $\mathbf{x} \leftrightarrow \mathbf{x}^0$), implies the following relations between the partial derivatives:
\begin{align}
    \frac{\partial}{\partial r} &=  \left(1 - \frac{\partial \Lambda}{\partial r} \right) \,\frac{\partial}{\partial r^0} \nonumber \\
    \frac{\partial}{\partial\theta} &= \frac{\partial}{\partial \theta^0} - \frac{\partial \Lambda}{\partial\theta} \,\frac{\partial}{\partial r^0} \\
    \frac{\partial}{\partial\phi} &= \frac{\partial}{\partial \phi^0} - \frac{\partial \Lambda}{\partial\phi} \,\frac{\partial}{\partial r^0} \;. \nonumber
\end{align}
The fact that $\Lambda$ is radial, however, guarantees that the angular coordinates of a test particle remain unchanged: $\theta = \theta^0$ and $\phi = \phi^0$.
Taking this into account, the analytical equations for the components of the magnetic field in the shell of a bubble become:
\begin{align}
    \label{eq:Brtp}
    B_r &= \left(\frac{r^0}{r}\right)^2 \, B^0_{~r} + \frac{r^0}{r} \, \mathbf{\nabla}_t \Lambda \cdot \mathbf{B}^0_{~t} \nonumber \\
    B_\theta &= \frac{r^0}{r}\,\frac{\partial r^0}{\partial r} \, B^0_{~\theta} \\
    B_\phi &= \frac{r^0}{r}\,\frac{\partial r^0}{\partial r} \, B^0_{~\phi} \;, \nonumber
\end{align}
where $\mathbf{\nabla}_t \Lambda$ and $\mathbf{B}^0_{~t}$ are the orthoradial components of the gradient of $\Lambda(r,\,\theta,\,\phi)$ and $\mathbf{B}^0$, that is,
\begin{align}
    &\mathbf{\nabla}_t = \frac{1}{r}\frac{\partial}{\partial\theta} \, \mathbf{e}_\theta + \frac{1}{r\,\sin\theta}\frac{\partial}{\partial\phi} \, \mathbf{e}_\phi \;, \;{\rm{and}} \\
    &\mathbf{B}^0_{~t} = B^0_{~\theta}\,\mathbf{e}_\theta + B^0_{~\phi}\,\mathbf{e}_\phi \;.
    \label{eq:orthoradialComp}
\end{align}
The expressions in Eq.~\eqref{eq:Brtp} are general and have been obtained without the need to specify any form for the displacement vector field except that we required it to be radial with respect to the explosion center. The present-time magnetic field in the shell of the bubble can be obtained as soon as the initial magnetic field, the displacement vector field, and its partial derivatives are known.
The expressions in Eq.~\eqref{eq:Brtp} show that the magnetic field lines remain in the same plane defined by the explosion center and the orthoradial component of the initial magnetic field ${\mathbf{B}^0}_t$, as measured in spherical coordinates centered on the explosion center. The radial component ($B_r$) adjusts so that magnetic field lines follow the bubble shape. It is worth noticing that, in general, the radial component of the magnetic field does not vanish in the bubble shell.
The main challenge in estimating Eq.~\eqref{eq:Brtp} resides in determining the partial derivatives of $\Lambda(r,\theta,\phi)$ at any location within the bubble shell.

\subsection{A linear model for the displacement vector field}
\label{sec:Dmodel}
To move forward, we need to particularize our equations with an expression for the displacement vector field $\Lambda(r,\theta,\phi)$, which could ideally be constrained from observation.
We consider that $\Lambda(r,\theta,\phi)$ is linear in $r$ and is such that all the matter located within the bubble at the initial time has been swept up in a shell defined by its inner and outer radius surfaces $R_{\rm{min}}$ and $R_{\rm{max}}$, respectively. By construction, we assume that $R_{\rm{min}}$ and $R_{\rm{max}}$ are both functions of the angular coordinates $(\theta,\,\phi)$ measured in the explosion-centric reference frame, and that the volume encompassed by the bubble is determined by $r \leq R_{\rm{max}}$. With these choices, the expression for the displacement along any radial direction may be written as:
\begin{align}\label{eq:Displ_LinModel}
    \Lambda(r,\theta,\phi) = \frac{R_{\rm{min}}}{R_{\rm{max}}-R_{\rm{min}}} \, (R_{\rm{max}} - r) \;,
\end{align}
where the angular dependence of $R_{\rm{min}}$ and $R_{\rm{max}}$ are implicit.
In this simple case, the partial derivatives of $\Lambda$ become:
\begin{align}
    \frac{\partial \Lambda}{\partial r} &= - \frac{R_{\rm{min}}}{R_{\rm{max}} - R_{\rm{min}}} \label{eq:dr_lambda}
\end{align}
\begin{align}
    \frac{\partial \Lambda}{\partial \theta} &= r \, \frac{R_{\rm{min}} \,\partial_\theta R_{\rm{max}} - R_{\rm{max}}\,\partial_\theta R_{\rm{min}}}{(R_{\rm{max}}-R_{\rm{min}})^2} \nonumber \\
    &\phantom{==} + \frac{{R_{\rm{max}}}^2 \, \partial_\theta R_{\rm{min}} - {R_{\rm{min}}}^2 \, \partial_\theta R_{\rm{max}}}{(R_{\rm{max}}-R_{\rm{min}})^2} \;,
    \label{eq:dxi_lambda}
\end{align}
where we have used the shorthand notation for the partial derivative: $\partial_\theta \equiv \partial / \partial \theta$. $\partial \Lambda / \partial \phi$ is given by an equivalent equation, replacing $\theta$ by $\phi$ in Eq.~\eqref{eq:dxi_lambda}.
As such, the partial derivatives of $\Lambda$ in the shell of the bubble are fully determined by the shape of the inner and outer surfaces of the bubble shell ($R_{\rm{min}}$, $R_{\rm{max}}$, and their angular derivatives), and $r$, the radial coordinate at which the magnetic field needs to be evaluated.
We recall that the magnetic field in the shell of the bubble given by Eq.~\eqref{eq:Brtp} is defined only for $r \in [R_{\rm{min}},\, R_{\rm{max}}]$.

Our choice for the displacement field implies that a test particle initially at the origin ($r^0 = 0$) is swept up to the position $r = R_{\rm{min}}$ and that the test particle initially at $r^0 = R_{\rm{max}}$ remains at its location with $r = R_{\rm{max}}$. With this model, the matter density and the magnetic field vanish inside the bubble ($r < R_{\rm{min}}$) and remain unchanged outside it ($r > R_{\rm{max}}$).
Mass conservation from before the explosion to the present-time implies
\begin{align}
    \rho(r) \, r^2 \, {\rm{d}}r = \rho^0(r^0)\, {r^0}^2 \, {\rm{d}}r^0 \;,
\end{align}
where $\rho^0$ is the initial matter density.
Assuming an initially homogeneous matter density and particularizing to our choice for $\Lambda$, the current matter density distribution is a function of $r$ and takes the form:
\begin{align}
    \rho(r) = \left\{
        \begin{array}{ll}
            0 & {\rm{if}} \; r < R_{\rm{min}} \\
            \rho^0 \, \left( \frac{{R_{\rm{max}}}}{{R_{\rm{max}}} - {R_{\rm{min}}}} \right)^3 \, \left(1-\frac{R_{\rm{min}}}{r} \right)^2 & {\rm{if}} \; r \in [R_{\rm{min}},\,R_{\rm{max}}]\\
            \rho^0 & {\rm{if}} \; r > R_{\rm{max}} \;.
        \end{array}
        \right.
    \label{eq:ShellDensity}
\end{align}
Matter density varies with $r$ inside the shell, and can also vary as a function of angular coordinates if $R_{\rm{min}}$ and $R_{\rm{max}}$ vary, that is, if the displacement vector field is not spherically symmetric.

For the sake of simplicity, we disregard any inhomogeneities or large-scale gradients in the initial matter density field. To deal with them, the initial density field and the displacement vector field would have to be modeled jointly, using the continuity equation, for example. We defer such a consideration to future work.
Therefore, in its current implementation, although it enables us to take into account complicated shapes for bubble shells that might result from inhomogeneities (\citealt{Kim2015}), our model best describes the regular component of the magnetic field in the bubble shell resulting from a single explosion in a homogeneous medium, but possibly with anisotropic energy deposition.

Below, we look into the properties of the present-time magnetic field for the case of spherical bubbles with explosion centers at their origins (Sect.~\ref{sec:SphBubble}), and then for the case where spherical symmetry is lost (Sect.\ref{sec:NonSphBubble}). In all cases, and as expected, the magnetic field in the shell of the bubble undergoes an amplification. We also demonstrate in Appendix~\ref{sec:Bflux} that our analytical model leads to magnetic flux conservation.

\subsection{The case of spherical bubbles}
\label{sec:SphBubble}
In this subsection, we particularize our equations to the case where the displacement vector field is spherically symmetric. In this case, the angular derivatives of $R_{\rm{min}}$ and $R_{\rm{max}}$ vanish, which simplifies the expression for $B_r$ in Eq.~\eqref{eq:Brtp}.
It is then possible to write a simple expression for the amplification of the magnetic field in the shell of the bubble as due to the compression of the field lines.
Substituting Eqs.~\eqref{eq:Displ_LinModel} and~\eqref{eq:dr_lambda} in Eq.~\eqref{eq:Brtp} where we have $\mathbf{\nabla}_t \Lambda = \mathbf{0}$ for spherical bubble, we obtain
\begin{align}
    \left( \frac{B}{B^0} \right) ^2 &= \left(\frac{r - R_{\rm{min}}}{r} \right)^2 \, \left( \frac{R_{\rm{max}}}{R_{\rm{max}} - R_{\rm{min}}} \right)^4 \nonumber \\
    & \phantom{==} \times \, \left(\sin^2\alpha + \cos^2\alpha \, \left( \frac{r - R_{\rm{min}}}{r} \right)^2 \right) \; ,
    \label{eq:BnormSphBubble}
\end{align}
where we introduced $\alpha$, the angle between the direction of the initial magnetic field ($\mathbf{B}^0$) and the unit radial vector $\mathbf{e}_r$ pointing away from the explosion center. We have $\alpha = \arccos(\mathbf{e}_r \cdot \mathbf{B}^0 / |\mathbf{B}^0|)$.

In the triptych of Fig.~\ref{fig:CC_Sph00}, we show the crosscuts through an example of a spherical bubble where the amplification of the magnetic field can be visualized along with the projections of $\mathbf{B}$ in the respective planes. In this example, where $\mathbf{B}^0$ belongs to the $XZ$ plane, the radial and zenith-angle dependence of the amplification can be appreciated.
Equation~\eqref{eq:BnormSphBubble} makes it clear that, for a fixed value of $\alpha$, the strength of the magnetic field ($B = |\mathbf{B}|$) increases from 0 at $r=R_{\rm{min}}$ to its maximum at $r=R_{\rm{max}}$.
We notice that this is a major difference as compared to models for the magnetic field in the shell of spherical bubbles constructed from geometrical arguments (e.g., \citealt{vanderLaan1962}; \citealt{Wolleben2007}; \citealt{Vidal2015}; \citealt{Korochkin2024}), where the strength of the magnetic field is radially constant across the shell.
Our model better matches results from magnetohydrodynamic (MHD) simulations of supernova explosions (e.g., \citealt{Kim2015}; \citealt{Maconi2023}).
In the outer surface of the bubble shell, the minimum value for $B$ occurs at the magnetic poles, for $\alpha = 0$ or $180^\circ$ and is such that $B = B^0$. That is, there is no amplification of the field parallel to the initial magnetic field. This can also be observed in the middle panel of Fig.~\ref{fig:CC_Sph00}.

The dependence of the amplification of the magnetic field as a function of $r$ and $\alpha$ are further illustrated in Fig.~\ref{fig:Bampl} for the same example spherical bubble with an inner radius of 100~pc and a shell thickness of 20~pc.
The maximum amplification happens in the directions perpendicular to $\mathbf{B}^0$, where the field lines are squeezed the most by the explosion. For $\alpha = 90^\circ$ and $r=R_{\rm{max}}$, the strength of the field at the current time is:
\begin{align}
    B = B^0 \, \left( \frac{R_{\rm{max}}}{R_{\rm{max}} - R_{\rm{min}}}\right) \;.
    \label{eq:BmaxSphBubble}
\end{align}
Therefore, the maximum strength of the magnetic field depends on the size of the bubble and the thickness of its shell. The larger the bubble and the thinner its shell, the larger the amplification of the magnetic field strength.

By construction, the magnetic field inside the bubble ($r<R_{\rm{min}}$) is set to zero while it is unchanged outside the bubble. Therefore, the amplification factor is respectively 0 and 1 in these regions. We note that, due to our choice for the displacement vector field, the final magnetic field is radially discontinuous at $r = R_{\rm{max}}$. This is true for any bubble shape. This discontinuity could be resolved by choosing a continuous form for the radial dependence of $\Lambda$, which we will address in future work.

Finally, we note that our model predicts a relation between the amplitude of the magnetic field and the density within the bubble shell. The comparison of Eq.~\eqref{eq:ShellDensity} and Eq.~\eqref{eq:BnormSphBubble} indicates that the $B - \rho$ relation depends on the angle $\alpha$. We have $B \propto \rho^{1/2}$ at the magnetic equator, when $\alpha = 90^\circ$. The exponent slowly increases as $\alpha$ decreases and asymptotically converges to the value of 1 at the magnetic poles ($\alpha = 0^\circ$ or $180^\circ$). As such, the explosion-induced compression of the matter density and amplification of the strength of the magnetic field imply a $B - \rho$ relation that is close to $B \propto \rho^{1/2}$ for most of the locations in the shell of the bubble. This relation corresponds to the case of lateral compression of elongated filaments of the ISM that are not parallel or perpendicular to their permeating magnetic field lines (e.g., \citealt{Tritsis2015}; \citealt{Seta2022}). It also agrees with results from MHD simulations of supernova explosions (e.g., \citealt{Maconi2023}).

\begin{figure*}
    \centering
    \includegraphics[trim={1.1cm 0.6cm 3.0cm 1.2cm},clip,height=.31\linewidth]{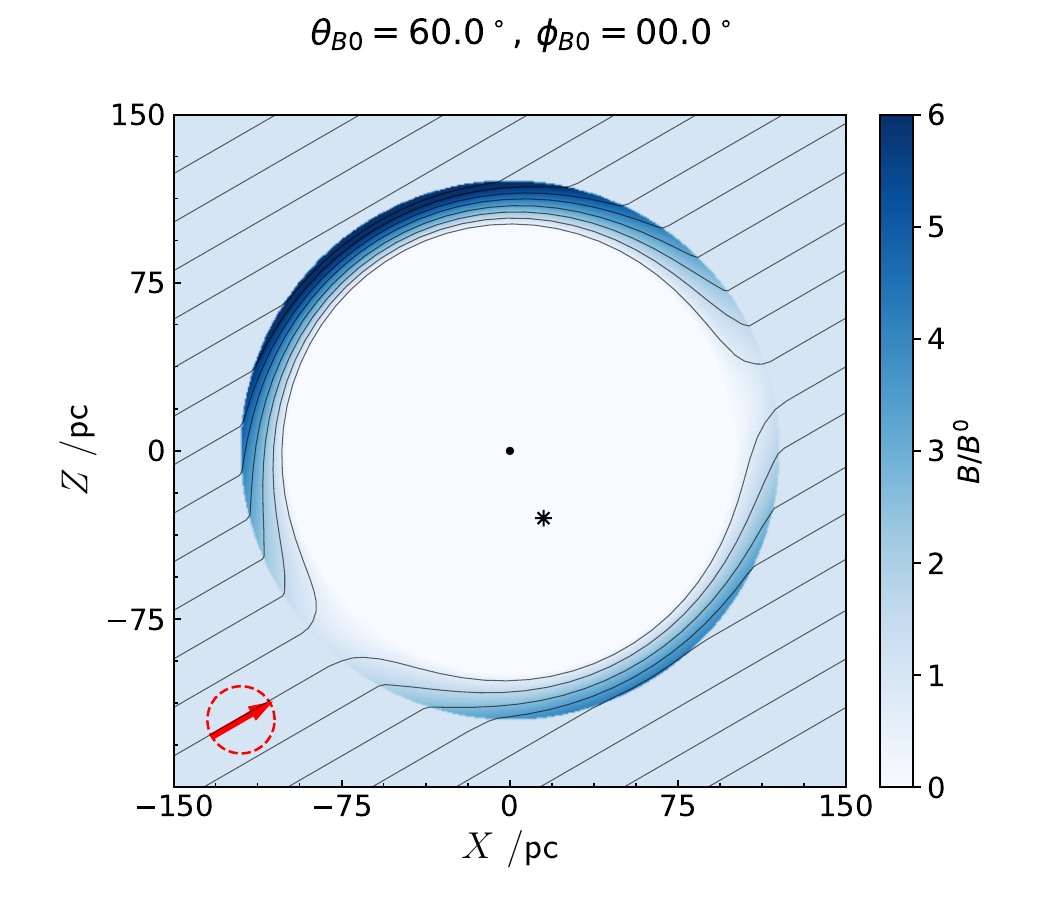}
    \includegraphics[trim={1.1cm 0.6cm 3.0cm 1.2cm},clip,height=.31\linewidth]{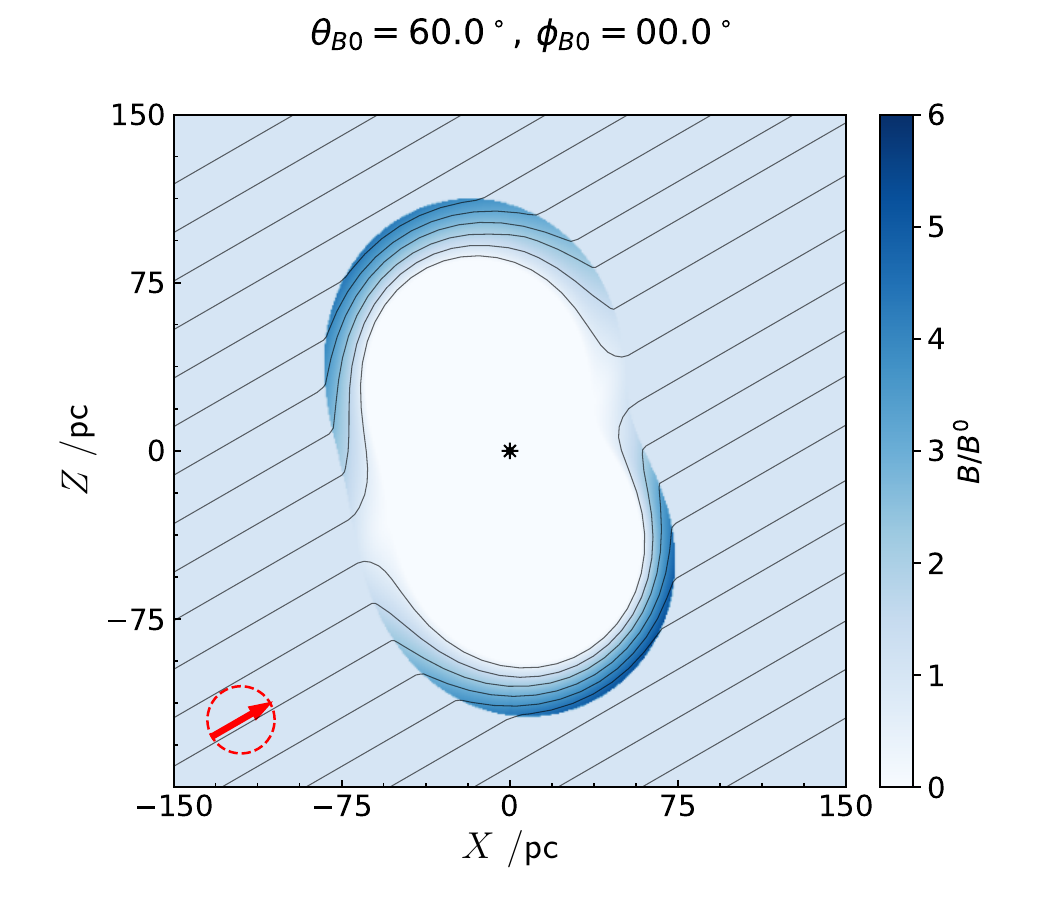}
    \includegraphics[trim={1.1cm 0.6cm 1.1cm 1.2cm},clip,height=.31\linewidth]{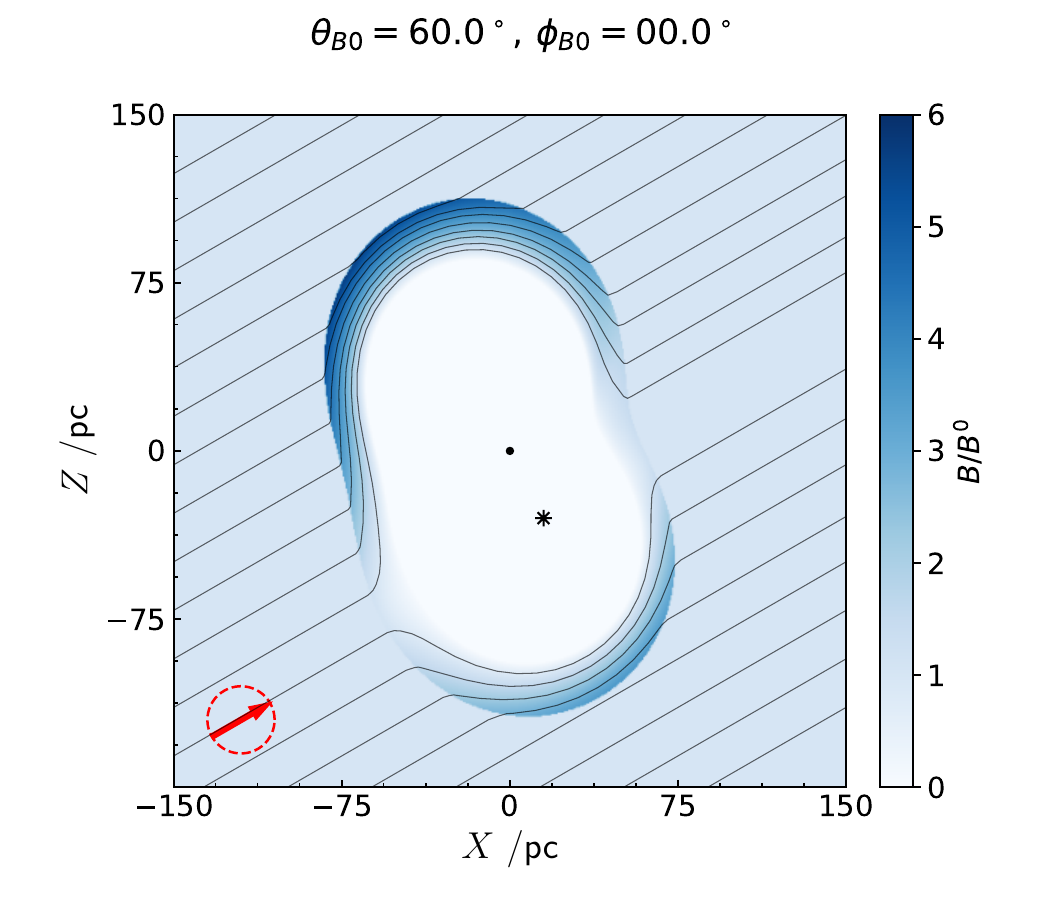}\\[-0.5ex]
    \caption{Crosscuts along the $XZ$ plane through three bubble cases: spherical bubble with off-centered explosion center (left) and non-spherical bubble with the explosion center at the origin of the axes (middle) and off-centered (right). The explosion center is marked with a black asterisk and the origin of the axes with a black dot.
    The bubble shape for the middle and right panels are the same. The direction of the initial magnetic field (also marked with the red arrow) is the same for every panel, and as in Fig.~\ref{fig:CC_Sph00}. $\mathbf{B}^0$ belongs to the $XZ$ plane. Colors and streamlines follow the same convention as in Fig.~\ref{fig:CC_Sph00}.
    }
    \label{fig:CC_otherBubbles}
\end{figure*}
\subsection{The non-spherical cases}
\label{sec:NonSphBubble}
Our equations for the magnetic field in the thick shell of bubbles make it possible to consider cases where the spherical symmetry is lost. This may happen for spherical bubbles but with an off-centered explosion center or, more generally, to non-spherical bubbles.
The general case should occur in the ISM due to a certain degree of anisotropy in the initial velocity field related to the explosion of the progenitor supernovae, to inhomogeneous initial density in which the supernovae remnant develops (\citealt{Kim2015}), or also to other feedback reactions from the compressed matter (e.g., \citealt{Ferriere1991a}; \citealt{Lopez2011}; \citealt{Orlando2016}).
To determine the magnetic field in such cases, it is necessary to estimate the angular derivatives of the inner and outer surfaces of the bubble shell.
For this purpose, it is worth noticing that the orthoradial components of the gradient of the surfaces are also given by the orthoradial components of the normal vectors to the surfaces. Introducing $\mathbf{n}(\theta,\phi)$, the unit normal vector to a close surface $R(\theta,\phi)$ of $\mathbb{R}^3$ which surrounds the origin, we have:
\begin{align}
    \mathbf{\nabla}_t \, R = \mathbf{e}_r - \frac{\mathbf{n}}{\mathbf{e}_r \cdot \mathbf{n}}
\end{align}
in the direction $\mathbf{e}_r$ given by $(\theta,\phi)$.
Therefore, the two orthoradial components of the gradient of $R$ in the direction $\mathbf{e}_r$ are related to the normal vector as:
\begin{align}
    \mathbf{\nabla}_t \Lambda \cdot \mathbf{e}_\theta &= \frac{1}{r}\,\frac{\partial \Lambda}{\partial \theta} = - \frac{\mathbf{n} \cdot \mathbf{e}_\theta}{\mathbf{n} \cdot \mathbf{e}_r} \\
    \mathbf{\nabla}_t \Lambda \cdot \mathbf{e}_\phi &= \frac{1}{r\sin\theta}\,\frac{\partial \Lambda}{\partial \phi} = - \frac{\mathbf{n} \cdot \mathbf{e}_\phi} {\mathbf{n} \cdot \mathbf{e}_r} \;.
\end{align}
If $R_{\rm{min}}$ and $R_{\rm{max}}$ can observationally be determined for a given bubble shell (e.g. see Appendix~\ref{sec:LBthickShell}), and if the explosion center is given, the magnetic field at a location $(r,\theta,\phi)$ in the shell can be estimated through Eq.~\eqref{eq:Brtp} using the normal vectors of $R_{\rm{min}}$ and $R_{\rm{max}}$ toward $(\theta,\phi)$ and substituting the above equations in Eq.~\eqref{eq:dxi_lambda}.
It is interesting to note that, even if the angular derivative of $R_{\rm{min}}$ and $R_{\rm{max}}$ differ for a given direction, what happens as soon as the shell thickness varies, the orthoradial component of the magnetic field (as seen from the explosion center) will keep the same throughout the shell thickness and is fully determined by the orthoradial component of the initial magnetic field. This is because $B_\phi / B_\theta = B^0_{~\phi} / B^0_{~\theta}$, as derived from Eq.~\eqref{eq:Brtp}.

In Fig.~\ref{fig:CC_otherBubbles} we show examples of magnetic fields obtained for cases where the spherical symmetry is lost. We consider the same spherical bubble as in Fig.~\ref{fig:CC_Sph00}, but with an off-centered explosion, and a more convoluted bubble shape for two possible positions of the explosion center. The comparison of these examples, also with the middle panel of Fig.~\ref{fig:CC_Sph00}, illustrates very well the leading dependence of the magnetic field amplification ratio with ($i$) the amount of swept-up matter (proportional to $R_{\rm{max}}$ as measured from the explosion center) and ($ii$) the thickness of the shell (inversely proportional to $R_{\rm{max}} - R_{\rm{min}}$). It is worth noting, however, that for non-spherical bubble shell, the gradient term ($\mathbf{\nabla}_t \Lambda
\, \cdot \mathbf{B}^0_{~t}$) in the expression for $B_r$ alter the simple picture given by Eqs.~\eqref{eq:BnormSphBubble} and~\eqref{eq:BmaxSphBubble}.
As a result, and as also clearly illustrated in Fig.~\ref{fig:CC_otherBubbles}, an off-centered explosion center may lead to significant asymmetry in the amplitude of the magnetic field. This asymmetry also depends on the geometry of the bubble shell.

\section{Application to the Local Bubble shell and its contribution to polarization observables}
\label{sec:LBcase}
Among all the bubbles and super-bubbles in the ISM, the Local Bubble is certainly a special case as the Solar system currently resides inside it.
In this second part of the paper, we use our analytical model to carry out a phenomenological study with the magnetized shell of the Local Bubble as a study case. In particular, we focus on the impact of the shape of the shell and the choice of the effective explosion center on predictions of the Local Bubble shell's contribution to sky maps of Faraday RM and synchrotron polarized emission.
In this theoretical paper, we do not intend to model the full sky based on this model, nor incorporate it into models for the large-scale GMF and perform a full exploration of the parameter space. Such an endeavor is deferred to future work.
Instead, we merely estimate the possible contribution of the Local Bubble's thick shell to observables for specific study cases.
To this end, and as explained below, we combine publicly available data on the shape of the Local Bubble from dust data, and reasonable values found in the literature for the free parameters to describe properties of the ISM and explosion.
Therefore throughout this section we assume that the Local Bubble as it stands today can be described by a single, effective explosion which happened in a homogeneous medium permeated by a uniform magnetic field. This is
a working assumption that only data can refute.
We ignore possible effects from the turbulent component of the magnetic field and from inhomogeneities in the initial matter density.
These are crude simplifications, the effects of which will need to be studied in future work.

\smallskip

To predict the contribution from the thick shell of the Local Bubble to the Faraday RM and polarized synchrotron emission using our analytical model (Eq.~\eqref{eq:Brtp} and~\eqref{eq:Displ_LinModel}), the following ingredients are needed.
Firstly, the strength and direction of the initial magnetic field (before the explosion) need to be set. This is done by fixing the three parameters $B^0$, $l_{B^0}$ and $b_{B^0}$. $l_{B^0}$ and $b_{B^0}$ are the Galactic longitude and latitude of the spherical coordinate system centered on the Sun. An initial magnetic field with $b_{B^0} = 90^\circ$ would point to the North Galactic pole and $(l_{B^0},b_{B^0}) = (0^\circ,\,0^\circ)$ would point toward the Galactic center. The two angles give the initial directions of the magnetic field and $B^0$ its amplitudes.
Secondly, the shapes of the inner and outer surfaces of the shell need to be given as input. As explained in Appendix~\ref{sec:LBthickShell}, we choose to use shell surfaces derived by \cite{Pelgrims2020} from the 3D dust map of \cite{Lallement2019}.
Thirdly, the location of the effective explosion center needs to be set. This is done by setting its Cartesian coordinates $(x_c,\,y_c,\,z_c)$ in the (Galactic) Heliocentric reference frame where the Galactic center (longitude 0) is toward positive $x$ and the Galactic North pole is toward positive $z$.
Using our analytical model, the three above inputs (six parameters plus inner and outer surfaces) make it possible to estimate the magnetic field anywhere in the present-time shell of the Local Bubble.
Finally, to produce maps of the Faraday RM and of the synchrotron Stokes parameters $Q$ and $U$, we need to adopt models for the thermal electron density, for the cosmic-ray electron (CRE) density, and for the energy-spectrum of the CRE. Our choices are described next.

\begin{table*}
    \centering
    Definition of Local Bubble scenarios\\[1.ex]
    \begin{tabular}{lcccccc}
    \hline \hline \\[-2ex]
         Names & \texttt{SCO} & \texttt{SCA} & \texttt{DCO} & \texttt{DCA} & \texttt{DDO} & \texttt{DDA} \\ 
         $R_{\rm{min}}$ & $R_{\rm{sph}}$ & $R_{\rm{sph}}$ & $R_{\rm{in}}^{\rm{O}} \coloneqq~$\texttt{O}$~\longleftarrow r_{\rm{inner}}^{\ell_{\rm{max}}=6}$ & \texttt{A}$~\longleftarrow R_{\rm{in}}^{\rm{O}}$ & \texttt{O}$~\longleftarrow r_{\rm{inner}}^{\ell_{\rm{max}}=6}$ & \texttt{A}$~\longleftarrow r_{\rm{inner}}^{\ell_{\rm{max}}=6}$ \\[.5ex]
         $R_{\rm{max}}$ & $R_{\rm{sph}} + \Delta_{\rm{sph}}$  & $R_{\rm{sph}} + \Delta_{\rm{sph}}$ & $R_{\rm{max}}^{\rm{O}} \coloneqq R_{\rm{in}}^{\rm{O}} + \Delta_{\rm{sph}}$ & \texttt{A}$~\longleftarrow R_M^{\rm{O}}$ & \texttt{O}$~\longleftarrow r_{\rm{outer}}^{\ell_{\rm{max}}=6}$ & \texttt{A}$~\longleftarrow r_{\rm{outer}}^{\ell_{\rm{max}}=6}$ \\[.5ex]
         $(x_c,y_c,z_c)$ & $(x_{\rm{sph}},y_{\rm{sph}},z_{\rm{sph}})$  & $(x_c^{\rm{P20}},\,y_c^{\rm{P20}},\,z_c^{\rm{P20}})$ & $(x_{\rm{sph}},y_{\rm{sph}},z_{\rm{sph}})$ & $(x_c^{\rm{P20}},\,y_c^{\rm{P20}},\,z_c^{\rm{P20}})$ & $(x_{\rm{sph}},y_{\rm{sph}},z_{\rm{sph}})$ & $(x_c^{\rm{P20}},\,y_c^{\rm{P20}},\,z_c^{\rm{P20}})$ \\[1.ex]
         \hline
    \end{tabular}\\[2.5ex]
    Model parameters\\[.5ex]
    \begin{tabular}{cccccccccccc}
    \hline \hline \\[-2ex]
         $x_{\rm{sph}}$ & $y_{\rm{sph}}$ & $z_{\rm{sph}}$ & $R_{\rm{sph}}$ & $\Delta$ & $\phantom{x}$ & $x_c^{\rm{P20}}$ & $y_c^{\rm{P20}}$ & $z_c^{\rm{P20}}$ & $\phantom{x}$ & $l_{B^0}$ & $b_{B^0}$ \\[.5ex]
         (pc) & (pc) & (pc) & (pc) & (pc) & $\phantom{x}$ & (pc) & (pc) & (pc) & $\phantom{x}$ & ($^\circ$) & ($^\circ$) \\[.5ex] \hline\\[-.5ex]
         -24.8 & -32.6 & -23.3 & 216.7 & 35 & $\phantom{x}$ & 23 & -34 & -122 & $\phantom{x}$ & 73 & 17 \\[.5ex]
         \hline\\[-2.5ex]
    \end{tabular}
    \caption{(top) Summary of the definition of the scenarios to compute the present-time magnetic field in the thick shell of the Local Bubble. (bottom) Values for the several parameters as defined in the text. In the top table, $R_{\rm{min}}$ and $R_{\rm{max}}$ are the inner and outer radius of the bubble shell as measured from the explosion center in $(x_c,y_c,z_c)$. The left arrow ($\longleftarrow$) indicates that a surface is shifted to the coordinate system with center \texttt{O}$~\coloneqq (x_{\rm{sph}},y_{\rm{sph}},z_{\rm{sph}})$ or \texttt{A}$~\coloneqq (x_c^{\rm{P20}},y_c^{\rm{P20}},z_c^{\rm{P20}})$, where the superscript P20 denotes the values derived in \cite{Pelgrims2020}
    (see their Fig.~9 and Table~1).
    }
    \label{tab:ScenarioAndParams}
\end{table*}
\subsection{Study cases}
\label{sec:LBscenarios}
As evidenced in Sect.~\ref{sec:AnalyticModel}, the present-time magnetic field in the shell of the Local Bubble may severely depend on the specific geometry of the shell (shape, size, and thickness) and the position of the explosion center.
We investigate the possible differences that specific choices imply on polarization observables by constructing six scenarios.

The first type consists of spherical bubble shells (with constant thickness). Their names start with \texttt{SC}.
For the second type, the bubble shells have their inner surface directly corresponding to the model derived by \cite{Pelgrims2020} from 3D dust data and obtained with a maximum multipole $l_{\rm{max}}=6$ (i.e. $r_{\rm{inner}}^{\ell_{\rm{max}}=6}$). Their names start with a \texttt{D}. For this second type, we consider the case where the shell thickness (as measured radially from a certain position) is constant (named \texttt{DC}) and the case where it varies (named \texttt{DD}). For this latter case, we adopt the outer surface of the shell ($r_{\rm{outer}}^{\ell_{\rm{max}}=6}$) as determined in Appendix~\ref{sec:LBthickShell}, also from the same 3D dust data.

To fix the center and the inner radius of the sphere for the spherical bubble-shell scenarios, we choose to fit a sphere to the 3D data points drawn from $r_{\rm{inner}}^{\ell_{\rm{max}}=6}$. As such, our spherical model for the Local Bubble is also derived from 3D dust data. In the Heliocentric Cartesian coordinate system, the center of the sphere is found at $(x_{\rm{sph}},\,y_{\rm{sph}},\,z_{\rm{sph}}) = (-24.8,\,-32.6,\,-23.3)$~pc and its radius is $R_{\rm{sph}} = 216.7$~pc.
For the constant thickness of the shell, we adopt the value of $\Delta = 35$~pc, in agreement with the model of \cite{Yao2017}, and which is not too far from the best-fit value of 30~pc obtained in \cite{Korochkin2024}. For the case named \texttt{DC} we choose $(x_{\rm{sph}},\,y_{\rm{sph}},\,z_{\rm{sph}})$ as the point from which the shell thickness, as measured radially, is constant.

\smallskip

\cite{Pelgrims2020} used $r_{\rm{inner}}^{\ell_{\rm{max}}=6}$ to model the high-Galactic latitudes ($|b| \geq 60^\circ$) of the dust polarized emission measured by the \textit{Planck} satellite at 353~GHz.
They obtained constraints on the position of the center of the effective explosion which has led to the Local Bubble, and on the orientation of the initial magnetic field. Because dust-polarized emission is not directly sensitive to the amplitude of the magnetic field, their solution for the position of the explosion center is degenerate along a line parallel to the initial direction of the magnetic field.
We adopt the solution which they found maximizes the likelihood (obtained using $r_{\rm{inner}}^{\ell_{\rm{max}}=6}$), and which approximately reads as $(x_c^{\rm{P20}},\,y_c^{\rm{P20}},\,z_c^{\rm{P20}}) = (23,\,-34,\,-122)$~pc and $(l_0^{\rm{P20}},\,b_0^{\rm{P20}}) = (73^\circ,\,17^\circ)$.

We notice that the explosion center found in \cite{Pelgrims2020} is significantly different than the center of the sphere that best fits $r_{\rm{inner}}^{\ell_{\rm{max}}=6}$.
Therefore, we consider both locations as possible explosion centers to build our study cases. The third letter of our scenarios' names indicates this choice. The letter \texttt{O} corresponds to explosion center at $(x_{\rm{sph}},\,y_{\rm{sph}},\,z_{\rm{sph}})$ and letter \texttt{A} corresponds to explosion center at $(x_c^{\rm{P20}},\,y_c^{\rm{P20}},\,z_c^{\rm{P20}})$.

In total, we are thus considering six scenarios to compute the present-time magnetic field in the thick shell of the Local Bubble. The definition of these scenarios, with our choices for their parameter values, are summarized in Table~\ref{tab:ScenarioAndParams}.

To fully determine the magnetic field in the shell of the Local Bubble for our scenarios, we need to set the value for the strength of the initial magnetic field, before the explosion happened.
In the ensemble of models for the large-scale Galactic magnetic field presented by \cite{Unger24}, the coherent component has a field strength which ranges from 0.2 to 0.5~$\mu$G at the Sun location. However, due to the turbulent component of the magnetic field, the actual strength for the initial magnetic field may have been much larger.
We adopt a value of  3~$\mu$G for the strength of the initial field ($B^0$). This is on the lower limit of the turbulent field strength of 3 to 6~$\mu$G found for the disk by \citealt{PlanckInt2016XLII} who updated the model of \citealt{Jansson2012b}. On the other hand, \cite{Korochkin2024} found a value of 3.5~$\mu$G for the strength of the GMF in the Solar neighborhood, which is close to our choice for $B^0$.
We emphasize on the fact that, in our simple model, the strength of the initial magnetic field acts only as a scaling factor. A scan over possible values for $B^0$ in the range from zero to 5~$\mu$G is also performed in Sect.~\ref{sec:DiscussionLBscenarios}.

\subsection{Faraday RM and synchrotron Stokes parameters}
\label{sec:RMQUobservables}
Free (thermal) electrons of the magnetized ISM induce a net rotation of the polarization plane of an electromagnetic wave passing through the medium before it reaches the observer. This Faraday rotation is proportional to the square of the wavelength and is characterized by the RM. The RM results from the line-of-sight integration, from the observer to the source at distance $d$, of the product of the thermal electron and the line-of-sight component of the magnetic field:
\begin{align}
    {\rm{RM}} \propto - \int_0^d {\rm{d}}r \, n_e(r) \, B_{||}(r) \; ,
    \label{eq:RMeq}
\end{align}
where $n_e$ is the number density of thermal electrons and $B_{||} = \mathbf{B}\cdot\mathbf{e}_r$ is the line-of-sight component of the magnetic field, where $\mathbf{e}_r$ is the radial unit vector pointing outward from the origin of the spherical coordinate system centered on the observer. To express the RM in rad/m$^2$, and if the magnetic field is expressed in microgauss, the distance in parsec, and the thermal electron in cm$^{-3}$, the proportionality factor takes the value of $\approx 0.81 \, ({\rm{rad/m^2}})({\rm{cm^3/pc}})(\mu{\rm{G}})^{-1}$.

To predict the contribution of the Local Bubble shell to the RM sky map, we thus need to adopt a model for the thermal electron density.
Thermal electron density models of \cite{Cordes2002} and \cite{Yao2017} predict a local value for $n_e$ of 0.011~cm$^{-3}$ and 0.016~cm$^{-3}$, respectively. These values were obtained for the diffuse ISM, without the density depletion within the Local Bubble as implemented in these models. The mid-plane density of the electron density in the local Galactic disk as determined by \cite{Ocker2020} is $n_e = 0.015\,{\rm{cm}}^{-3}$. It is within the range of the two above models and we adopt it as our default value for the initial, homogeneous thermal electron density ($n_e^0$), that is, for the thermal electron density before the effective supernovae explosion which has led to the Local Bubble.
According to our simple model for the formation of the Local Bubble, the thermal electron density distribution is modified by the explosion. We assume that the thermal electron density follows Eq.~\eqref{eq:ShellDensity} and that the fraction of ionized to total gas remains constant so that the shell properties as inferred from 3D dust map apply for the thermal electron density.
We notice that in this framework, changing the explosion center inside a given geometry for a bubble shell leads to different thermal electron densities in the shell.

\medskip

Relativistic electrons in the ISM spiral about the ambient magnetic field lines and loose energy through synchrotron emission which is linearly polarized perpendicular to the orthoradial component of the magnetic field (Eq.~\ref{eq:orthoradialComp}). Relativistic electrons with energies on the order of 10~GeV are responsible for the Galactic synchrotron emission observed in the frequency range of 10 to 30~GHz, well covered by cosmic microwave background experiments, and frequencies at which Faraday rotation is negligible.

We consider the simplified assumption that CRE follows approximately a power-law energy distribution of the form:
\begin{align}
    N(E)\,{\rm{d}}E &= \kappa \, E^{-p} \, {\rm{d}}E
        = n_{10} \, \left(\frac{E}{10 \, {\rm{GeV}}}\right)^{-p} {\rm{d}}E \;,
\end{align}
where $p$ is the electron spectral index, and where we adopt a normalization at 10~GeV energy. In this case, the energy spectrum of CRE is related to the cosmic-ray flux by:
\begin{align}
    n_{10} \, \left(\frac{E}{10\,\rm{GeV}}\right)^{-p} = \frac{4 \pi}{c} \, \Phi_{10} \, \left(\frac{E}{10\,\rm{GeV}}\right)^{-p} \;,
\end{align}
where $\Phi_{10} = 10^{-0.55}\,{\rm{Gev}}^{-1}\,{\rm{m}}^{-2}\,{\rm{s}}^{-1}\,{\rm{sr}}^{-1}$ at 10~GeV.
Therefore, we have $n_{10} = 1.2 \times 10^{-23} {\rm cm^{-3} eV^{-1}}$ and $\kappa \approx 3.03 \times 10^{-25} \, {\rm{J^2 \, m^{-3}}}$ for $p=3$.
These numbers were obtained by \cite{Unger24} from the flux in the local ISM (i.e., outside of the heliosphere) as derived from AMS2 data (\citealt{Aguilar2014,Aguilar2015,Aguilar2019}).
Following \cite{Longair2011}, assuming $p=3$ and a uniform CRE density, the simulated Stokes parameters $Q$ and $U$ for the linearly polarized synchrotron emission at a given frequency $\nu$ can be written in terms of line-of-sight integration over solid angle of the plane-of-sky components of the magnetic field as
\begin{align}
    Q_\nu &= - \frac{S_\nu}{4 \pi} \int_0^\infty {\rm{d}}r \, \left( (B_\theta(r)^2 - B_\phi(r)^2 ) \right) \nonumber \\
    U_\nu &= - \frac{S_\nu}{4 \pi} \int_0^\infty {\rm{d}}r \, 2\, B_\theta(r) \, B_\phi(r)  \;,
    \label{eq:QandUeq}
\end{align}
and can be converted to antenna temperature for comparison with data.
For observation at 30~GHz, and if we express the line-of-sight integration in parsec and the magnetic field in microgauss, the amplitude $S_\nu$ in Eq.~\eqref{eq:QandUeq} is $\approx 2.8 \times 10^{-27} \,{\rm{J / (pc \, m^2 \, \mu G^{2})}}$. The "$-$" signs in the above expressions are such that the Stokes parameters are given in the HEALPix convention (positive $Q$ to the South, negative $Q$ to the East, positive $U$ to the South-East, negative $U$ to the North-East), also referred to as COSMO convention (\citealt{Gorski2005}).

Although cosmic rays are not passive during the formation of a supernova remanent (e.g., \citealt{vanderLaan1962}), and may even trigger deformation of the bubble shell due to pressure balance (e.g., \citealt{Ferriere1991b}), we do not consider these effects in this exploratory study. Instead, we assume that, at the present time, the CRE are distributed uniformly as described above. Further work will be carried out to go beyond this simplifying assumption. If CRE density has been enhanced in the shell of the Local Bubble, our calculations provide a lower limit to the possible contribution to the synchrotron polarized emission.

\begin{table}
    \centering
    \begin{tabular}{cccc}
    \hline\hline\\[-1.5ex]
        $B^0$   & $n_e^0$   &   $p$   &   $n_{10}$\\[.5ex]
        ($\mu$G)    & (cm$^{-3}$) & --- & (cm$^{-3}$ eV$^{-1}$)\\[.5ex]
        \hline\\[-1.5ex]
        3   & 0.015 &   3   & $1.2\times10^{-23}$\\[.5ex]
        \hline
    \end{tabular}
    \caption{Fixed parameters used to compute the contribution from the thick shell of the Local Bubble to the Faraday RM and synchrotron Stokes parameters $Q$ and $U$. The parameters are defined in the text.}
    \label{tab:FiducialParams}
\end{table}

\subsection{Results}
\label{sec:ResultsLBscenarios}
\begin{figure*}
    \centering
    \begin{tabular}{lccc}
     $\,$ & RM & $Q$ & $U$ \\
    \rotatebox{90}{\hspace{1.25cm}\texttt{SCO}} &
        \includegraphics[trim={0.4cm 2.4cm 0.4cm 1.9cm},clip,width=.3\linewidth]{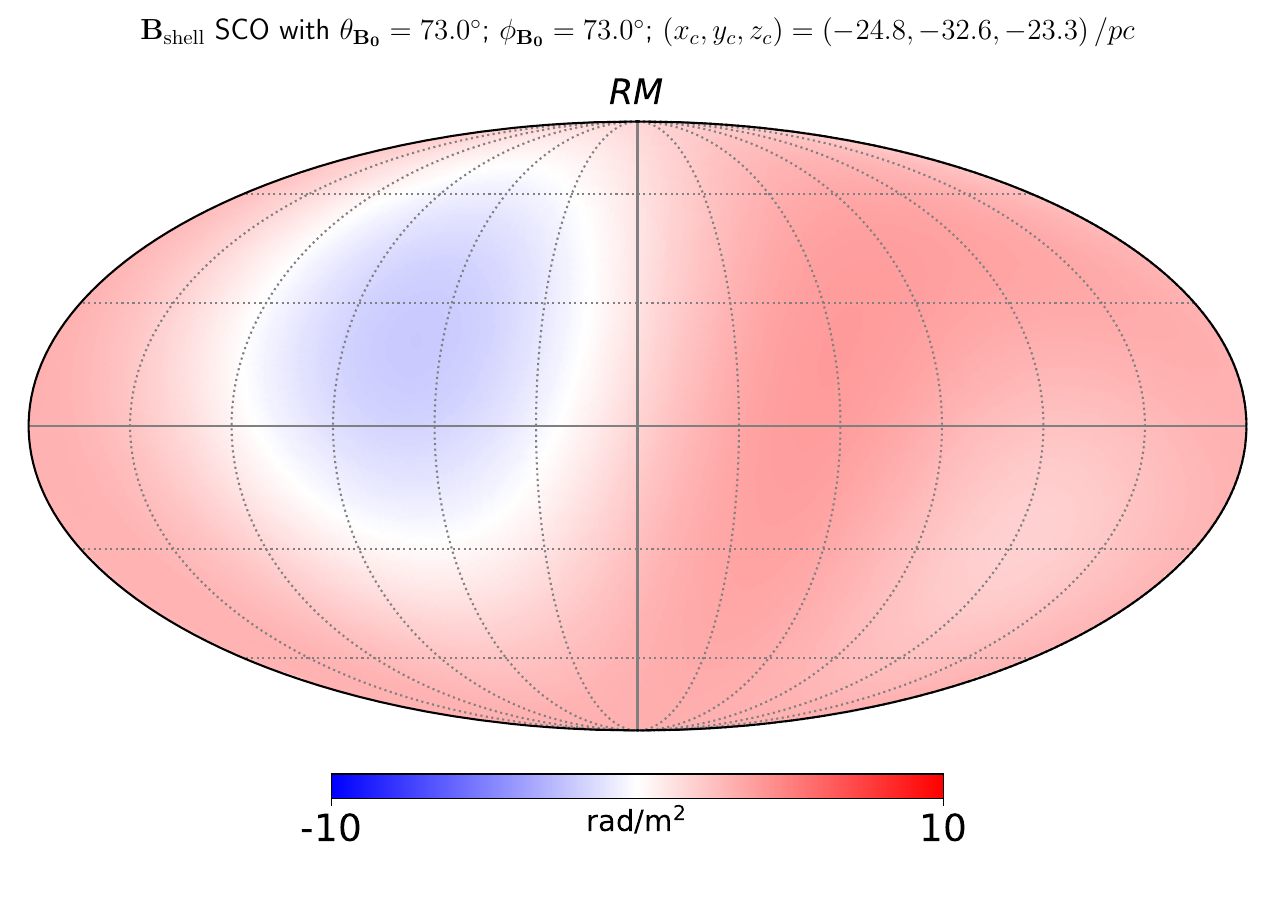} &
        \includegraphics[trim={0.4cm 2.4cm 0.4cm 2.cm},clip,width=.3\linewidth]{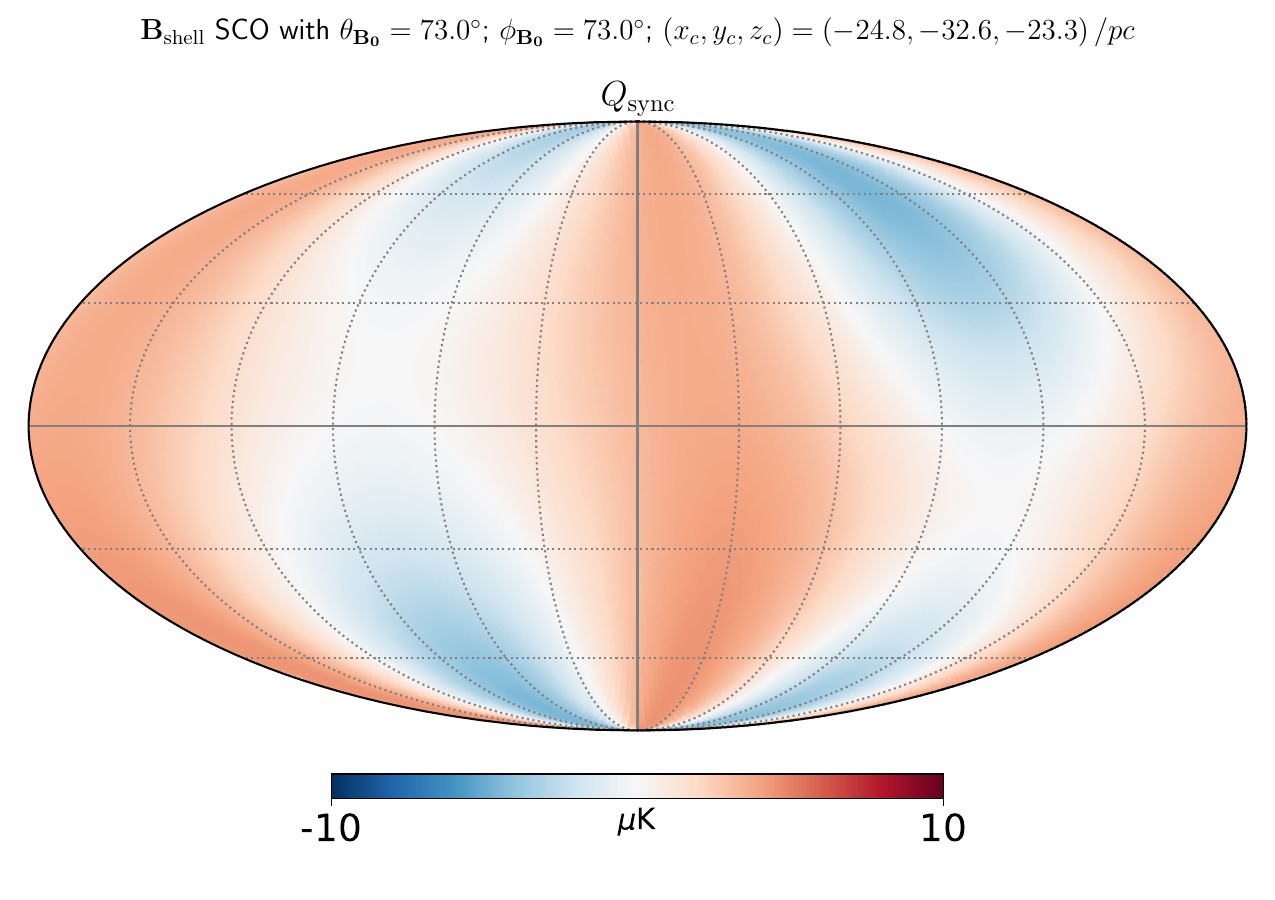} &
            \includegraphics[trim={0.4cm 2.4cm 0.4cm 2.cm},clip,width=.3\linewidth]{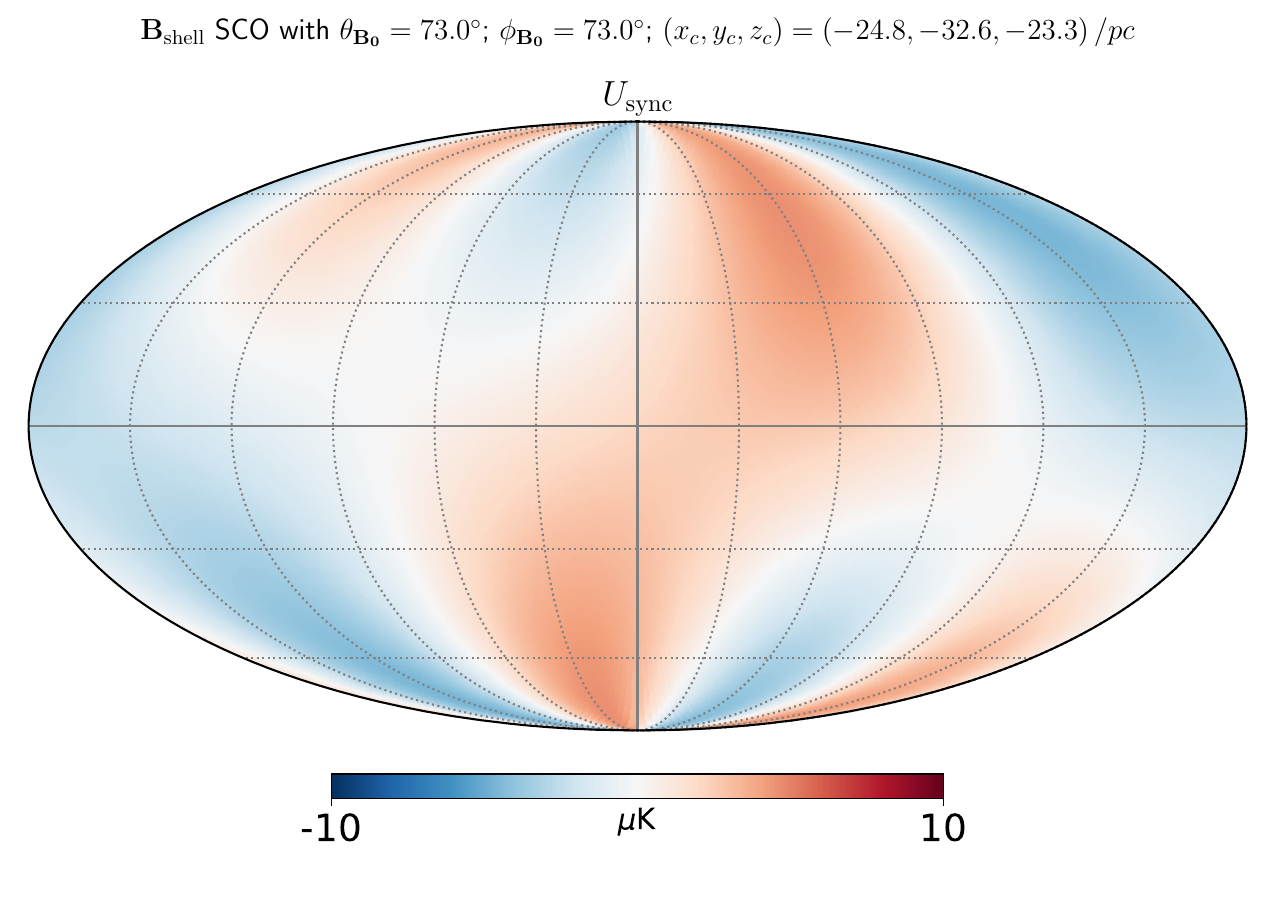} \\
    \rotatebox{90}{\hspace{1.25cm}\texttt{SCA}} &
        \includegraphics[trim={0.4cm 2.4cm 0.4cm 1.9cm},clip,width=.3\linewidth]{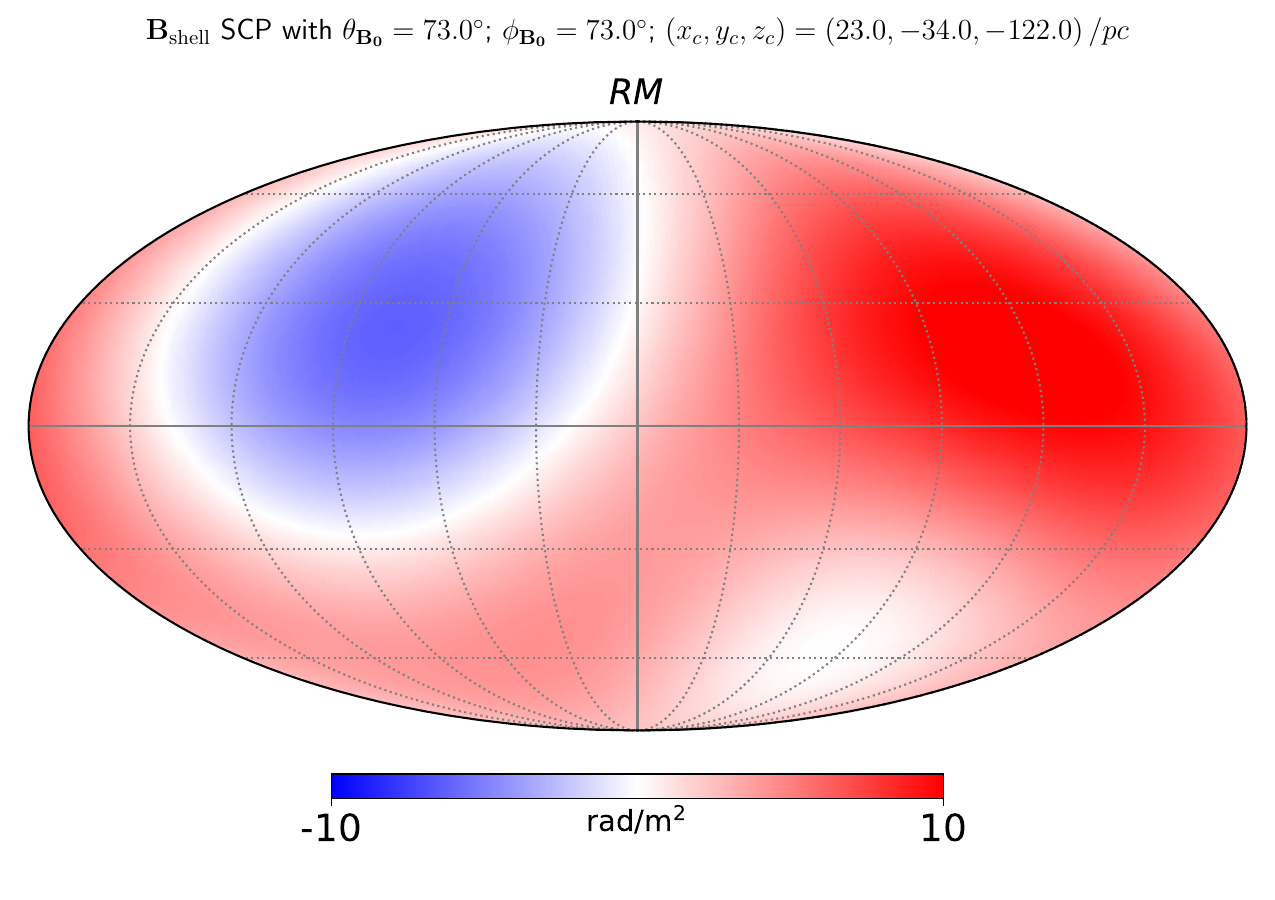} &
        \includegraphics[trim={0.4cm 2.4cm 0.4cm 2.cm},clip,width=.3\linewidth]{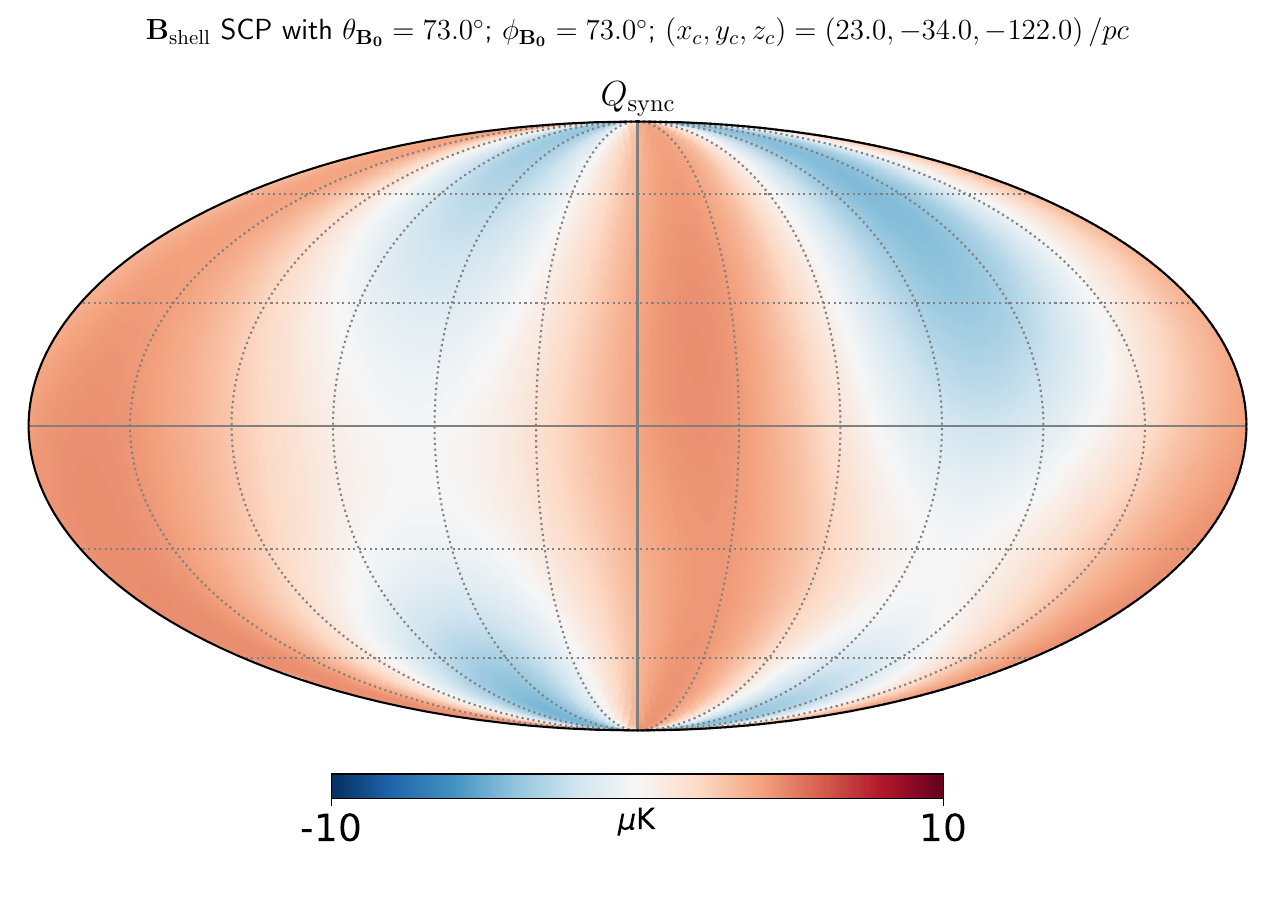} &
            \includegraphics[trim={0.4cm 2.4cm 0.4cm 2.cm},clip,width=.3\linewidth]{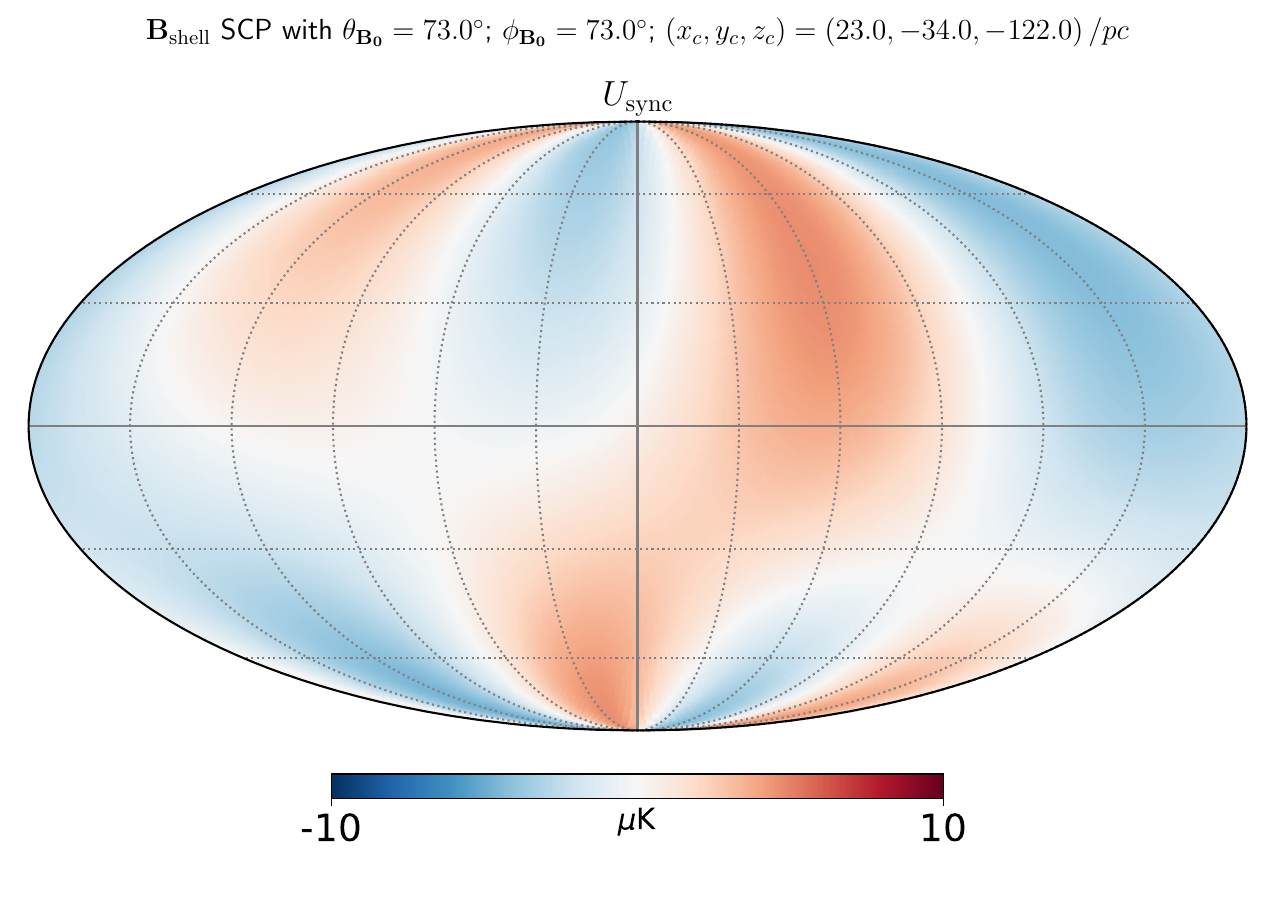} \\
    \rotatebox{90}{\hspace{1.25cm}\texttt{DCO}} &
        \includegraphics[trim={0.4cm 2.4cm 0.4cm 1.9cm},clip,width=.3\linewidth]{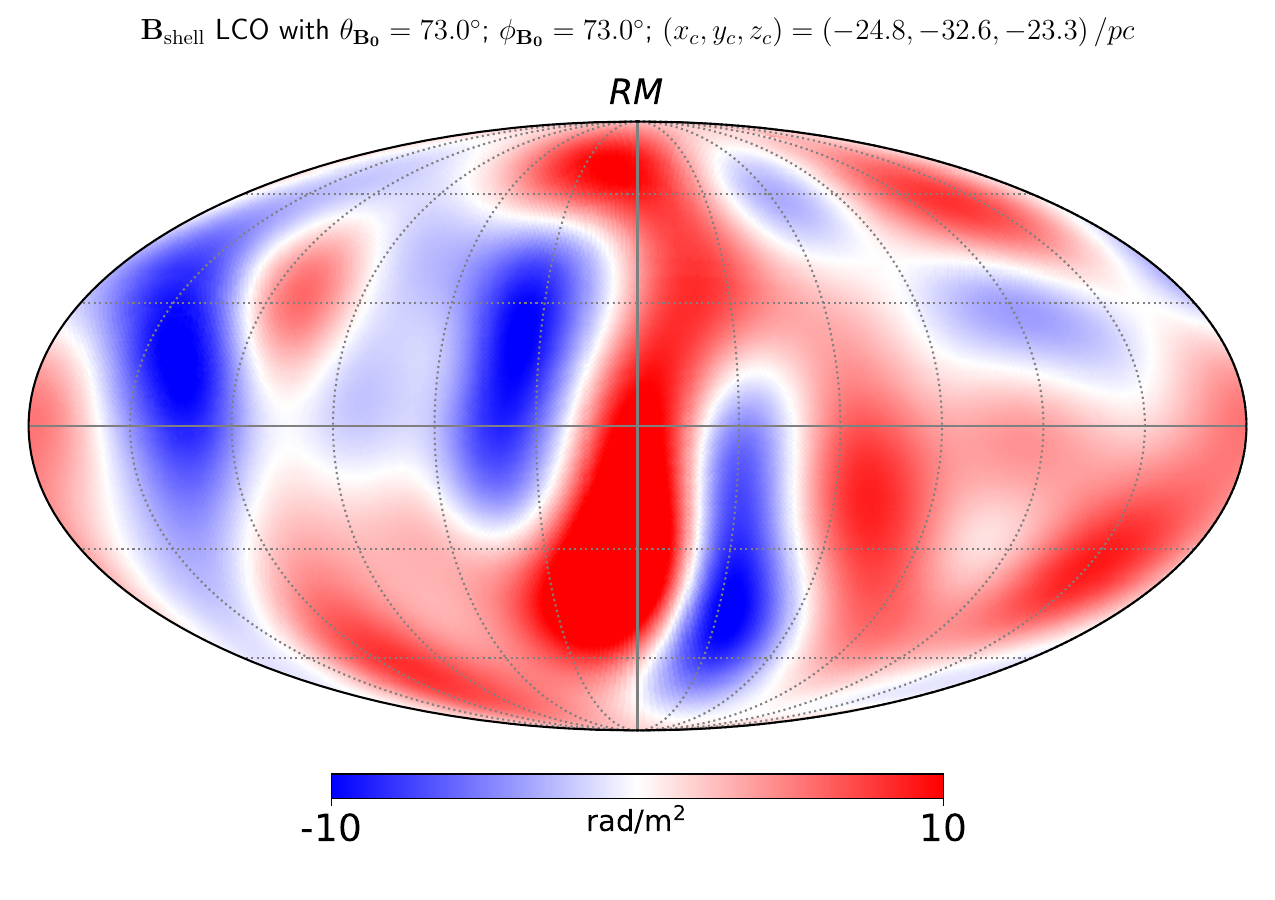} &
        \includegraphics[trim={0.4cm 2.4cm 0.4cm 2.cm},clip,width=.3\linewidth]{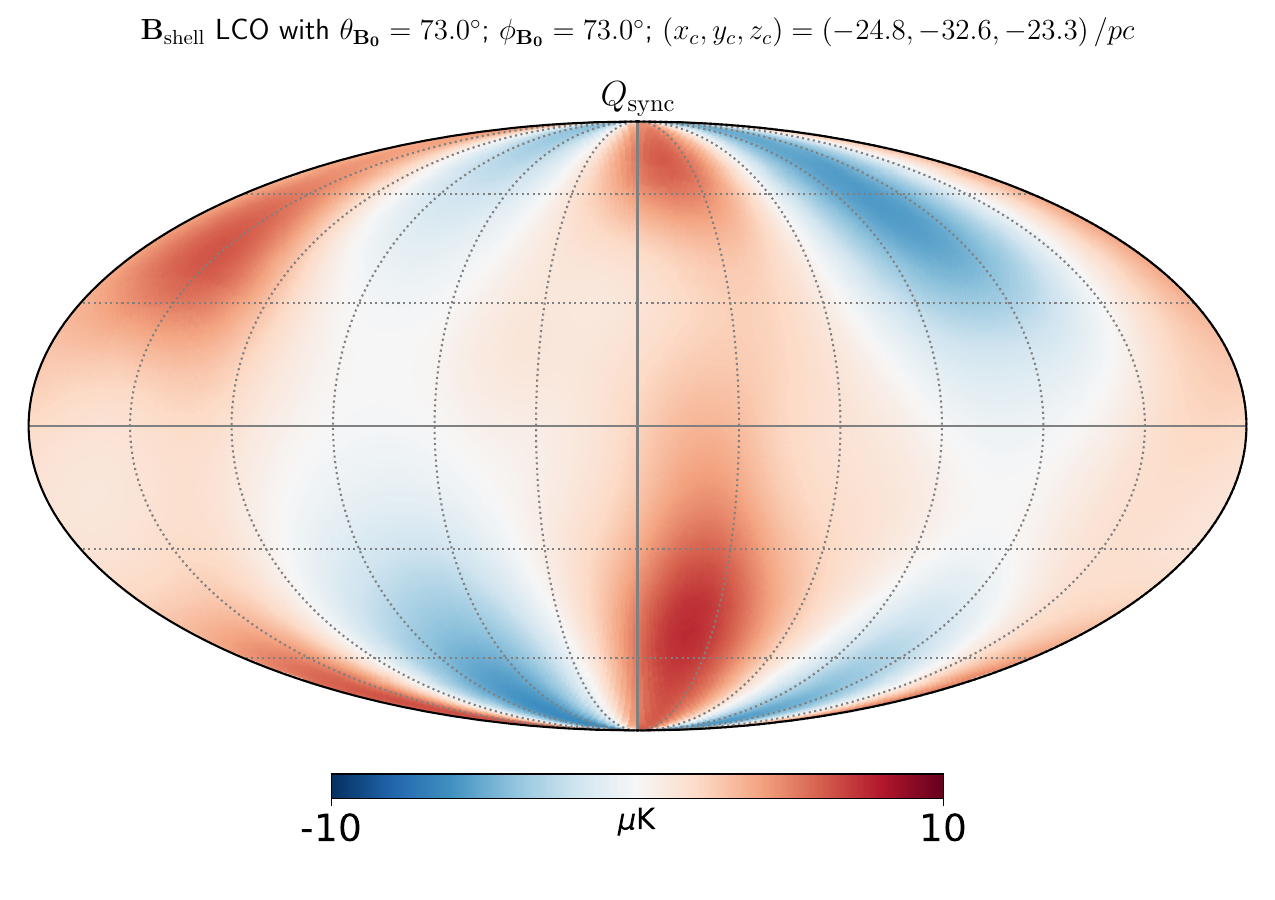} &
            \includegraphics[trim={0.4cm 2.4cm 0.4cm 2.cm},clip,width=.3\linewidth]{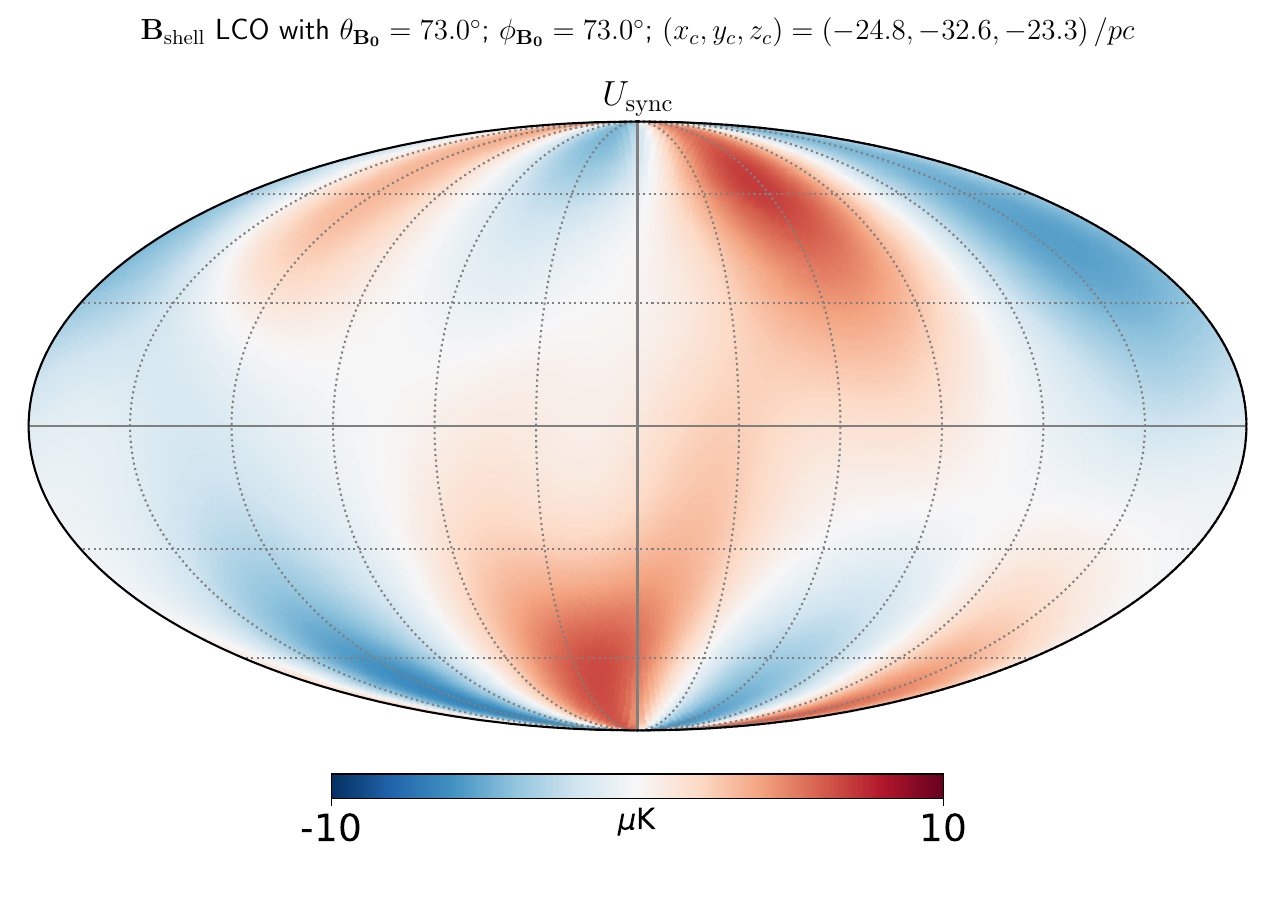} \\
    \rotatebox{90}{\hspace{1.25cm}\texttt{DCA}} &
        \includegraphics[trim={0.4cm 2.4cm 0.4cm 1.9cm},clip,width=.3\linewidth]{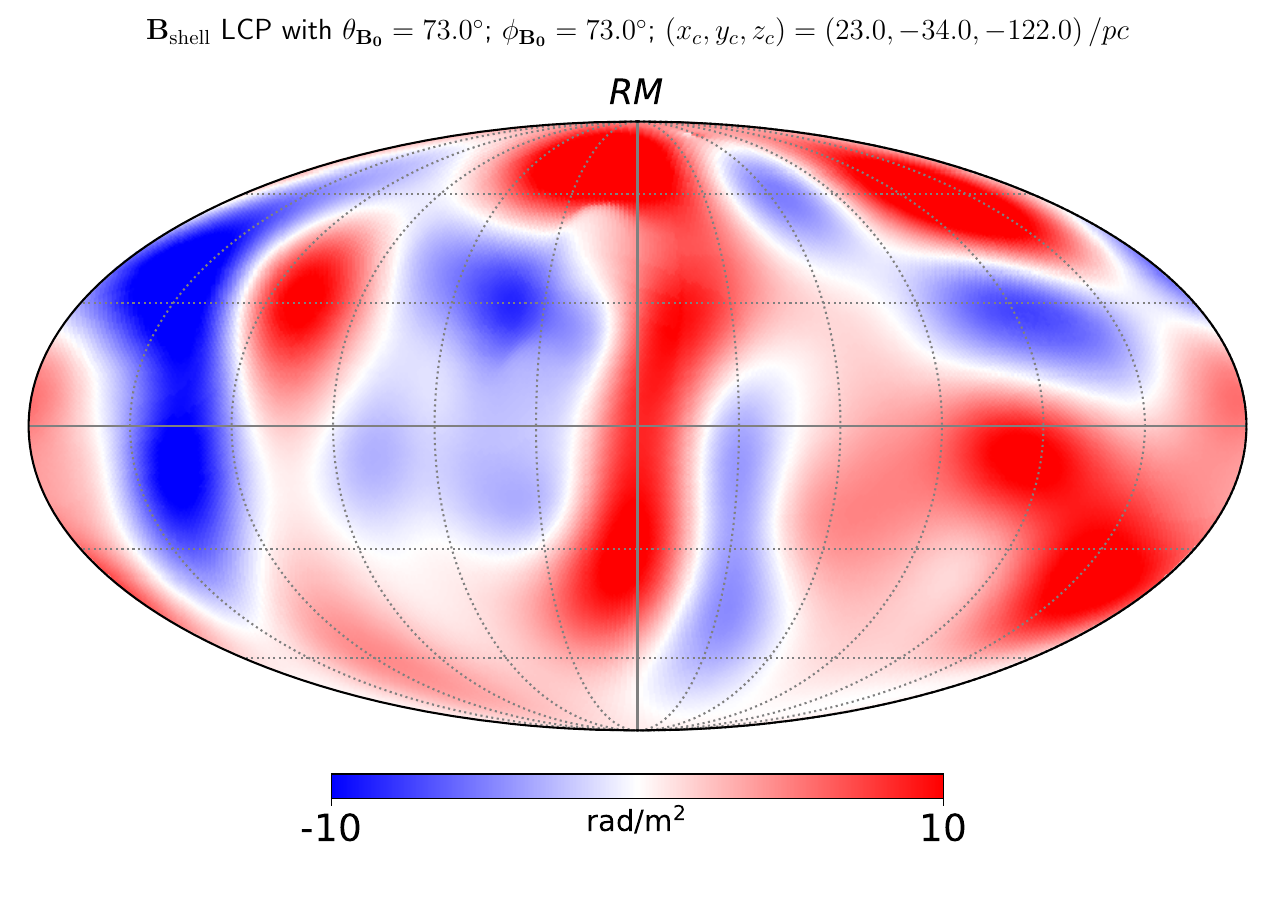} &
        \includegraphics[trim={0.4cm 2.4cm 0.4cm 2.cm},clip,width=.3\linewidth]{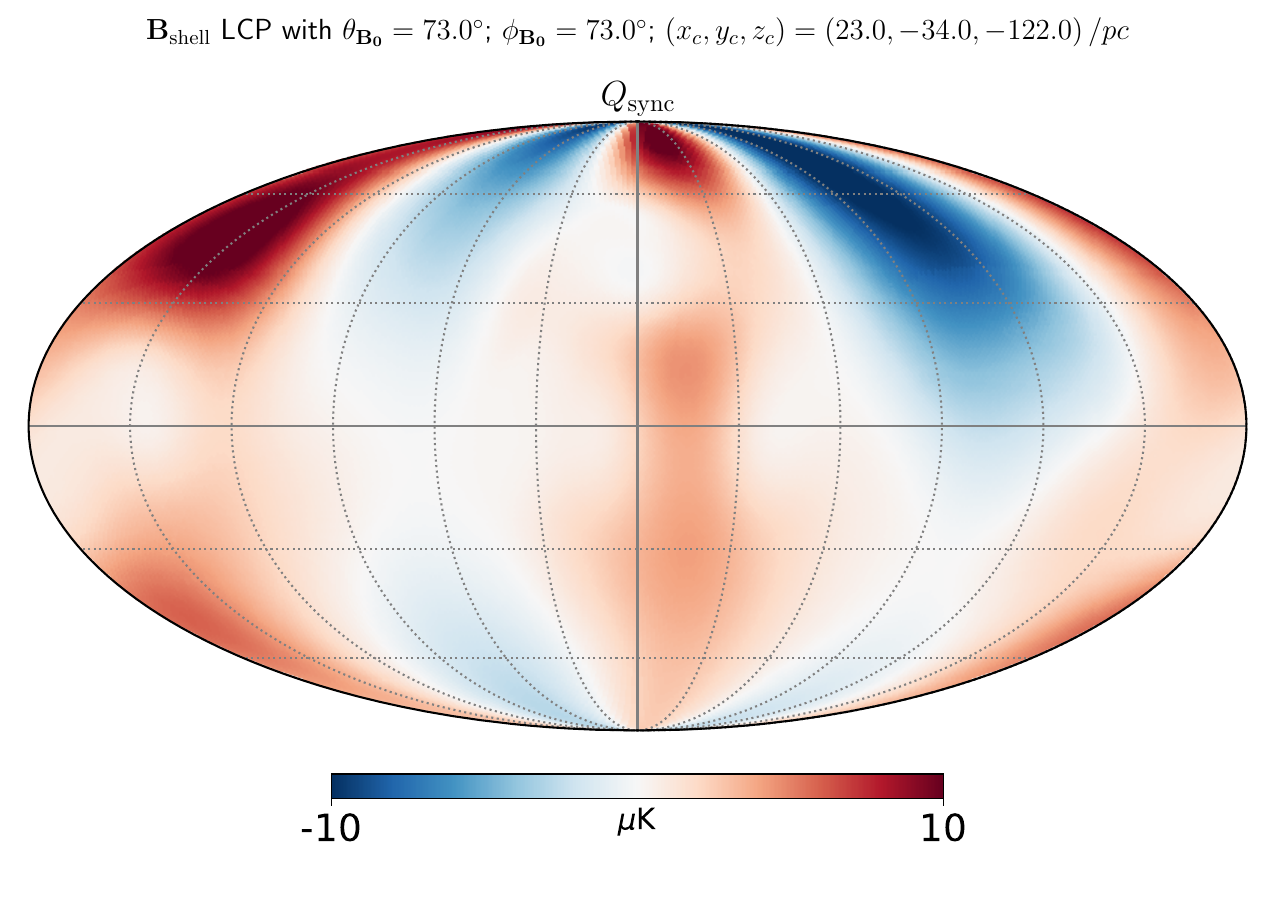} &
            \includegraphics[trim={0.4cm 2.4cm 0.4cm 2.cm},clip,width=.3\linewidth]{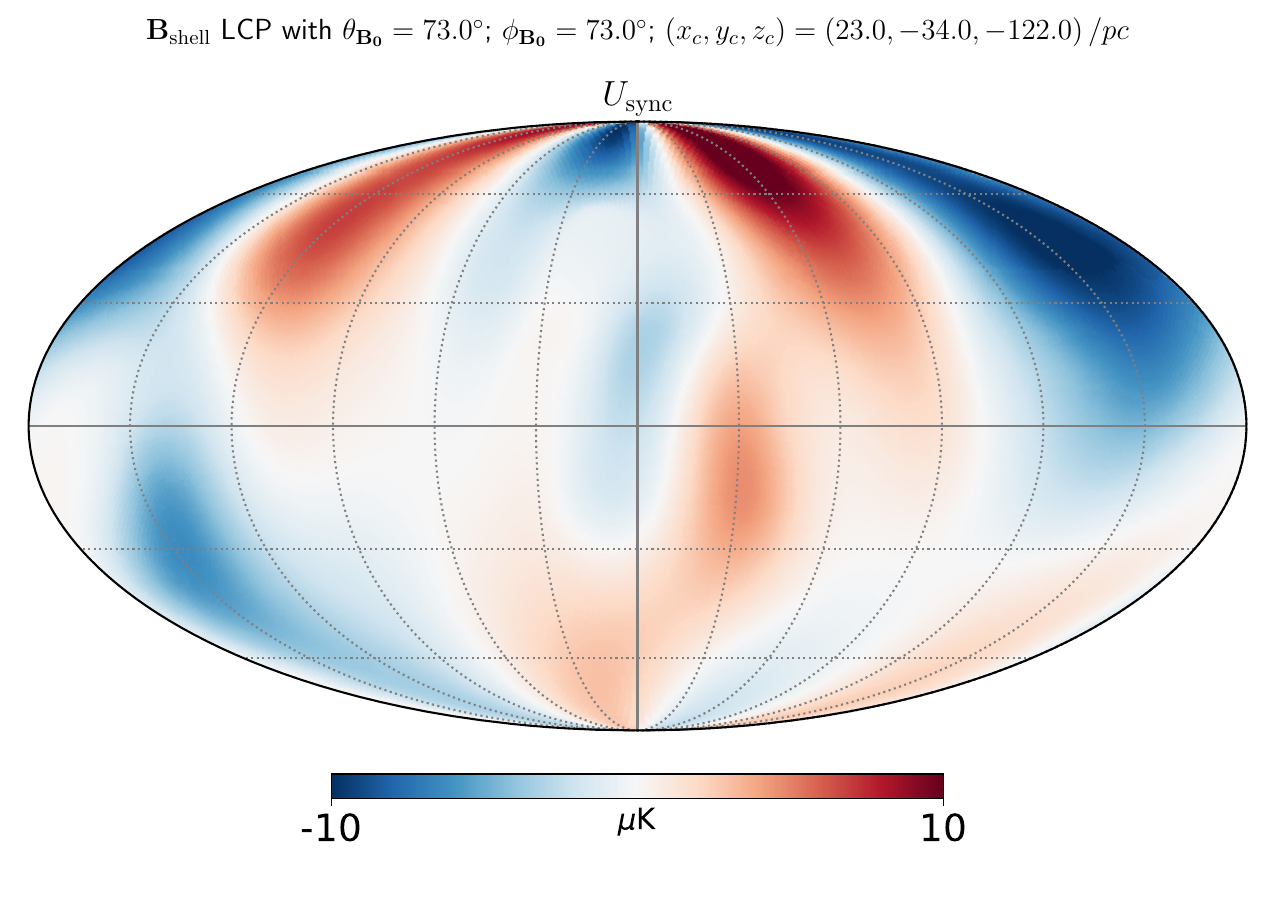} \\
    \rotatebox{90}{\hspace{1.25cm}\texttt{DDO}} &
        \includegraphics[trim={0.4cm 2.4cm 0.4cm 1.9cm},clip,width=.3\linewidth]{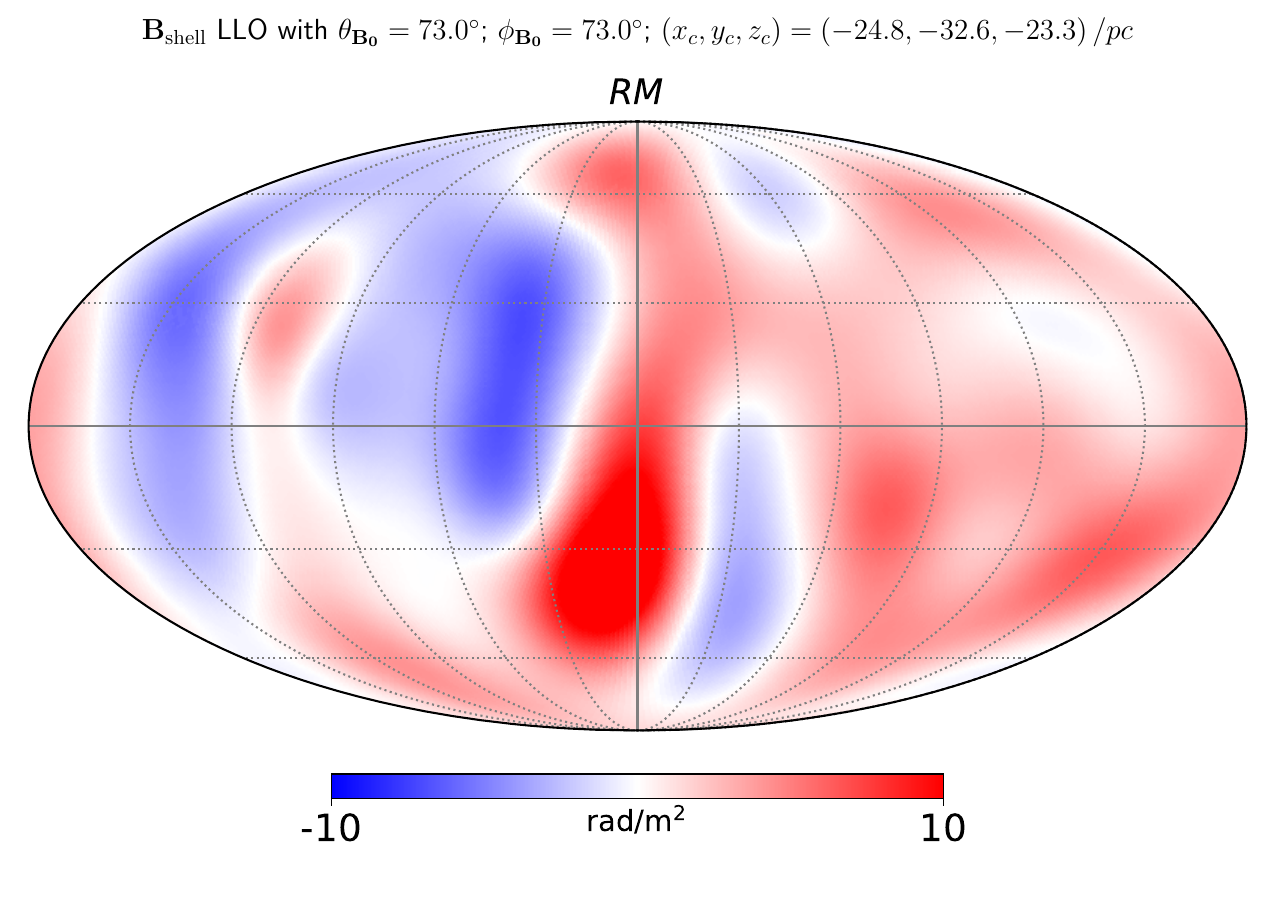} &
        \includegraphics[trim={0.4cm 2.4cm 0.4cm 2.cm},clip,width=.3\linewidth]{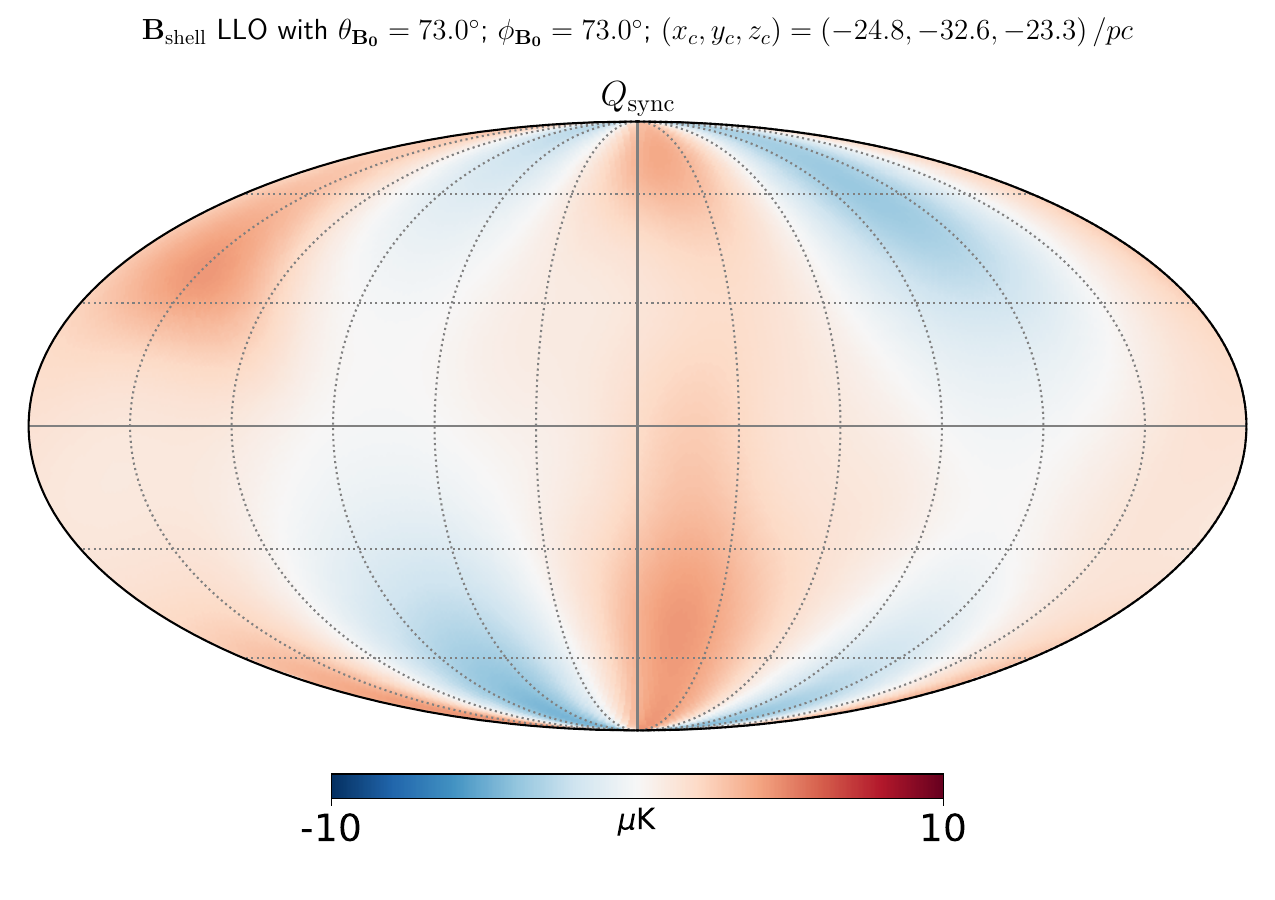} &
            \includegraphics[trim={0.4cm 2.4cm 0.4cm 2.cm},clip,width=.3\linewidth]{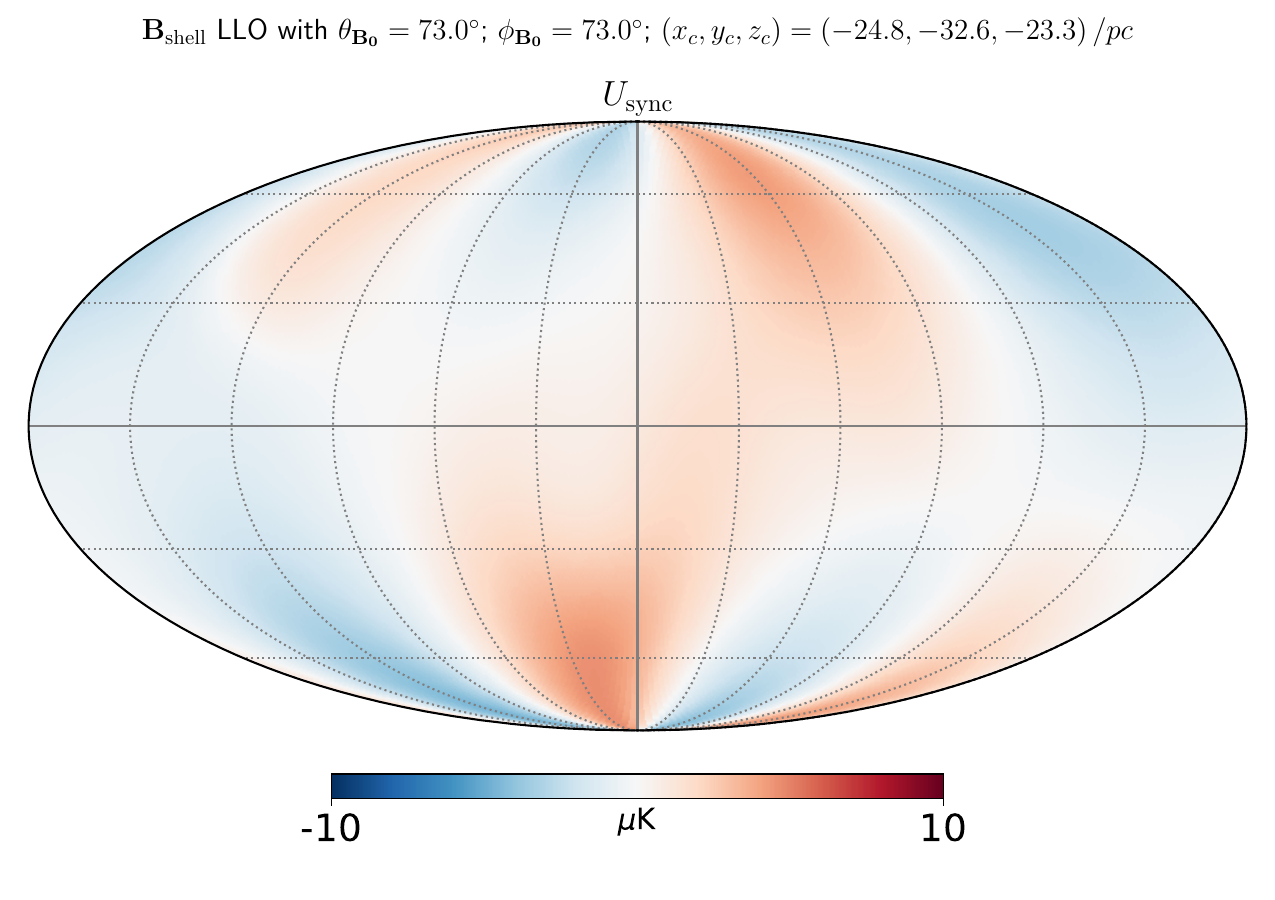} \\
    \rotatebox{90}{\hspace{1.85cm}\texttt{DDA}} &
        \includegraphics[trim={0.4cm 0.6cm 0.4cm 2.cm},clip,width=.3\linewidth]{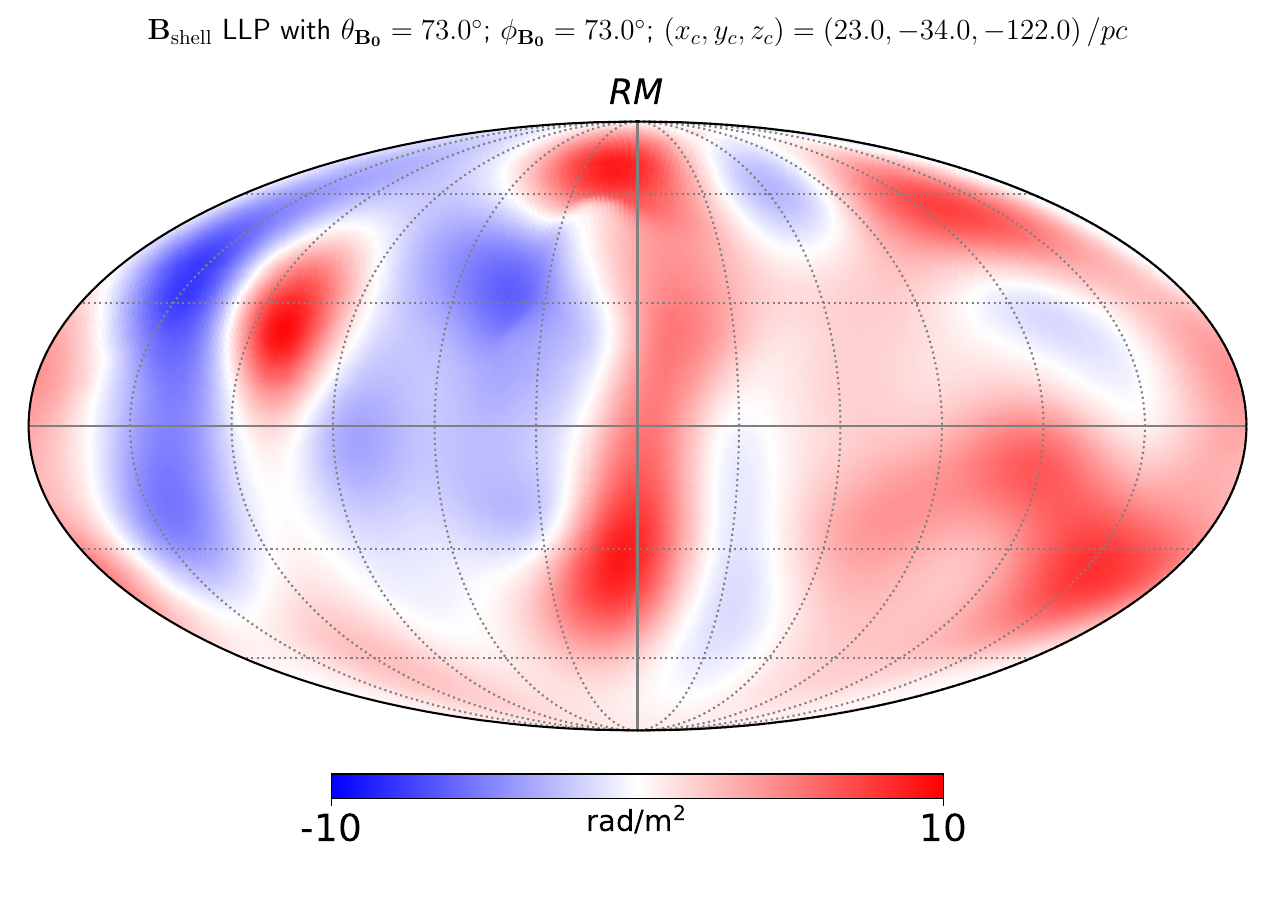} &
        \includegraphics[trim={0.4cm 0.6cm 0.4cm 2.cm},clip,width=.3\linewidth]{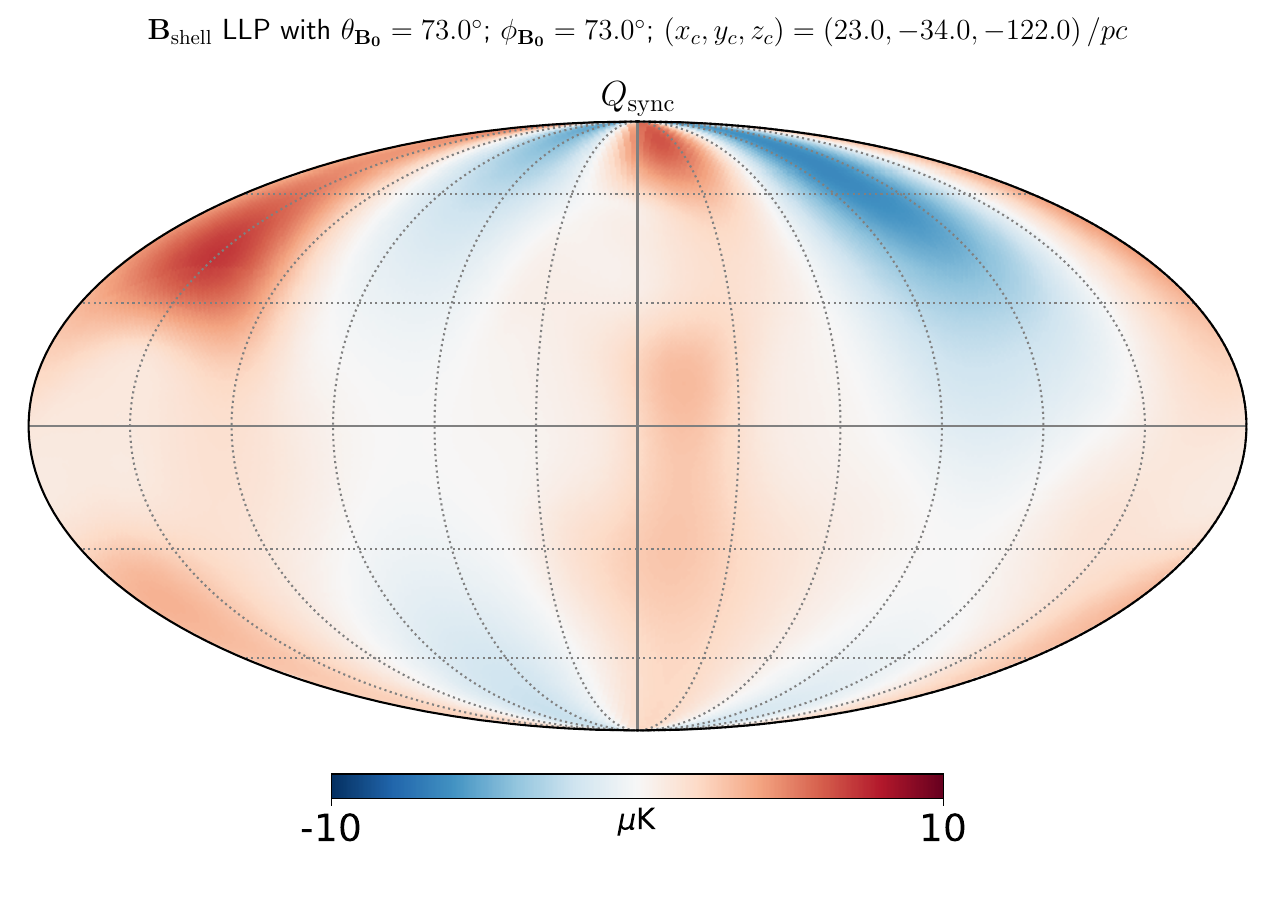} &
            \includegraphics[trim={0.4cm 0.6cm 0.4cm 2.cm},clip,width=.3\linewidth]{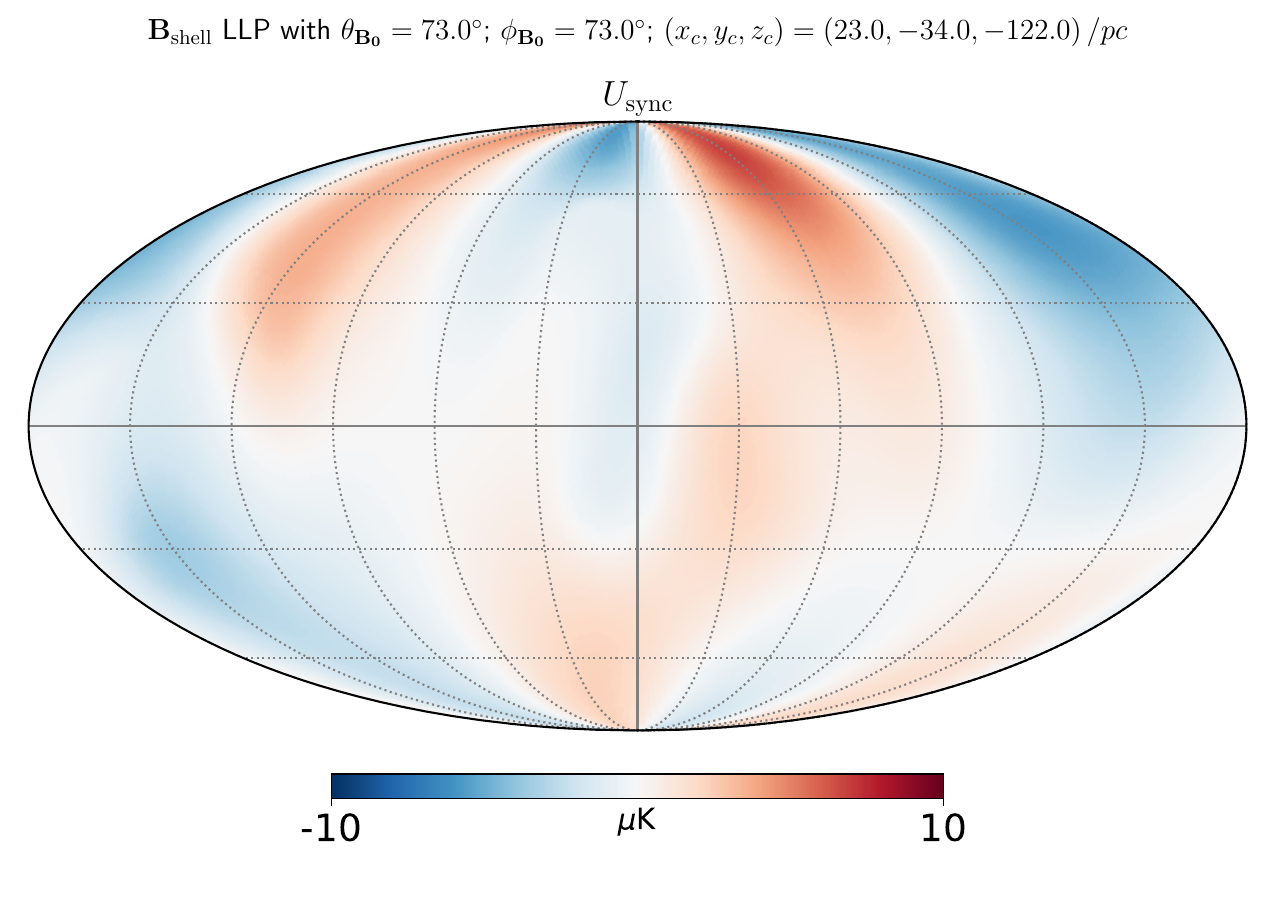}
    \end{tabular}
    \caption{Mollweide projection of the full-sky maps of the contributions from the shell of the Local Bubble to the RM, $Q$, and $U$ signal (from left to right) as predicted for our different scenarios (rows). Maps are given in Galactic coordinates with the Galactic center at the center of the maps, Galactic longitude increases to the left and the Galactic North pole is at the top. Color scales are shared in columns. They range from -10 to 10 rad/m$^2$ for the RM maps and from -10 to 10 $\mu$K for $Q$ and $U$ maps.
    }
    \label{fig:RMQUmaps_LBscenaros}
\end{figure*}
The full-sky maps of the predicted contribution of our bubble-shell scenarios to the Faraday RM and $Q$ and $U$ Stokes parameters of the polarized synchrotron emission are shown in Fig.~\ref{fig:RMQUmaps_LBscenaros}.
The maps were produced using an HEALPix angular tesselation with $N_{\rm{side}} = 64$ (\citealt{Gorski2005}). For the radial sampling, we use 100 grid points through the thick shell of the Local Bubble as seen from the Sun. The line-of-sight integrations (Eqs.~\eqref{eq:RMeq} and~\eqref{eq:QandUeq}) were performed using the simple midpoint rule and using the parameter values given in the previous subsections and summarized in Tables~\ref{tab:ScenarioAndParams} and~\ref{tab:FiducialParams}. We checked that increasing the radial sampling does not affect our result.

\begin{figure*}
    \centering
    \begin{tabular}{lc}
    \rotatebox{90}{\hspace{.6cm}{\small $75^\circ \leq b \leq 85^\circ$}} &
        \includegraphics[trim={0.4cm 1.8cm .5cm 0.cm},clip,width=.94\linewidth]{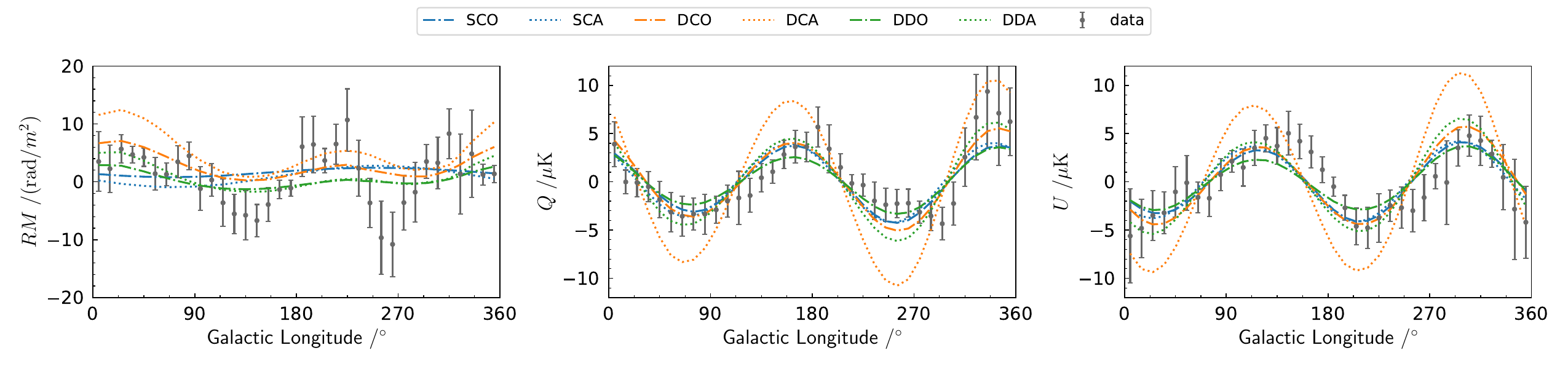}\\
    \rotatebox{90}{\hspace{.6cm}{\small $65^\circ \leq b \leq 75^\circ$}} &
        \includegraphics[trim={0.4cm 1.8cm .5cm 1.6cm},clip,width=.94\linewidth]{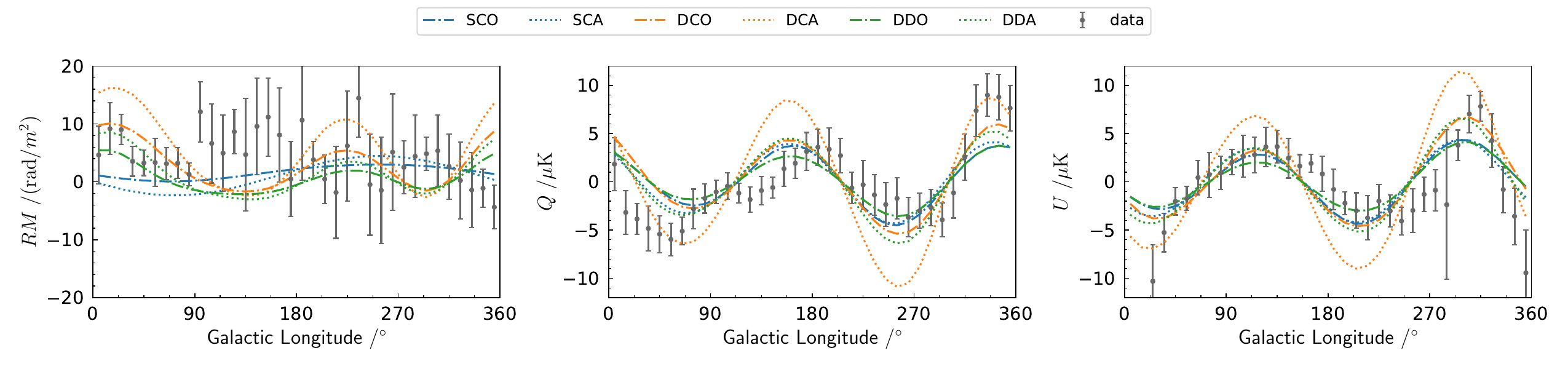}\\
    \rotatebox{90}{\hspace{.6cm}{\small $55^\circ \leq b \leq 65^\circ$}} &
        \includegraphics[trim={0.4cm 1.8cm .5cm 1.6cm},clip,width=.94\linewidth]{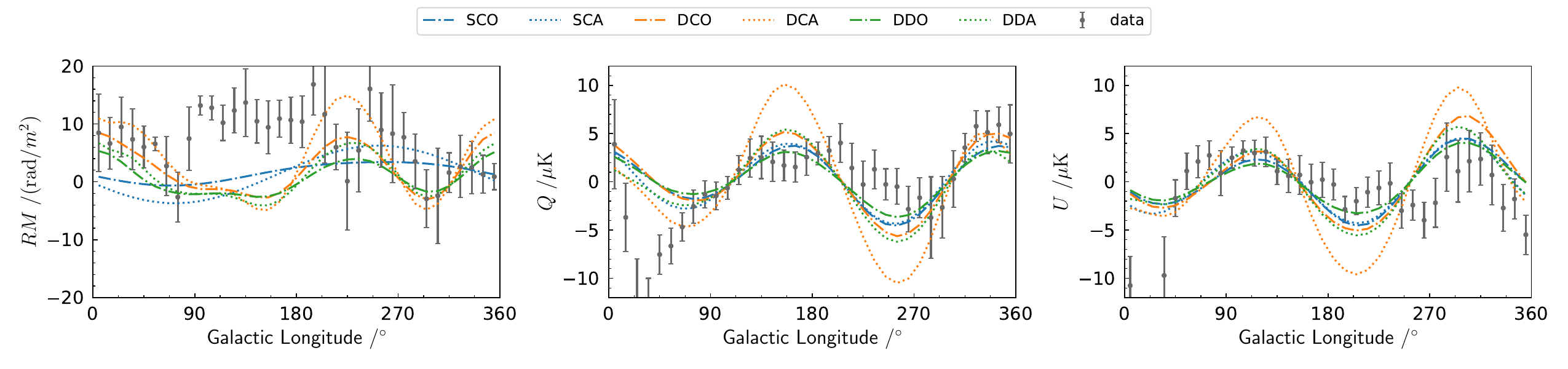}\\[2.ex]
    \rotatebox{90}{\hspace{.4cm}{\small $-65^\circ \leq b \leq -55^\circ$}} &
        \includegraphics[trim={0.4cm 1.8cm .5cm 1.6cm},clip,width=.94\linewidth]{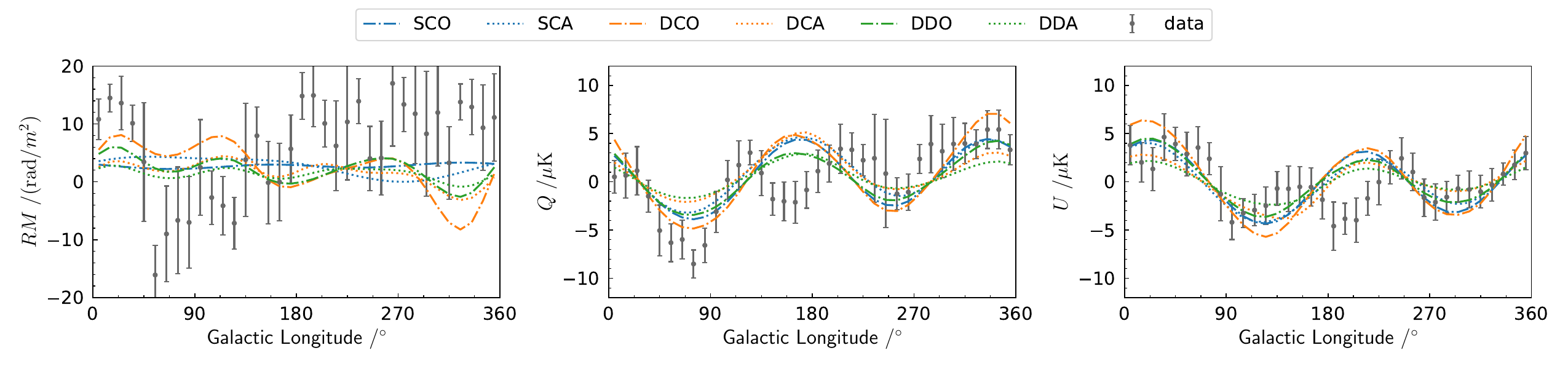}\\
    \rotatebox{90}{\hspace{.4cm}{\small $-75^\circ \leq b \leq -65^\circ$}} &
        \includegraphics[trim={0.4cm 1.8cm .5cm 1.6cm},clip,width=.94\linewidth]{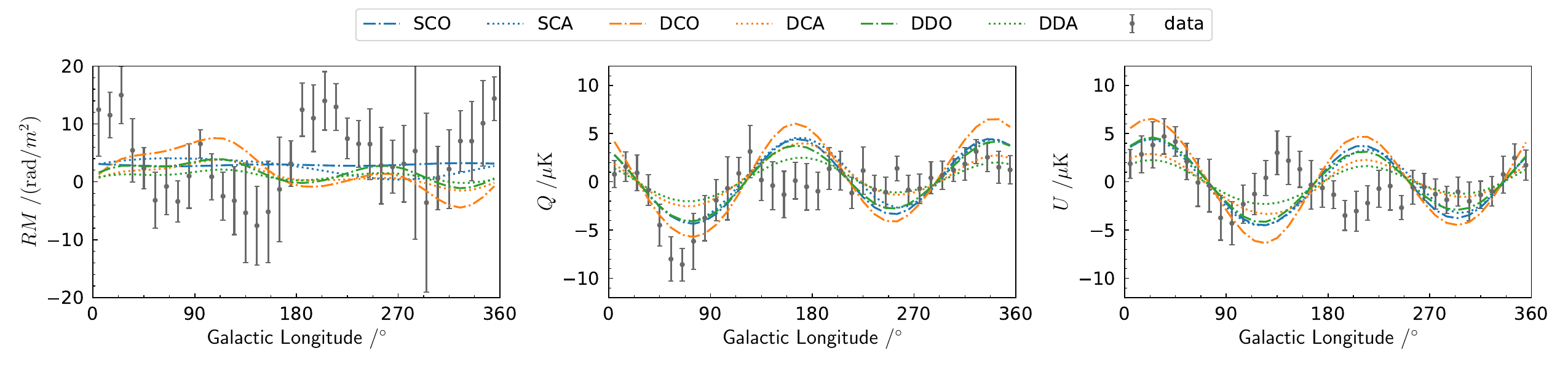}\\
    \rotatebox{90}{\hspace{.8cm}{\small $-85^\circ \leq b \leq -75^\circ$}} &
        \includegraphics[trim={0.4cm .5cm .5cm 1.6cm},clip,width=.94\linewidth]{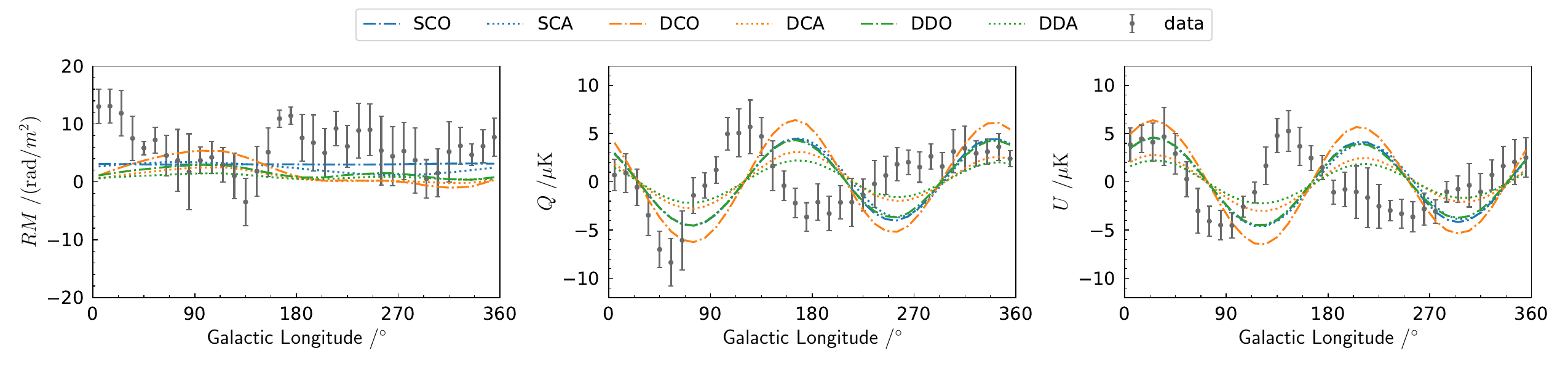}
    \end{tabular}
    \caption{Longitude profiles for constant latitude stripes of 10$^\circ$ width for $|b| \in [55^\circ,\,85^\circ]$, as described in the text. Predictions for the different scenarios for the Local Bubble magnetic field are shown as continuous lines according to the legend, for the RM, $Q$, and $U$ signal (from left to right). For reference, observational data is also shown as gray points with symmetric errorbars.
    }
    \label{fig:LonProfiles}
\end{figure*}
As seen in Fig.~\ref{fig:RMQUmaps_LBscenaros}, and as expected, the convoluted shape of the Local Bubble shell leads to a certain number of anisotropies in the maps, the details of which depend on both, the choice of the explosion center and the thickness of the shell (in addition to its shape).
The shape and exact position of these anisotropies in the sky also depend on the orientation of the initial magnetic field (not shown in the figure).
The substantial North-South asymmetry seen in the maps of the \texttt{DCA} and \texttt{DDA} cases as compared to \texttt{DCO} and \texttt{DDO} cases is due to the fact that in \texttt{A} cases, the explosion center is located significantly more below the Galactic plane (-122~pc as compared to -23.3~pc). As a consequence, the amount of matter and magnetic field lines that have been swept up to the northern part of the Bubble is much larger than to the South. This is reminiscent of what was discussed in Sect.~\ref{sec:AnalyticModel} and could, to some extent, be at the origin of the asymmetry observed in RM data (\citealt{Dickey2022}).

\smallskip

To compare the model predictions between them and also compare them with observations, we show longitude profiles for several stripes of constant latitudes in Fig.~\ref{fig:LonProfiles}. We consider latitude stripes of the width of 10$^\circ$ with centers at 60$^\circ$, 70$^\circ$, and 80$^\circ$, both in the northern and southern hemispheres. The longitude profiles are constructed by taking the average of the model predictions, or observations, in longitude bins of 10$^\circ$ width. The error bars corresponding to the data show the standard deviation of the observation in each bin. It is indicative of the scatter in the data that primarily results from small-scale fluctuations in the magnetized ISM (e.g., \citealt{Jaffe10}; \citealt{Jansson12}; \citealt{Pelgrims2021b}; \citealt{Unger24}; \citealt{Korochkin2024}), plus the source contribution in case of Faraday RM.
For this comparison, we use the RM data from \cite{Unger24} at $N_{\rm{side}}=32$, and the synchrotron $Q$ and $U$ data at 30~GHz from \textit{Planck} processed data by the \texttt{Commander} component separation method (\citealt{PlanckIV2020}), smoothed at a resolution of 1.2$^\circ$, and downgraded at $N_{\rm{side}}=64$. We do not mask the data to build these longitude profiles. For example, the North polar spur is well seen in the region with $l\lesssim90^\circ$ and $55^\circ<b<75^\circ$ in both $Q$ and $U$ data.

\smallskip

In Fig.~\ref{fig:LonProfiles}, it is seen that the different scenarios lead to variations of the RM signal whose number and amplitude are model-dependent. They depend on the shape of the shell and on its thickness.
Overall, the contribution from the Local Bubble shell to the RM sky is smaller (in amplitude) than the observed signal. This indicates that the Local Bubble shell is on average only a subdominant component of the RM integrated over the full path length through our Galaxy.
This result is at odds with that of \cite{Reissl2023} who found, based on MHD simulations of Milky Way-like galaxies, that ISM bubble shells surrounding an observer contribute significantly to the RM signal at high latitudes. This may indicates that we are either underestimating the value of $n_e^0$ or missing a contribution to the line-of-sight component of the magnetic field in the shell which could come from the compressed turbulent component.
However, it is interesting to note that in some parts of the sky, such as at $b>75^\circ$ for $l<90^\circ$, the signal from the shell of the Local Bubble that is predicted for some of our scenarios reaches the amplitude in the data and, furthermore, that the RM data are not very constraining at these latitudes. Overall, the significance of the discrepancy between our model predictions and the data is relatively low, as further discussed in Sect.~\ref{sec:DiscussionLBscenarios}.

The shape and thickness of the Local Bubble shell, as well as the choice for the location of the explosion center, also induce changes in the predictions for the $Q$ and $U$ Stokes parameters of the synchrotron polarized emission. Overall, the different scenarios lead to the same phase for the sinusoidal variation of the signal with longitude. This is because we assume the same direction for the initial magnetic field. The different scenarios, however, predict variations in the amplitude of this sinusoidal signal and even departure from a perfect sines function. These differences primarily depend on the assumed shape and thickness of the shell.

Although the contribution from the Local Bubble shell is only subdominant for the RM at those latitudes, this is not the case for the synchrotron $Q$ and $U$, as seen in Fig.~\ref{fig:LonProfiles}. Surprisingly, the contribution from the Local Bubble shell to the synchrotron sky is very significant for $|b| \gtrsim 55^\circ$. Of course, alone the shell of the Local Bubble cannot account for the entirety of the signal and its variations. However, it is very striking that our simple model with our a priori choice for the model parameters leads to a shell contribution to the $Q$ and $U$ sky that has the same overall amplitude and phase as that of the observation.
It should be emphasized that the models do not result from a fit to RM or synchrotron data. The result of a fit would likely improve the agreement seen in Fig.~\ref{fig:LonProfiles}, especially for $Q$ and $U$.
Further analysis will be needed to better understand why the agreements between data and model predictions are different for Faraday RM than for synchrotron polarized emission. Such an analysis might require the inclusion of the effects of the turbulent component of the magnetic field and inhomogeneities and will require the full exploration of the parameter space. This is beyond the scope of this exploratory study.
\begin{sidewaysfigure*}
    \centering
    \begin{tabular}{lccc}
     $\,$ & RM & $Q$ & $U$ \\[1.ex]
    \rotatebox{90}{\hspace{1.5cm}\texttt{data}} &
        \includegraphics[trim={0.2cm 2.4cm 0.2cm 0.4cm},clip,width=.3\linewidth]{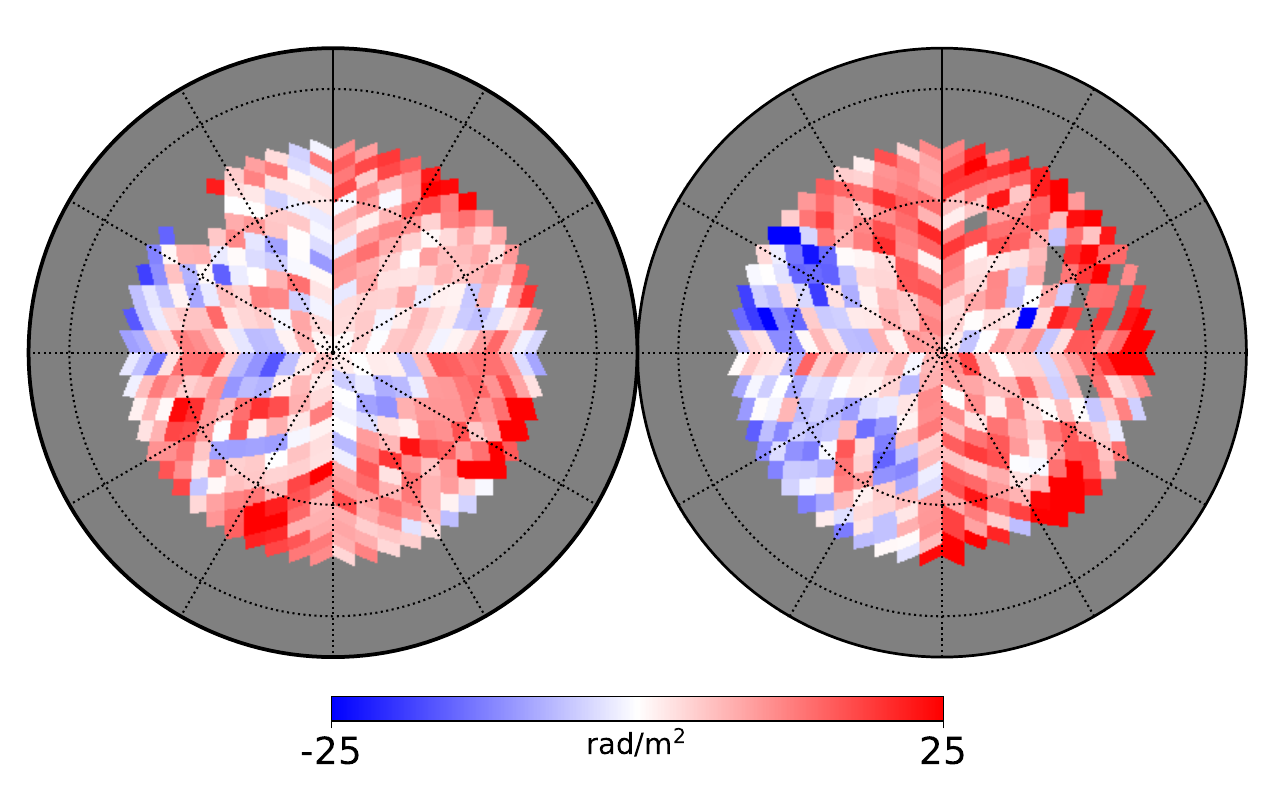} &
        \includegraphics[trim={0.2cm 2.4cm 0.2cm 0.4cm},clip,width=.3\linewidth]{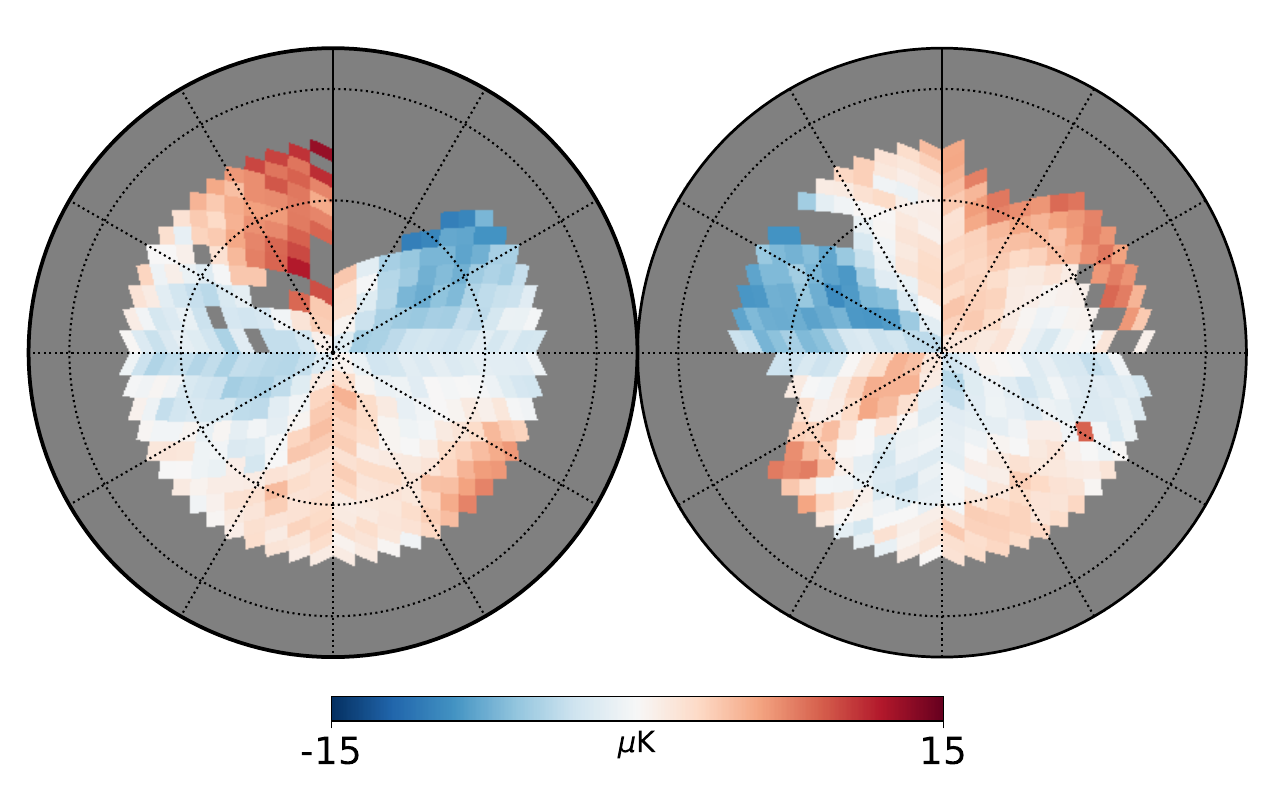} &
            \includegraphics[trim={0.2cm 2.4cm 0.2cm 0.4cm},clip,width=.3\linewidth]{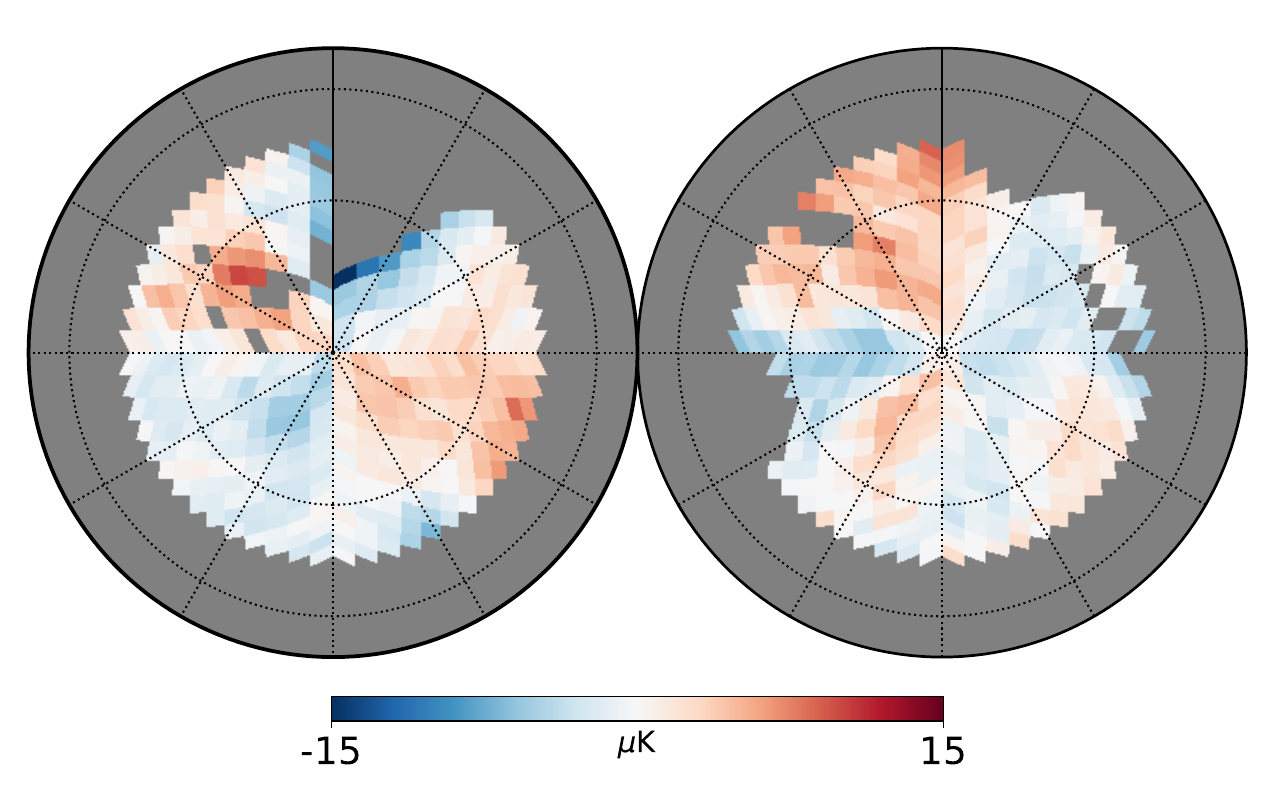} \\
    \rotatebox{90}{\hspace{2.3cm}\texttt{SCA}} &
        \includegraphics[trim={0.2cm .4cm 0.2cm 0.4cm},clip,width=.3\linewidth]{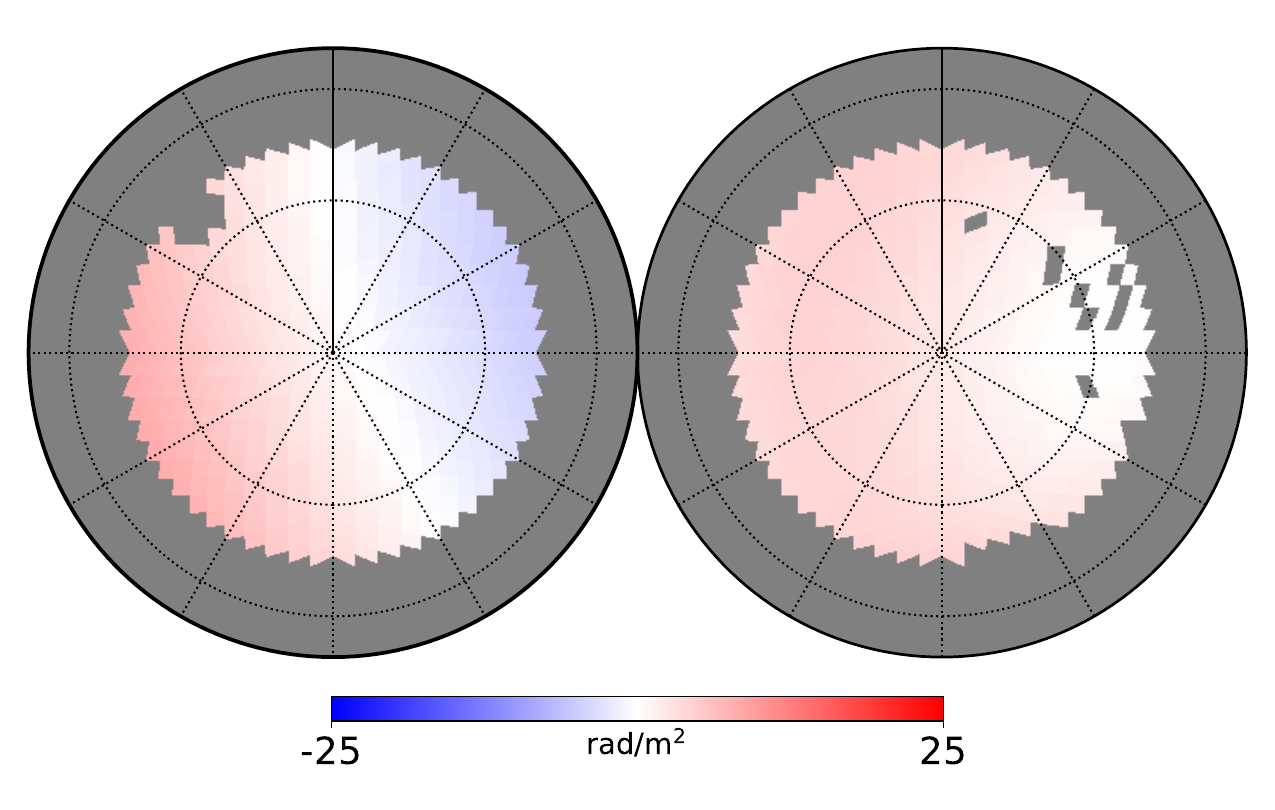} &
        \includegraphics[trim={0.2cm .4cm 0.2cm 0.4cm},clip,width=.3\linewidth]{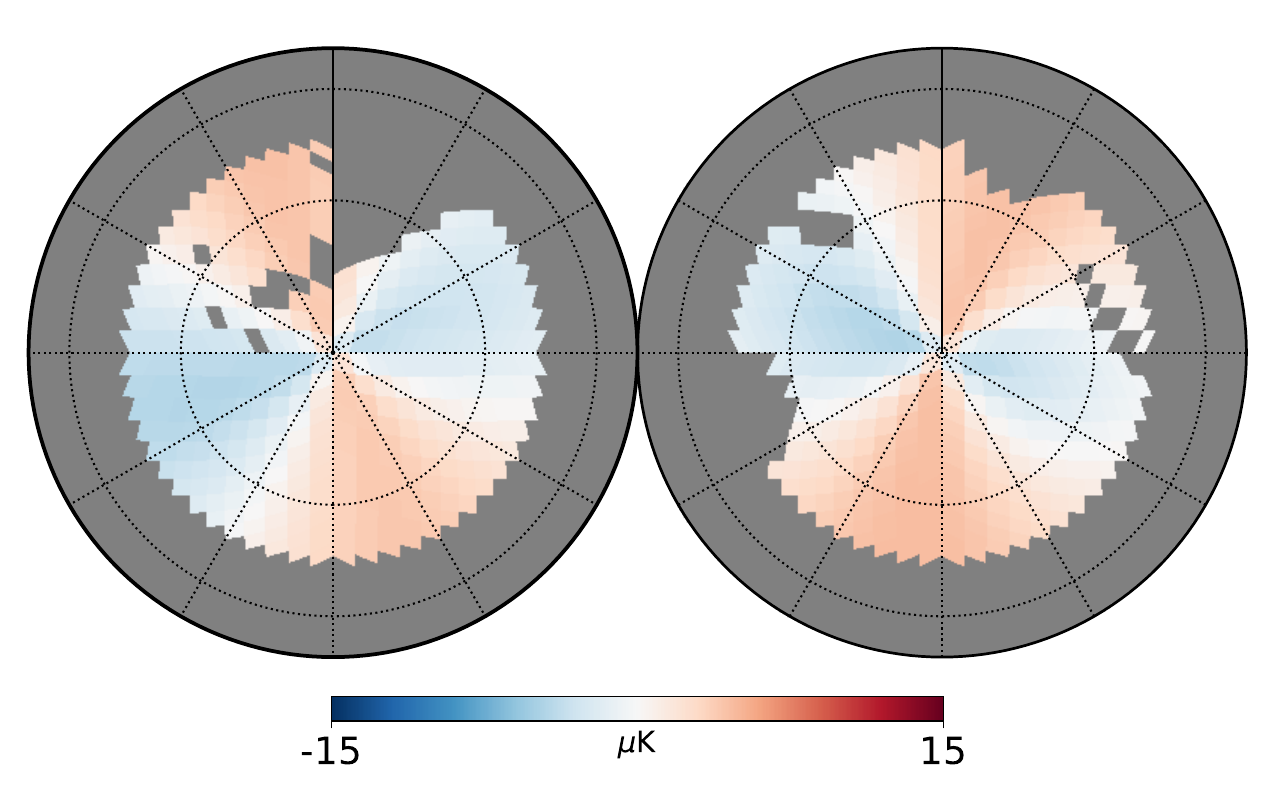} &
            \includegraphics[trim={0.2cm .4cm 0.2cm 0.4cm},clip,width=.3\linewidth]{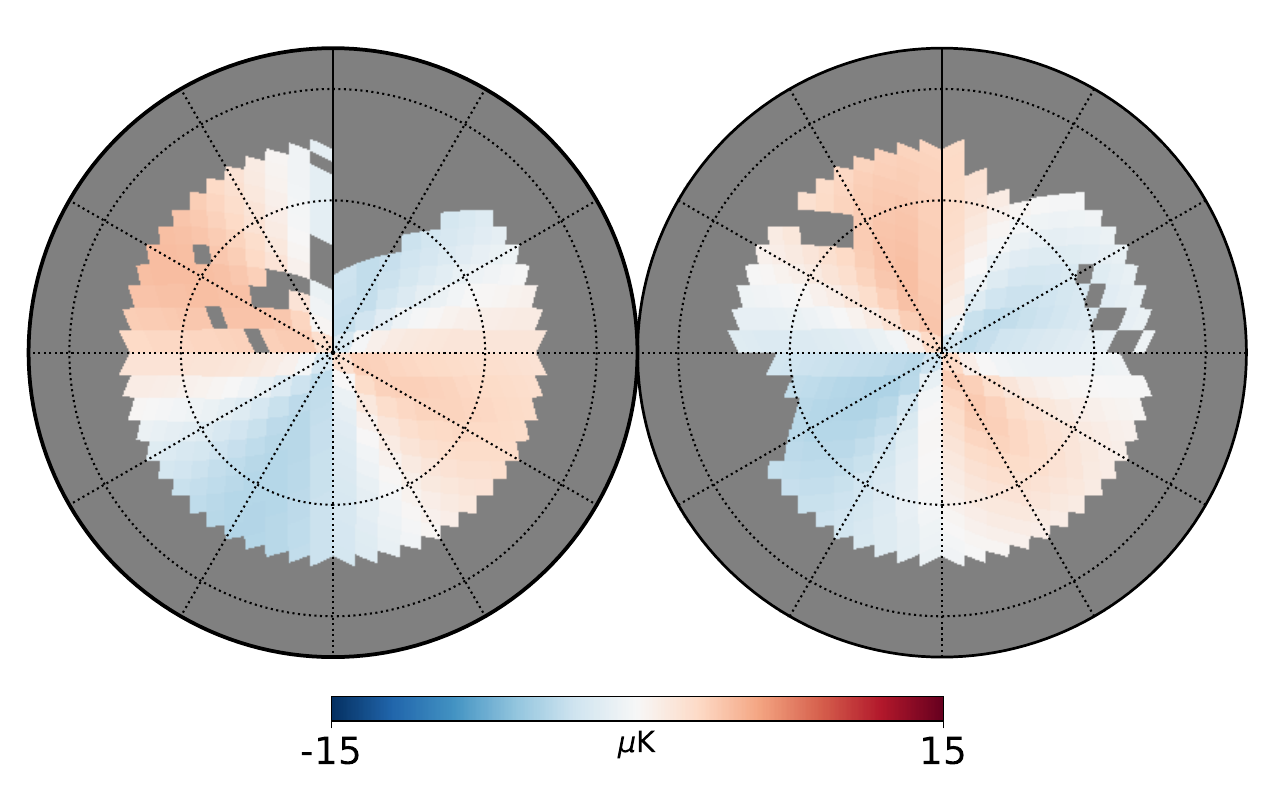} \\[1.ex]
    \rotatebox{90}{\hspace{2.2cm}\texttt{pull}} &
        \includegraphics[trim={0.2cm .4cm 0.2cm 0.4cm},clip,width=.3\linewidth]{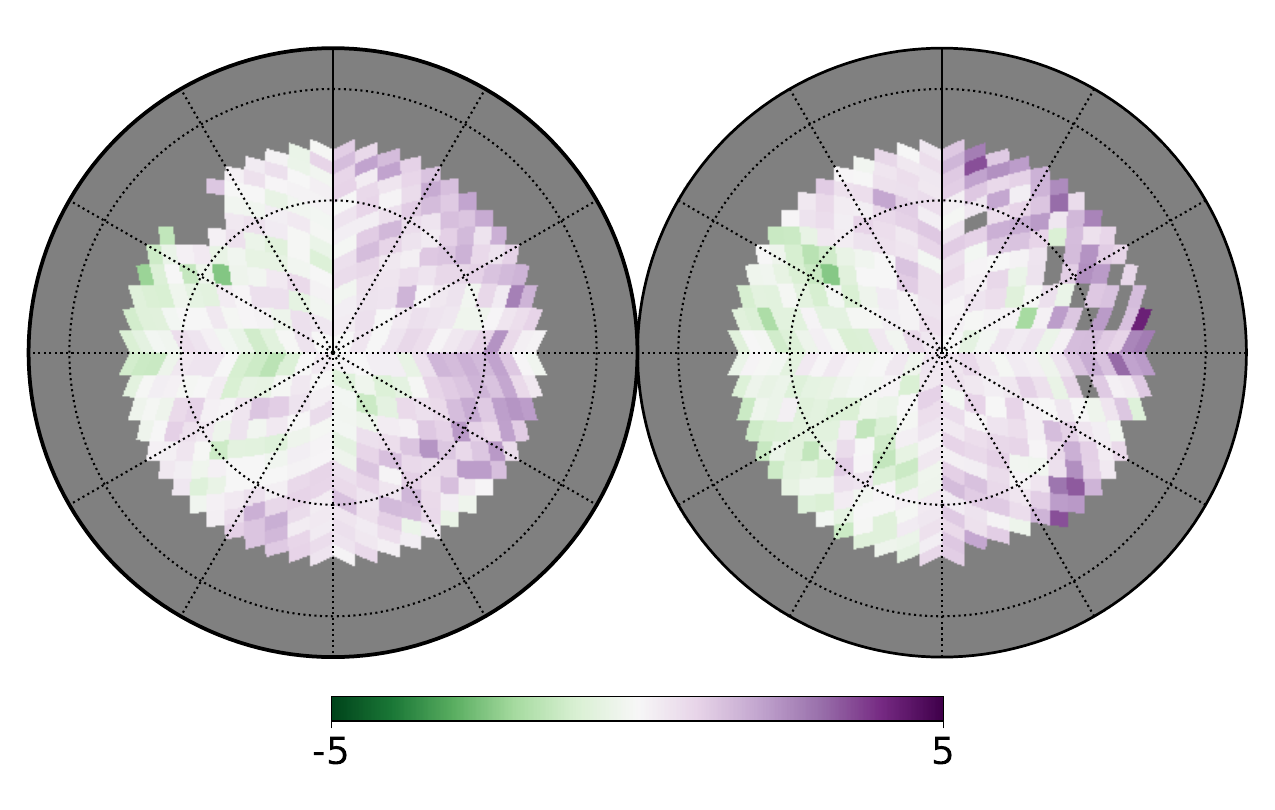} &
        \includegraphics[trim={0.2cm 0.4cm 0.2cm 0.4cm},clip,width=.3\linewidth]{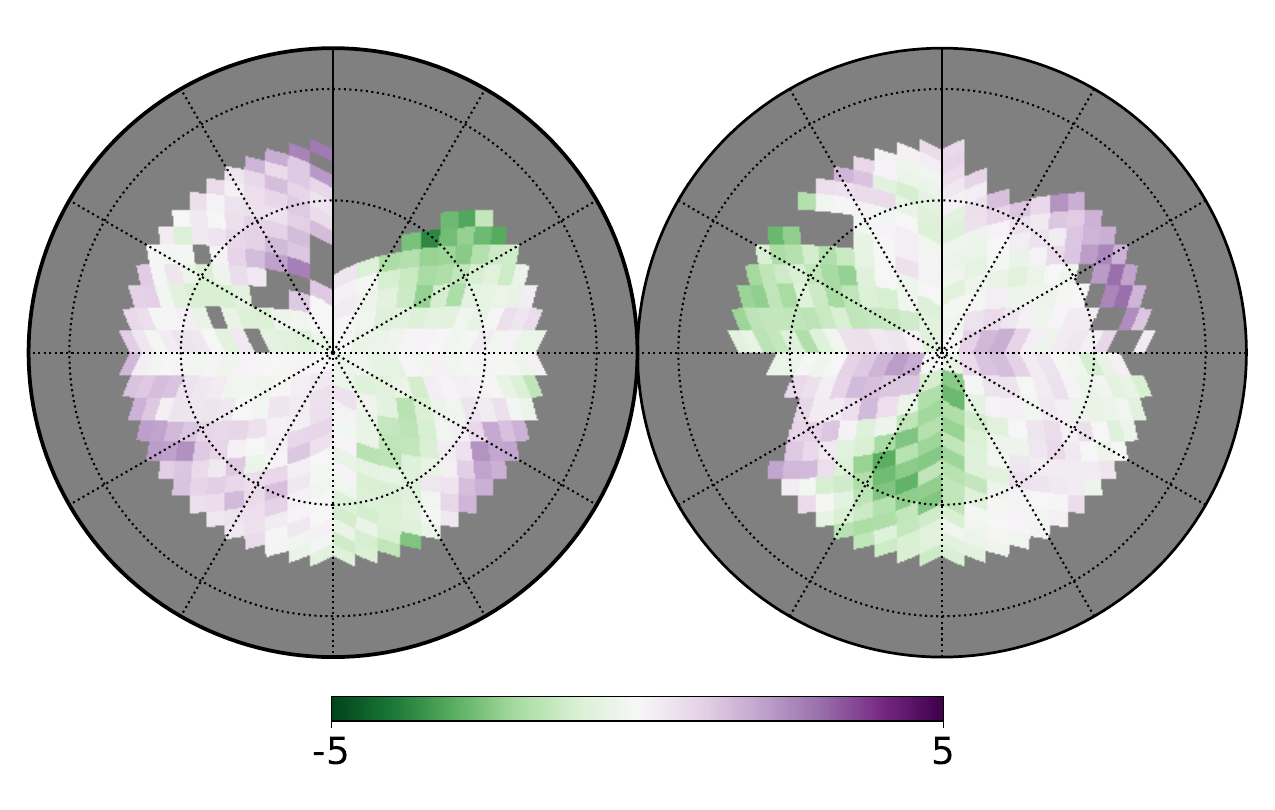} &
            \includegraphics[trim={0.2cm 0.4cm 0.2cm 0.4cm},clip,width=.3\linewidth]{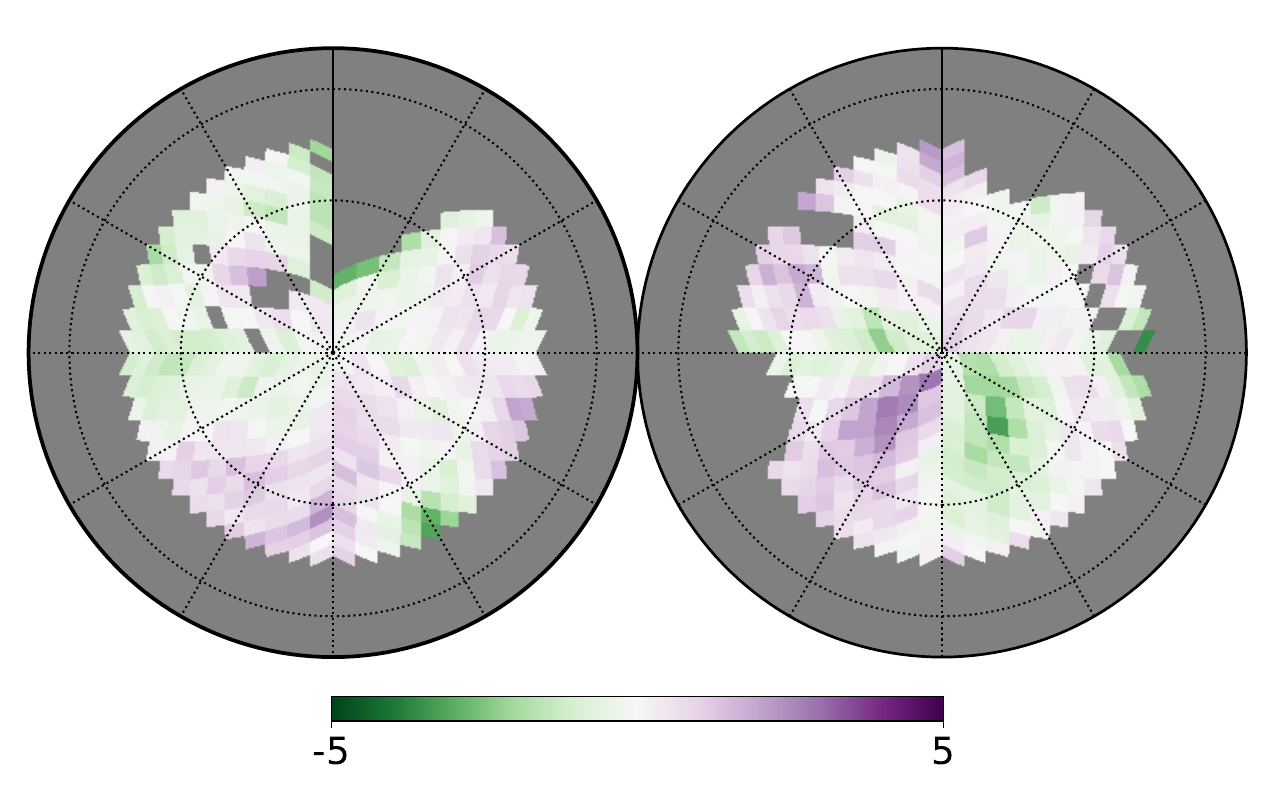}\\
    \end{tabular}
    \caption{Orthographic projection of the same data used in \cite{Unger24}, the model predictions for the \texttt{SCA} scenario, and pulls, with applied masked (gray area) as explained in the text.
    The North Pole is at the center of the left disk and the South to the right. The longitude zero is marked by the vertical thick lines starting from the poles upward. Longitude increases clockwise on the left and counterclockwise on the right.
    }
    \label{fig:dataAndSCPmodel}
\end{sidewaysfigure*}

\subsection{Discussion}
\label{sec:DiscussionLBscenarios}
Although we do not attempt a full adjustment of the model to the data, we proceed to a better quantification of the comparison between the maps from the different scenarios of the Local Bubble shell and the data.
We consider the same data as used by \cite{Unger24} to constrain models of the large-scale regular component of the Galactic magnetic field. We used their RM, $Q$, and $U$ maps at the $N_{\rm{side}} = 16$ resolution, along with their uncertainties, and their masks.
For our comparison, we choose to keep only all unmasked pixels with Galactic latitude $|b| > 45^\circ$ to avoid sky regions where the signal is clearly dominated by other Galactic components than the wall of the Local Bubble, such as the Galactic disk.
We are left with 751 pixels for the $Q$ and $U$ maps and with 819 pixels for the RM map. The masked data are shown at the top row of Fig.~\ref{fig:dataAndSCPmodel}.
Accordingly, we downgrade the modeled maps shown in Fig.~\ref{fig:RMQUmaps_LBscenaros} from $N_{\rm{side}} = 64$ to $N_{\rm{side}} = 16$ and apply the same mask than to the data.
If $d_X^i$, $m_X^i$, and $\sigma_X^i$ are the data, model and uncertainty for the observable $X$ in pixel $i$, where $X$ is either RM, $Q$, or $U$, we then compute the residual ($d_X^i - m_X^i$) and pull ($\chi_X^i \coloneqq (d_X^i - m_X^i)/\sigma_X^i$) for each pixel.
Using these values, we can compute the reduced $\chi^2$ of each map and for each scenario.
The values are given in Table~\ref{tab:Chi2-B0_3}, along with a total value that results from the combination of RM, $Q$, and $U$ maps.
\begin{table}
    \centering
    \begin{tabular}{lcccccc}
    \hline\hline\\[-1.5ex]
          & \texttt{SCO} & \texttt{SCA}& \texttt{DCO}& \texttt{DCA}& \texttt{DDO}& \texttt{DDA}   \\[.5ex] \hline\\[-1.5ex]
         $\chi^2_{~RM} /{\rm{ndf}}$ & 0.79 & 0.92 & 1.0 & 0.96 & 0.9 & 0.91 \\[.5ex]
         $\chi^2_{~Q} /{\rm{ndf}}$  & 1.26 & 1.15 & 1.38 & 2.44 & 1.25 & 1.46 \\[.5ex]
         $\chi^2_{~U} /{\rm{ndf}}$  & 0.85 & 0.78 & 1.04 & 1.75 & 0.75 & 0.86 \\[1.5ex]
         $\chi^2_{~{\rm{tot}}} /{\rm{ndf}}$ & 0.96 & 0.95 & 1.14 & 1.7 & 0.97 & 1.07 \\[.5ex]
         \hline
    \end{tabular}
    \caption{Values for the reduced $\chi^2$ corresponding to the three maps and for the six scenarios. $\chi^2_{~{\rm{tot}}} /{\rm{ndf}}$ results from the combination of the three maps.
    }
    \label{tab:Chi2-B0_3}
\end{table}
The maps of the model and pull for the scenario \texttt{SCA} are shown in Fig.~\ref{fig:dataAndSCPmodel} at the second and third row, respectively.
\begin{figure}
    \centering
    \includegraphics[trim={0.4cm 1.cm 1.5cm 0.cm},clip,width=.98\linewidth]{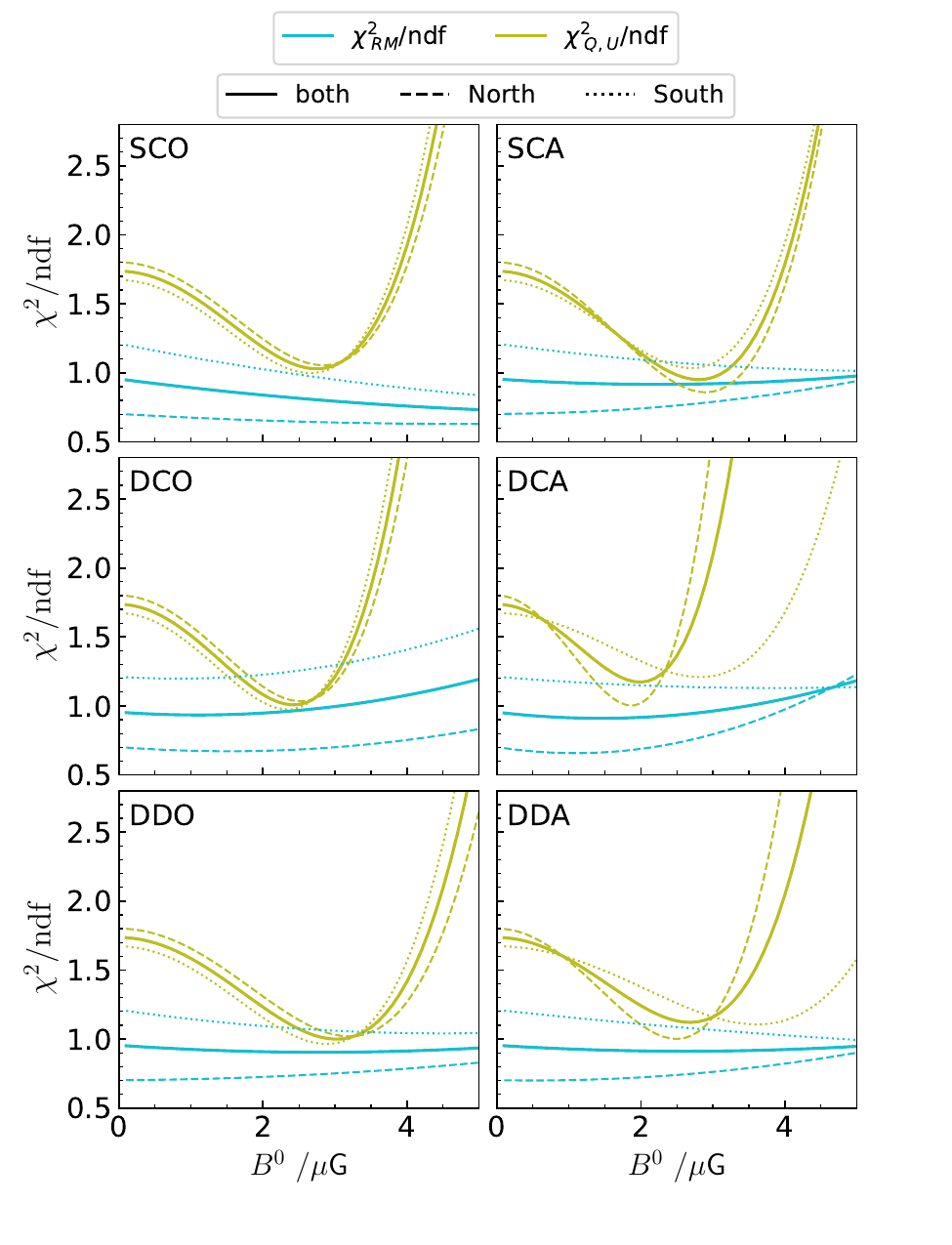}\\[-1.5ex]
    \caption{Contribution to the reduced $\chi^2$ of the different observables as a function of the strength of the initial magnetic field for the six scenarios. The contribution from $Q$ and $U$ are combined. The contributions from the northern and southern hemispheres are also shown with dashed and dotted lines, respectively.
    }
    \label{fig:ScanB0}
\end{figure}

The qualitative match found between the data and the six scenarios in Fig.~\ref{fig:LonProfiles} is confirmed. Among them, the \texttt{SCA} scenario provides the best performances and the \texttt{DCA} scenario the worst, with a total reduced $\chi^2$ of 0.94 and 1.47, respectively.
These $\chi^2/{\rm{ndf}}$ are obtained without any minimization.
In addition, the low values found for the reduced $\chi^2$ are due to the relatively large uncertainties assigned to the data at those latitudes, in particular for the Faraday RM as discussed below.
We notice that the pull maps of $Q$ and $U$ shown in Fig.~\ref{fig:dataAndSCPmodel} show very well-structured patterns that are unlikely to be the result of random noise. This is evidence that other large-scale components need to be included to model the data at those latitudes, such as a halo field for example.

\smallskip

To investigate further the relevance of the contribution of the shell of the Local Bubble to the RM, $Q$, and $U$ maps, we vary the strength of the initial magnetic field ($B^0$) from 0~$\mu$G, where the Local Bubble shell has no contribution to the observable, to a value of 5~$\mu$G. For each value of $B^0$, we compute the reduced $\chi^2$ for RM separately and $Q$ and $U$ combined, and also consider the northern and southern hemispheres at a time.
If the data has sufficient constraining power, and if a model contributes substantially to the data with a relatively good match, we expect to observe a well-defined minimum in the curve of reduced $\chi^2$ versus $B^0$. If, instead, the model contribution is negligible for all $B^0$ values, we expect to see a relatively flat curve.

The results for the six scenarios under consideration are shown in Fig.~\ref{fig:ScanB0}.
On the one hand, it is seen that Faraday rotation contributes only marginally to the total reduced $\chi^2$.
For all tested values of $B^0$, the reduced $\chi^2_{\rm{RM}}$ varies only marginally and remains close to the value where the contribution from the Local Bubble shell is assumed to be zero (i.e., when $B^0 = 0$). This value is found to be $\approx 0.9$, showing that the S/N of RM data for $|b|>45^\circ$ is low on average. This demonstrates that the constraining power of the RM data at those latitudes is rather weak.
On the other hand, for all six scenarios, we observe a clear minimum in the curves corresponding to $Q$ and $U$, and the reached minima are significantly lower than in the case of no contribution from the shell of the Local Bubble. For all scenarios, the minima are found for a value of $B^0$ in the range of 2 to 4~$\mu$G. For the cases with the explosion center at the position \texttt{O} (see Table~\ref{tab:ScenarioAndParams}), the curves for the northern and southern hemispheres are very similar. However, for the cases with the explosion center at the position \texttt{A}, the agreement between the model and the data is better in the northern hemisphere (dashed line) than in the southern hemisphere (dotted line). Indeed, the minima are more pronounced in the northern part than in the southern part of the sky. There is also a tension between the favored value for $B^0$ from both hemispheres.
To understand better these features, the magnetic field in the shell of the Local Bubble needs to be included in global models for the regular component of the GMF, and a full exploration of the parameter space is needed. This task goes well beyond the scope of this paper.
Overall, this latter test confirms that the shell of the Local Bubble likely makes a significant contribution to the observed $Q$ and $U$ Stokes parameter of the Galactic synchrotron emission whereas it only makes a marginal contribution to the Faraday RM integrated over the full path length across the Galaxy. This is obtained for Galactic latitudes $|b| > 45^\circ$, considering the masks of \cite{Unger24} which, in particular, masks out bright features of the synchrotron sky, and without having explored the parameter space of our model other than the value of $B^0$. Our latest test also enabled us to test explicitly the robustness of our conclusion against the specific choice for the value of the strength of the initial magnetic field.
In fact, for a fixed bubble-shell geometry, fixed explosion center, and fixed orientation of the initial magnetic field, and according to our model assumptions, the contribution of the Local Bubble shell to the RM scales with $n_e^0 \, B^0$, while its contribution to the $Q$ and $U$ scales with $n_{10} \, (B^0)^2$.
Consequently, one way to increase the RM contribution of the Local Bubble shell, and thus better match the RM data without deteriorating the qualitative agreement observed for the synchrotron polarized emission, would be to increase substantially the ratio $n_e^0/n_{10}$ as compared to the one we used in this work. Either $n_e^0$ needs to be higher, or $n_{10}$ to be smaller. If $n_{10}$ overestimates the true density of CRE in the shell of the Local Bubble, then a larger value for the magnetic field would be needed, which would allow for an increased contribution to the RM.

\smallskip

To conclude this discussion section, we emphasize that, although it improves on the current state-of-the-art, our model remains simple and is based on several assumptions, which may not apply satisfactorily to the case of the Local Bubble, and not all the effects of which are known.
For instance, variations in both the magnetic field and the initial matter density are completely disregarded in our model.
It could be argued that fluctuations, which are then stretched and squeezed during the bubble formation, could change the balance between the predicted amplitudes of the Faraday RM and synchrotron polarized emission (e.g., \citealt{Beck2003}). This could possibly reconcile our results with those from MHD simulations of Milky Way-like galaxies where the observer is located inside super-bubbles (\citealt{Reissl2023}), or not (\citealt{Pakmor2018}).
Addressing such open questions is beyond the scope of this paper and will be the subject of follow-up studies.

\section{Summary and concluding remarks}
\label{sec:conclusion}
An analytical prescription for the coherent magnetic field in the thick shell of ISM (super-)bubbles has been derived.
We constructed our analytical model starting from the induction equation without magnetic diffusion and the basic assumption that ISM bubbles result from a single explosion which induces a purely radial displacement of the surrounding gas, but not necessarily with spherical symmetry.
Our model assumes uniform initial conditions.
Relying on the frozen-in approximation, we obtained equations for the magnetic field components expressed in spherical coordinates centered on the explosion center. The solution makes it possible to estimate the magnetic vector field anywhere in the thick shell of the bubble. By construction, the solution leads to a magnetic field that is divergence-free.
To study and illustrate the properties of our analytical solution, we further proposed a linear model for the radial displacement vector field which relates the initial locations of matter particles to their present locations in the thick shell of the bubble.
For the bubble cases with spherical symmetry, we demonstrated that our analytical model leads to the general expectations in terms of geometry for the magnetic field lines and amplification.
Our solution also makes it possible to study other cases where the spherical symmetry is broken. We showed that strong anisotropies in magnetic field amplification can occur, not only due to the magnitude of the radial displacement but also due to the circumvolutions in the shell geometry.

Assuming that the inner and outer radius of ISM bubbles can be extracted from observation, our analytical prescription makes it possible to model the magnetic field in their shell, with some limitations and caveats. For example, due to the choice of the radially discontinuous, linear model for the radial displacement field, our magnetic field undergoes a sharp discontinuity at the outer radius of the bubble shell. This discontinuity could be overcome by adopting a continuous model for the radial displacement and will be addressed in future works.
The main limitation of our model, in its general form or with the linear displacement field, is that it does not explicitly account for the dynamics of the bubble formation, but solely depends on the initial and final (present-time) conditions.
While this is the main limitation from the point of view of the physical modeling of ISM bubbles, we emphasize that it is also the main asset of our model for all practical uses.

\smallskip

We further used our model to explore the possible contribution of the shell of the Local Bubble to the extragalactic Faraday RM and synchrotron polarization data at 30~GHz, which our model makes it possible to do since it predicts the amplitude of the magnetic field in the thick shell.
For this study, we relied on 3D geometrical models for the shell of the Local Bubble informed from a 3D dust density map and adopted published values for all free parameters in our model. We considered six scenarios with varying complexity for the bubble shell, considered either constant or varying shell thickness, and explored two possible explosion centers. We found that all these geometrical parameters lead to (sometimes large) variations in the maps of observables. For the six scenarios, we found that the shell of the Local Bubble may be a substantial contributor to the synchrotron data at high Galactic latitudes ($|b|>45^\circ$), whereas its contribution to the integrated Faraday RM is generally small.
In that respect, our study reinforces the recent result of \cite{Korochkin2024} which indicates that, at high Galactic latitudes, the shell of the Local Bubble is an important foreground to the magnetized Galaxy and that its contribution to the synchrotron polarization data is significant while subdominant for the RM data. This comes with an improved model for the magnetic field in the shell of the Local Bubble and an improved model for the shell itself. In addition, we showed that the specific geometry of the shell and the location of the explosion center inside the bubble are crucial parameters that may lead to very different predictions.
A spherical model can be used to approximate the overall patterns in the RM, $Q$, and $U$ maps, but only the solution for the irregular Bubble can lead to detailed fine-structures seen in observation.
When applied to the case of the Local Bubble, the main limitations of our model possibly come from the fact that we assume the Local Bubble to result from a single explosion that happened in a homogeneous medium permeated by a uniform magnetic field. Given the size of the Local Bubble, these are crude assumptions that will have to be move past in the future.

Our studies underline the need to include the Local Bubble and its magnetized shell in models of the regular component of the GMF.
In fact, focusing on the high Galactic latitudes, we find that the predictions from the Local Bubble shell have signal amplitude comparable to the predictions from the halo components of the regular Galactic magnetic field used in \cite{Unger24} and shown in their Fig.~10. Only a combined fit to the data, including the shell of the Local Bubble into the description of the large-scale GMF, with a full exploration of the parameter space will make it possible to quantitatively judge on their respective contribution. We will address such an endeavor in future work and explore the possible impact on the studies of the propagation of ultra-high energy cosmic rays across the Galaxy.
In this context, the analytical expression for the magnetic field in the Local Bubble shell obtained in this paper will prove very useful, as it enables us to determine the magnetic field with only a handful number of parameters. Of course, this model relies on several working assumptions and does not explicitly account for all the physics involved in a supernovae explosion such as pressure balance, feedback from ISM matter onto the development of the shell (e.g., \citealt{vanderLaan1962}; \citealt{Ferriere1991a,Ferriere1991b}) and other phase transitions which may take place over the time in such peculiar environments.
However, our model is general enough to effectively accommodate a diverse range of complexities and to be useful for better understanding magnetized bubbles and super-bubbles of the ISM, also other than the Local Bubble itself. Modeling magnetized supernova remnants may also prove useful in deciding between competing large-scale GMF models (\citealt{West2016}), a context in which our analytical model could present advantages.

\begin{acknowledgements}
We are grateful to the anonymous referee who provided us with a constructive report that helped us improve our manuscript.
The authors would like to thank Katia Ferri{\`e}re, Alexander Korochkin, and Peter Tinyakov for helpful discussions.
VP acknowledges funding from a Marie Curie Action of the European Union (grant agreement No. 101107047).
\end{acknowledgements}

\bibliographystyle{aa}
\bibliography{myBiblio}

\begin{thebibliography}{73}
\expandafter\ifx\csname natexlab\endcsname\relax\def\natexlab#1{#1}\fi

\bibitem[{{Aguilar} {et~al.}(2014){Aguilar}, {Aisa}, {Alpat}, {Alvino},
  {Ambrosi}, {Andeen}, {Arruda}, {Attig}, {Azzarello}, {Bachlechner}, {Barao},
  {Barrau}, {Barrin}, {Bartoloni}, {Basara}, {Battarbee}, {Battiston}, {Bazo},
  {Becker}, {Behlmann}, {Beischer}, {Berdugo}, {Bertucci}, {Bigongiari},
  {Bindi}, {Bizzaglia}, {Bizzarri}, {Boella}, {de Boer}, {Bollweg},
  {Bonnivard}, {Borgia}, {Borsini}, {Boschini}, {Bourquin}, {Burger}, {Cadoux},
  {Cai}, {Capell}, {Caroff}, {Casaus}, {Cascioli}, {Castellini}, {Cernuda},
  {Cervelli}, {Chae}, {Chang}, {Chen}, {Chen}, {Cheng}, {Chen}, {Cheng},
  {Chikanian}, {Chou}, {Choumilov}, {Choutko}, {Chung}, {Clark}, {Clavero},
  {Coignet}, {Consolandi}, {Contin}, {Corti}, {Coste}, {Crispoltoni}, {Cui},
  {Dai}, {Delgado}, {Della Torre}, {Demirk{\"o}z}, {Derome}, {Di Falco}, {Di
  Masso}, {Dimiccoli}, {D{\'\i}az}, {von Doetinchem}, {Donnini}, {Du},
  {Duranti}, {D'Urso}, {Eline}, {Eppling}, {Eronen}, {Fan}, {Farnesini},
  {Feng}, {Fiandrini}, {Fiasson}, {Finch}, {Fisher}, {Galaktionov}, {Gallucci},
  {Garc{\'\i}a}, {Garc{\'\i}a-L{\'o}pez}, {Gargiulo}, {Gast}, {Gebauer},
  {Gervasi}, {Ghelfi}, {Gillard}, {Giovacchini}, {Goglov}, {Gong}, {Goy},
  {Grabski}, {Grandi}, {Graziani}, {Guandalini}, {Guerri}, {Guo}, {Habiby},
  {Haino}, {Han}, {He}, {Heil}, {Hoffman}, {Hsieh}, {Huang}, {Huh}, {Incagli},
  {Ionica}, {Jang}, {Jinchi}, {Kanishev}, {Kim}, {Kim}, {Kirn}, {Kossakowski},
  {Kounina}, {Kounine}, {Koutsenko}, {Krafczyk}, {Kunz}, {La Vacca}, {Laudi},
  {Laurenti}, {Lazzizzera}, {Lebedev}, {Lee}, {Lee}, {Leluc}, {Li}, {Li}, {Li},
  {Li}, {Li}, {Li}, {Li}, {Li}, {Li}, {Lim}, {Lin}, {Lipari}, {Lippert}, {Liu},
  {Liu}, {Lomtadze}, {Lu}, {Lu}, {Luebelsmeyer}, {Luo}, {Luo}, {Lv}, {Majka},
  {Malinin}, {Ma{\~n}{\'a}}, {Mar{\'\i}n}, {Martin}, {Mart{\'\i}nez}, {Masi},
  {Maurin}, {Menchaca-Rocha}, {Meng}, {Mo}, {Morescalchi}, {Mott},
  {M{\"u}ller}, {Ni}, {Nikonov}, {Nozzoli}, {Nunes}, {Obermeier}, {Oliva},
  {Orcinha}, {Palmonari}, {Palomares}, {Paniccia}, {Papi}, {Pauluzzi},
  {Pedreschi}, {Pensotti}, {Pereira}, {Pilo}, {Piluso}, {Pizzolotto},
  {Plyaskin}, {Pohl}, {Poireau}, {Postaci}, {Putze}, {Quadrani}, {Qi},
  {R{\"a}ih{\"a}}, {Rancoita}, {Rapin}, {Ricol}, {Rodr{\'\i}guez},
  {Rosier-Lees}, {Rozhkov}, {Rozza}, {Sagdeev}, {Sandweiss}, {Saouter},
  {Sbarra}, {Schael}, {Schmidt}, {Schuckardt}, {Schulz von Dratzig},
  {Schwering}, {Scolieri}, {Seo}, {Shan}, {Shan}, {Shi}, {Shi}, {Shi},
  {Siedenburg}, {Son}, {Spada}, {Spinella}, {Sun}, {Sun}, {Tacconi}, {Tang},
  {Tang}, {Tang}, {Tao}, {Tescaro}, {Ting}, {Ting}, {Tomassetti}, {Torsti},
  {T{\"u}rko{\v{g}}lu}, {Urban}, {Vagelli}, {Valente}, {Vannini}, {Valtonen},
  {Vaurynovich}, {Vecchi}, {Velasco}, {Vialle}, {Wang}, {Wang}, {Wang}, {Wang},
  {Wang}, {Weng}, {Whitman}, {Wienkenh{\"o}ver}, {Wu}, {Xia}, {Xie}, {Xie},
  {Xiong}, {Xin}, {Xu}, {Xu}, {Yan}, {Yang}, {Yang}, {Ye}, {Yi}, {Yu}, {Yu},
  {Zeissler}, {Zhang}, {Zhang}, {Zhang}, {Zhang}, {Zheng}, {Zhuang}, {Zhukov},
  {Zichichi}, {Zimmermann}, {Zuccon}, {Zurbach}, \& {AMS
  Collaboration}}]{Aguilar2014}
{Aguilar}, M., {Aisa}, D., {Alpat}, B., {et~al.} 2014, \prl, 113, 221102

\bibitem[{{Aguilar} {et~al.}(2015){Aguilar}, {Aisa}, {Alpat}, {Alvino},
  {Ambrosi}, {Andeen}, {Arruda}, {Attig}, {Azzarello}, {Bachlechner}, {Barao},
  {Barrau}, {Barrin}, {Bartoloni}, {Basara}, {Battarbee}, {Battiston}, {Bazo},
  {Becker}, {Behlmann}, {Beischer}, {Berdugo}, {Bertucci}, {Bigongiari},
  {Bindi}, {Bizzaglia}, {Bizzarri}, {Boella}, {de Boer}, {Bollweg},
  {Bonnivard}, {Borgia}, {Borsini}, {Boschini}, {Bourquin}, {Burger}, {Cadoux},
  {Cai}, {Capell}, {Caroff}, {Casaus}, {Cascioli}, {Castellini}, {Cernuda},
  {Cerreta}, {Cervelli}, {Chae}, {Chang}, {Chen}, {Chen}, {Cheng}, {Chen},
  {Cheng}, {Chou}, {Choumilov}, {Choutko}, {Chung}, {Clark}, {Clavero},
  {Coignet}, {Consolandi}, {Contin}, {Corti}, {Gil}, {Coste}, {Creus},
  {Crispoltoni}, {Cui}, {Dai}, {Delgado}, {Della Torre}, {Demirk{\"o}z},
  {Derome}, {Di Falco}, {Di Masso}, {Dimiccoli}, {D{\'\i}az}, {von Doetinchem},
  {Donnini}, {Du}, {Duranti}, {D'Urso}, {Eline}, {Eppling}, {Eronen}, {Fan},
  {Farnesini}, {Feng}, {Fiandrini}, {Fiasson}, {Finch}, {Fisher},
  {Galaktionov}, {Gallucci}, {Garc{\'\i}a}, {Garc{\'\i}a-L{\'o}pez},
  {Gargiulo}, {Gast}, {Gebauer}, {Gervasi}, {Ghelfi}, {Gillard}, {Giovacchini},
  {Goglov}, {Gong}, {Goy}, {Grabski}, {Grandi}, {Graziani}, {Guandalini},
  {Guerri}, {Guo}, {Haas}, {Habiby}, {Haino}, {Han}, {He}, {Heil}, {Hoffman},
  {Hsieh}, {Huang}, {Huh}, {Incagli}, {Ionica}, {Jang}, {Jinchi}, {Kanishev},
  {Kim}, {Kim}, {Kirn}, {Kossakowski}, {Kounina}, {Kounine}, {Koutsenko},
  {Krafczyk}, {La Vacca}, {Laudi}, {Laurenti}, {Lazzizzera}, {Lebedev}, {Lee},
  {Lee}, {Leluc}, {Levi}, {Li}, {Li}, {Li}, {Li}, {Li}, {Li}, {Li}, {Li}, {Li},
  {Lim}, {Lin}, {Lipari}, {Lippert}, {Liu}, {Liu}, {Lolli}, {Lomtadze}, {Lu},
  {Lu}, {Lu}, {Luebelsmeyer}, {Luo}, {Lv}, {Majka}, {Ma{\~n}{\'a}},
  {Mar{\'\i}n}, {Martin}, {Mart{\'\i}nez}, {Masi}, {Maurin}, {Menchaca-Rocha},
  {Meng}, {Mo}, {Morescalchi}, {Mott}, {M{\"u}ller}, {Ni}, {Nikonov},
  {Nozzoli}, {Nunes}, {Obermeier}, {Oliva}, {Orcinha}, {Palmonari},
  {Palomares}, {Paniccia}, {Papi}, {Pauluzzi}, {Pedreschi}, {Pensotti},
  {Pereira}, {Picot-Clemente}, {Pilo}, {Piluso}, {Pizzolotto}, {Plyaskin},
  {Pohl}, {Poireau}, {Postaci}, {Putze}, {Quadrani}, {Qi}, {Qin}, {Qu},
  {R{\"a}ih{\"a}}, {Rancoita}, {Rapin}, {Ricol}, {Rodr{\'\i}guez},
  {Rosier-Lees}, {Rozhkov}, {Rozza}, {Sagdeev}, {Sandweiss}, {Saouter},
  {Sbarra}, {Schael}, {Schmidt}, {von Dratzig}, {Schwering}, {Scolieri}, {Seo},
  {Shan}, {Shan}, {Shi}, {Shi}, {Shi}, {Siedenburg}, {Son}, {Spada},
  {Spinella}, {Sun}, {Sun}, {Tacconi}, {Tang}, {Tang}, {Tang}, {Tao},
  {Tescaro}, {Ting}, {Ting}, {Tomassetti}, {Torsti}, {T{\"u}rko{\v{g}}lu},
  {Urban}, {Vagelli}, {Valente}, {Vannini}, {Valtonen}, {Vaurynovich},
  {Vecchi}, {Velasco}, {Vialle}, {Vitale}, {Vitillo}, {Wang}, {Wang}, {Wang},
  {Wang}, {Wang}, {Wang}, {Weng}, {Whitman}, {Wienkenh{\"o}ver}, {Wu}, {Wu},
  {Xia}, {Xie}, {Xie}, {Xiong}, {Xin}, {Xu}, {Xu}, {Yan}, {Yang}, {Yang}, {Ye},
  {Yi}, {Yu}, {Yu}, {Zeissler}, {Zhang}, {Zhang}, {Zhang}, {Zhang}, {Zheng},
  {Zhuang}, {Zhukov}, {Zichichi}, {Zimmermann}, {Zuccon}, {Zurbach}, \& {AMS
  Collaboration}}]{Aguilar2015}
{Aguilar}, M., {Aisa}, D., {Alpat}, B., {et~al.} 2015, \prl, 114, 171103

\bibitem[{{Aguilar} {et~al.}(2019){Aguilar}, {Ali Cavasonza}, {Ambrosi},
  {Arruda}, {Attig}, {Azzarello}, {Bachlechner}, {Barao}, {Barrau}, {Barrin},
  {Bartoloni}, {Basara}, {Ba{\c{s}}e{\v{g}}mez-du Pree}, {Battiston}, {Becker},
  {Behlmann}, {Beischer}, {Berdugo}, {Bertucci}, {Bindi}, {de Boer}, {Bollweg},
  {Borgia}, {Boschini}, {Bourquin}, {Bueno}, {Burger}, {Burger}, {Cai},
  {Capell}, {Caroff}, {Casaus}, {Castellini}, {Cervelli}, {Chang}, {Chen},
  {Chen}, {Chen}, {Cheng}, {Chou}, {Choutko}, {Chung}, {Clark}, {Coignet},
  {Consolandi}, {Contin}, {Corti}, {Crispoltoni}, {Cui}, {Dadzie}, {Dai},
  {Datta}, {Delgado}, {Della Torre}, {Demirk{\"o}z}, {Derome}, {Di Falco},
  {Dimiccoli}, {D{\'\i}az}, {von Doetinchem}, {Dong}, {Donnini}, {Duranti},
  {Egorov}, {Eline}, {Eronen}, {Feng}, {Fiandrini}, {Fisher}, {Formato},
  {Galaktionov}, {Garc{\'\i}a-L{\'o}pez}, {Gargiulo}, {Gast}, {Gebauer},
  {Gervasi}, {Giovacchini}, {G{\'o}mez-Coral}, {Gong}, {Goy}, {Grabski},
  {Grandi}, {Graziani}, {Guo}, {Haino}, {Han}, {He}, {Heil}, {Hsieh}, {Huang},
  {Huang}, {Incagli}, {Jia}, {Jinchi}, {Kanishev}, {Khiali}, {Kirn}, {Konak},
  {Kounina}, {Kounine}, {Koutsenko}, {Kulemzin}, {La Vacca}, {Laudi},
  {Laurenti}, {Lazzizzera}, {Lebedev}, {Lee}, {Lee}, {Leluc}, {Li}, {Li}, {Li},
  {Li}, {Light}, {Lin}, {Lippert}, {Liu}, {Liu}, {Liu}, {Lu}, {Lu},
  {Luebelsmeyer}, {Luo}, {Luo}, {Luo}, {Lyu}, {Machate}, {Ma{\~n}{\'a}},
  {Mar{\'\i}n}, {Martin}, {Mart{\'\i}nez}, {Masi}, {Maurin}, {Menchaca-Rocha},
  {Meng}, {Mo}, {Molero}, {Mott}, {Mussolin}, {Nelson}, {Ni}, {Nikonov},
  {Nozzoli}, {Oliva}, {Orcinha}, {Palermo}, {Palmonari}, {Paniccia}, {Pashnin},
  {Pauluzzi}, {Pensotti}, {Perrina}, {Phan}, {Picot-Clemente}, {Plyaskin},
  {Pohl}, {Poireau}, {Popkow}, {Quadrani}, {Qi}, {Qin}, {Qu}, {Rancoita},
  {Rapin}, {Conde}, {Rosier-Lees}, {Rozhkov}, {Rozza}, {Sagdeev}, {Solano},
  {Schael}, {Schmidt}, {Schulz von Dratzig}, {Schwering}, {Seo}, {Shan}, {Shi},
  {Siedenburg}, {Song}, {Sun}, {Tacconi}, {Tang}, {Tang}, {Tian}, {Ting},
  {Ting}, {Tomassetti}, {Torsti}, {Urban}, {Vagelli}, {Valente}, {Valtonen},
  {V{\'a}zquez Acosta}, {Vecchi}, {Velasco}, {Vialle}, {Viz{\'a}n}, {Wang},
  {Wang}, {Wang}, {Wang}, {Wang}, {Wang}, {Wei}, {Weng}, {Wu}, {Xiong}, {Xu},
  {Yan}, {Yang}, {Yi}, {Yu}, {Yu}, {Zannoni}, {Zeissler}, {Zhang}, {Zhang},
  {Zhang}, {Zhang}, {Zhao}, {Zheng}, {Zhuang}, {Zhukov}, {Zichichi},
  {Zimmermann}, {Zuccon}, \& {AMS Collaboration}}]{Aguilar2019}
{Aguilar}, M., {Ali Cavasonza}, L., {Ambrosi}, G., {et~al.} 2019, \prl, 122,
  041102

\bibitem[{{Alves} {et~al.}(2018){Alves}, {Boulanger}, {Ferri{\`e}re}, \&
  {Montier}}]{Alves2018}
{Alves}, M.~I.~R., {Boulanger}, F., {Ferri{\`e}re}, K., \& {Montier}, L. 2018,
  \aap, 611, L5

\bibitem[{{Andersson} \& {Potter}(2005)}]{Andersson2005}
{Andersson}, B.~G. \& {Potter}, S.~B. 2005, \mnras, 356, 1088

\bibitem[{{Andersson} \& {Potter}(2006)}]{Andersson2006}
{Andersson}, B.~G. \& {Potter}, S.~B. 2006, \apjl, 640, L51

\bibitem[{{Beck}(2015)}]{Beck2015}
{Beck}, R. 2015, \aapr, 24, 4

\bibitem[{{Beck} {et~al.}(2003){Beck}, {Shukurov}, {Sokoloff}, \&
  {Wielebinski}}]{Beck2003}
{Beck}, R., {Shukurov}, A., {Sokoloff}, D., \& {Wielebinski}, R. 2003, \aap,
  411, 99

\bibitem[{{Berdyugin} {et~al.}(2014){Berdyugin}, {Piirola}, \&
  {Teerikorpi}}]{Berdyugin2014}
{Berdyugin}, A., {Piirola}, V., \& {Teerikorpi}, P. 2014, \aap, 561, A24

\bibitem[{{Breitschwerdt} {et~al.}(2016){Breitschwerdt}, {Feige}, {Schulreich},
  {Avillez}, {Dettbarn}, \& {Fuchs}}]{Breitschwerdt2016}
{Breitschwerdt}, D., {Feige}, J., {Schulreich}, M.~M., {et~al.} 2016, \nat,
  532, 73

\bibitem[{{Cordes} \& {Lazio}(2002)}]{Cordes2002}
{Cordes}, J.~M. \& {Lazio}, T.~J.~W. 2002, arXiv e-prints, arXiv:0207156

\bibitem[{{Cox} \& {Reynolds}(1987)}]{Cox1987}
{Cox}, D.~P. \& {Reynolds}, R.~J. 1987, \araa, 25, 303

\bibitem[{{Dickey} {et~al.}(2022){Dickey}, {West}, {Thomson}, {Landecker},
  {Bracco}, {Carretti}, {Han}, {Hill}, {Ma}, {Mao}, {Ordog}, {Brown},
  {Douglas}, {Erceg}, {Jeli{\'c}}, {Kothes}, \& {Wolleben}}]{Dickey2022}
{Dickey}, J.~M., {West}, J., {Thomson}, A. J.~M., {et~al.} 2022, \apj, 940, 75

\bibitem[{{Edenhofer} {et~al.}(2023){Edenhofer}, {Zucker}, {Frank}, {Saydjari},
  {Speagle}, {Finkbeiner}, \& {En{\ss}lin}}]{Edenhofer2023}
{Edenhofer}, G., {Zucker}, C., {Frank}, P., {et~al.} 2023, arXiv e-prints,
  arXiv:2308.01295

\bibitem[{{Erceg} {et~al.}(2024{\natexlab{a}}){Erceg}, {Jeli{\'c}},
  {Haverkorn}, {Bracco}, {Ceraj}, {Turi{\'c}}, \& {Soler}}]{Erceg2024a}
{Erceg}, A., {Jeli{\'c}}, V., {Haverkorn}, M., {et~al.} 2024{\natexlab{a}},
  \aap, 687, A23

\bibitem[{{Erceg} {et~al.}(2024{\natexlab{b}}){Erceg}, {Jeli{\'c}},
  {Haverkorn}, {Gajovi{\'c}}, {Hardcastle}, {Shimwell}, \&
  {Tasse}}]{Erceg2024b}
{Erceg}, A., {Jeli{\'c}}, V., {Haverkorn}, M., {et~al.} 2024{\natexlab{b}},
  \aap, 688, A200

\bibitem[{{Ferri{\`e}re} {et~al.}(1991){Ferri{\`e}re}, {Mac Low}, \&
  {Zweibel}}]{Ferriere1991a}
{Ferri{\`e}re}, K.~M., {Mac Low}, M.-M., \& {Zweibel}, E.~G. 1991, \apj, 375,
  239

\bibitem[{{Ferri{\`e}re} \& {Zweibel}(1991)}]{Ferriere1991b}
{Ferri{\`e}re}, K.~M. \& {Zweibel}, E.~G. 1991, \apj, 383, 602

\bibitem[{{Frisch} {et~al.}(2012){Frisch}, {Andersson}, {Berdyugin}, {Piirola},
  {DeMajistre}, {Funsten}, {Magalhaes}, {Seriacopi}, {McComas}, {Schwadron},
  {Slavin}, \& {Wiktorowicz}}]{Frisch2012}
{Frisch}, P.~C., {Andersson}, B.~G., {Berdyugin}, A., {et~al.} 2012, \apj, 760,
  106

\bibitem[{{Gaia Collaboration}(2016)}]{Gaia2016}
{Gaia Collaboration}. 2016, \aap, 595, A1

\bibitem[{{Gontcharov} \& {Mosenkov}(2019)}]{Gontcharov2019}
{Gontcharov}, G.~A. \& {Mosenkov}, A.~V. 2019, \mnras, 483, 299

\bibitem[{{G{\'o}rski} {et~al.}(2005){G{\'o}rski}, {Hivon}, {Banday},
  {Wandelt}, {Hansen}, {Reinecke}, \& {Bartelmann}}]{Gorski2005}
{G{\'o}rski}, K.~M., {Hivon}, E., {Banday}, A.~J., {et~al.} 2005, \apj, 622,
  759

\bibitem[{{Green} {et~al.}(2019){Green}, {Schlafly}, {Zucker}, {Speagle}, \&
  {Finkbeiner}}]{Green2019}
{Green}, G.~M., {Schlafly}, E., {Zucker}, C., {Speagle}, J.~S., \&
  {Finkbeiner}, D. 2019, \apj, 887, 93

\bibitem[{{Halal} {et~al.}(2024){Halal}, {Clark}, \& {Tahani}}]{Halal2024}
{Halal}, G., {Clark}, S.~E., \& {Tahani}, M. 2024, \apj, 973, 54

\bibitem[{{Haverkorn}(2015)}]{Haverkorn2015}
{Haverkorn}, M. 2015, in Astrophysics and Space Science Library, Vol. 407,
  Magnetic Fields in Diffuse Media, ed. A.~{Lazarian}, E.~M. {de Gouveia Dal
  Pino}, \& C.~{Melioli}, 483

\bibitem[{{Heiles}(1998)}]{Heiles1998}
{Heiles}, C. 1998, in IAU Colloq. 166: The Local Bubble and Beyond, ed.
  D.~{Breitschwerdt}, M.~J. {Freyberg}, \& J.~{Truemper}, Vol. 506, 229--238

\bibitem[{{Jaffe}(2019)}]{Jaffe2019}
{Jaffe}, T.~R. 2019, Galaxies, 7, 52

\bibitem[{{Jaffe} {et~al.}(2010){Jaffe}, {Leahy}, {Banday}, {Leach}, {Lowe}, \&
  {Wilkinson}}]{Jaffe10}
{Jaffe}, T.~R., {Leahy}, J.~P., {Banday}, A.~J., {et~al.} 2010, \mnras, 401,
  1013

\bibitem[{{Jansson} \& {Farrar}(2012{\natexlab{a}})}]{Jansson12}
{Jansson}, R. \& {Farrar}, G.~R. 2012{\natexlab{a}}, \apj, 757, 14

\bibitem[{{Jansson} \& {Farrar}(2012{\natexlab{b}})}]{Jansson2012b}
{Jansson}, R. \& {Farrar}, G.~R. 2012{\natexlab{b}}, \apjl, 761, L11

\bibitem[{{Kim} \& {Ostriker}(2015)}]{Kim2015}
{Kim}, C.-G. \& {Ostriker}, E.~C. 2015, \apj, 802, 99

\bibitem[{{Korochkin} {et~al.}(2024){Korochkin}, {Semikoz}, \&
  {Tinyakov}}]{Korochkin2024}
{Korochkin}, A., {Semikoz}, D., \& {Tinyakov}, P. 2024, arXiv e-prints,
  arXiv:2407.02148

\bibitem[{{Krause} \& {Diehl}(2014)}]{Krause2014}
{Krause}, M. G.~H. \& {Diehl}, R. 2014, \apjl, 794, L21

\bibitem[{{Lallement} {et~al.}(2019){Lallement}, {Babusiaux}, {Vergely},
  {Katz}, {Arenou}, {Valette}, {Hottier}, \& {Capitanio}}]{Lallement2019}
{Lallement}, R., {Babusiaux}, C., {Vergely}, J.~L., {et~al.} 2019, \aap, 625,
  A135

\bibitem[{{Lallement} {et~al.}(2022){Lallement}, {Vergely}, {Babusiaux}, \&
  {Cox}}]{Lallement2022}
{Lallement}, R., {Vergely}, J.~L., {Babusiaux}, C., \& {Cox}, N.~L.~J. 2022,
  \aap, 661, A147

\bibitem[{{Lallement} {et~al.}(2003){Lallement}, {Welsh}, {Vergely}, {Crifo},
  \& {Sfeir}}]{Lallement2003}
{Lallement}, R., {Welsh}, B.~Y., {Vergely}, J.~L., {Crifo}, F., \& {Sfeir}, D.
  2003, \aap, 411, 447

\bibitem[{{Leike} \& {En{\ss}lin}(2019)}]{Leike2019}
{Leike}, R.~H. \& {En{\ss}lin}, T.~A. 2019, \aap, 631, A32

\bibitem[{{Leroy}(1999)}]{Leroy1999}
{Leroy}, J.~L. 1999, \aap, 346, 955

\bibitem[{{Liu} {et~al.}(2017){Liu}, {Chiao}, {Collier}, {Cravens}, {Galeazzi},
  {Koutroumpa}, {Kuntz}, {Lallement}, {Lepri}, {McCammon}, {Morgan}, {Porter},
  {Snowden}, {Thomas}, {Uprety}, {Ursino}, \& {Walsh}}]{Liu2017}
{Liu}, W., {Chiao}, M., {Collier}, M.~R., {et~al.} 2017, \apj, 834, 33

\bibitem[{{Longair}(2011)}]{Longair2011}
{Longair}, M.~S. 2011, {High Energy Astrophysics}

\bibitem[{{Lopez} {et~al.}(2011){Lopez}, {Ramirez-Ruiz}, {Huppenkothen},
  {Badenes}, \& {Pooley}}]{Lopez2011}
{Lopez}, L.~A., {Ramirez-Ruiz}, E., {Huppenkothen}, D., {Badenes}, C., \&
  {Pooley}, D.~A. 2011, \apj, 732, 114

\bibitem[{{Maconi} {et~al.}(2023){Maconi}, {Soler}, {Reissl}, {Girichidis},
  {Klessen}, {Hennebelle}, {Molinari}, {Testi}, {Smith}, {Sormani}, {Teh}, \&
  {Traficante}}]{Maconi2023}
{Maconi}, E., {Soler}, J.~D., {Reissl}, S., {et~al.} 2023, \mnras, 523, 5995

\bibitem[{{Magalh{\~a}es} {et~al.}(2005){Magalh{\~a}es}, {Pereyra},
  {Melgarejo}, {de Matos}, {Carciofi}, {Benedito}, {Valentim}, {Vidotto}, {da
  Silva}, {de Souza}, {Faria}, \& {Gabriel}}]{Magalhaes2005}
{Magalh{\~a}es}, A.~M., {Pereyra}, A., {Melgarejo}, R., {et~al.} 2005, in
  Astronomical Society of the Pacific Conference Series, Vol. 343, Astronomical
  Polarimetry: Current Status and Future Directions, ed. A.~{Adamson},
  C.~{Aspin}, C.~{Davis}, \& T.~{Fujiyoshi}, 305

\bibitem[{{Ma{\'\i}z-Apell{\'a}niz}(2001)}]{MaizApell2001}
{Ma{\'\i}z-Apell{\'a}niz}, J. 2001, \apjl, 560, L83

\bibitem[{{Medan} \& {Andersson}(2019)}]{Medan2019}
{Medan}, I. \& {Andersson}, B.~G. 2019, \apj, 873, 87

\bibitem[{{Ocker} {et~al.}(2020){Ocker}, {Cordes}, \& {Chatterjee}}]{Ocker2020}
{Ocker}, S.~K., {Cordes}, J.~M., \& {Chatterjee}, S. 2020, \apj, 897, 124

\bibitem[{{O'Neill} {et~al.}(2024{\natexlab{a}}){O'Neill}, {Goodman}, {Soler},
  {Zucker}, \& {Han}}]{ONeill2024b}
{O'Neill}, T.~J., {Goodman}, A.~A., {Soler}, J.~D., {Zucker}, C., \& {Han},
  J.~J. 2024{\natexlab{a}}, arXiv e-prints, arXiv:2410.17341

\bibitem[{{O'Neill} {et~al.}(2024{\natexlab{b}}){O'Neill}, {Zucker}, {Goodman},
  \& {Edenhofer}}]{ONeill2024}
{O'Neill}, T.~J., {Zucker}, C., {Goodman}, A.~A., \& {Edenhofer}, G.
  2024{\natexlab{b}}, \apj, 973, 136

\bibitem[{{Orlando} {et~al.}(2016){Orlando}, {Miceli}, {Pumo}, \&
  {Bocchino}}]{Orlando2016}
{Orlando}, S., {Miceli}, M., {Pumo}, M.~L., \& {Bocchino}, F. 2016, \apj, 822,
  22

\bibitem[{{Pakmor} {et~al.}(2018){Pakmor}, {Guillet}, {Pfrommer}, {G{\'o}mez},
  {Grand}, {Marinacci}, {Simpson}, \& {Springel}}]{Pakmor2018}
{Pakmor}, R., {Guillet}, T., {Pfrommer}, C., {et~al.} 2018, \mnras, 481, 4410

\bibitem[{{Parker}(1970)}]{Parker1970}
{Parker}, E.~N. 1970, \apj, 162, 665

\bibitem[{{Pelgrims} {et~al.}(2020){Pelgrims}, {Ferri{\`e}re}, {Boulanger},
  {Lallement}, \& {Montier}}]{Pelgrims2020}
{Pelgrims}, V., {Ferri{\`e}re}, K., {Boulanger}, F., {Lallement}, R., \&
  {Montier}, L. 2020, \aap, 636, A17

\bibitem[{{Pelgrims} {et~al.}(2021){Pelgrims}, {Mac{\'\i}as-P{\'e}rez}, \&
  {Ruppin}}]{Pelgrims2021b}
{Pelgrims}, V., {Mac{\'\i}as-P{\'e}rez}, J.~F., \& {Ruppin}, F. 2021, \aap,
  652, A130

\bibitem[{{Pelgrims} {et~al.}(2024){Pelgrims}, {Mandarakas}, {Skalidis},
  {Tassis}, {Panopoulou}, {Pavlidou}, {Blinov}, {Kiehlmann}, {Clark},
  {Hensley}, {Romanopoulos}, {Basyrov}, {Eriksen}, {Falalaki}, {Ghosh},
  {Gjerl{\o}w}, {Kypriotakis}, {Maharana}, {Papadaki}, {Pearson}, {Potter},
  {Ramaprakash}, {Readhead}, \& {Wehus}}]{Pelgrims2024}
{Pelgrims}, V., {Mandarakas}, N., {Skalidis}, R., {et~al.} 2024, \aap, 684,
  A162

\bibitem[{{Pelgrims} {et~al.}(2023){Pelgrims}, {Panopoulou}, {Tassis},
  {Pavlidou}, {Basyrov}, {Blinov}, {Gjerl{\ensuremath{\varnothing}}w},
  {Kiehlmann}, {Mandarakas}, {Papadaki}, {Skalidis}, {Tsouros}, {Anche},
  {Eriksen}, {Ghosh}, {Kypriotakis}, {Maharana}, {Ntormousi}, {Pearson},
  {Potter}, {Ramaprakash}, {Readhead}, \& {Wehus}}]{Pelgrims2023}
{Pelgrims}, V., {Panopoulou}, G.~V., {Tassis}, K., {et~al.} 2023, \aap, 670,
  A164

\bibitem[{{Planck Collaboration Int. XLII}(2016)}]{PlanckInt2016XLII}
{Planck Collaboration Int. XLII}. 2016, \aap, 596, A103

\bibitem[{{Planck Collaboration IV}(2020)}]{PlanckIV2020}
{Planck Collaboration IV}. 2020, \aap, 641, A4

\bibitem[{{Reissl} {et~al.}(2023){Reissl}, {Klessen}, {Pellegrini}, {Rahner},
  {Pakmor}, {Grand}, {G{\'o}mez}, {Marinacci}, \& {Springel}}]{Reissl2023}
{Reissl}, S., {Klessen}, R.~S., {Pellegrini}, E.~W., {et~al.} 2023, Nature
  Astronomy, 7, 1295

\bibitem[{{Santos} {et~al.}(2011){Santos}, {Corradi}, \& {Reis}}]{Santos2011}
{Santos}, F.~P., {Corradi}, W., \& {Reis}, W. 2011, \apj, 728, 104

\bibitem[{{Schulreich} {et~al.}(2023){Schulreich}, {Feige}, \&
  {Breitschwerdt}}]{Schulreich2023}
{Schulreich}, M.~M., {Feige}, J., \& {Breitschwerdt}, D. 2023, \aap, 680, A39

\bibitem[{{Seta} \& {Federrath}(2022)}]{Seta2022}
{Seta}, A. \& {Federrath}, C. 2022, \mnras, 514, 957

\bibitem[{{Shimwell} {et~al.}(2022){Shimwell}, {Hardcastle}, {Tasse}, {Best},
  {R{\"o}ttgering}, {Williams}, {Botteon}, {Drabent}, {Mechev}, {Shulevski},
  {van Weeren}, {Bester}, {Br{\"u}ggen}, {Brunetti}, {Callingham}, {Chy{\.z}y},
  {Conway}, {Dijkema}, {Duncan}, {de Gasperin}, {Hale}, {Haverkorn}, {Hugo},
  {Jackson}, {Mevius}, {Miley}, {Morabito}, {Morganti}, {Offringa}, {Oonk},
  {Rafferty}, {Sabater}, {Smith}, {Schwarz}, {Smirnov}, {O'Sullivan},
  {Vedantham}, {White}, {Albert}, {Alegre}, {Asabere}, {Bacon}, {Bonafede},
  {Bonnassieux}, {Brienza}, {Bilicki}, {Bonato}, {Calistro Rivera}, {Cassano},
  {Cochrane}, {Croston}, {Cuciti}, {Dallacasa}, {Danezi}, {Dettmar}, {Di
  Gennaro}, {Edler}, {En{\ss}lin}, {Emig}, {Franzen}, {Garc{\'\i}a-Vergara},
  {Grange}, {G{\"u}rkan}, {Hajduk}, {Heald}, {Heesen}, {Hoang}, {Hoeft},
  {Horellou}, {Iacobelli}, {Jamrozy}, {Jeli{\'c}}, {Kondapally}, {Kukreti},
  {Kunert-Bajraszewska}, {Magliocchetti}, {Mahatma}, {Ma{\l}ek}, {Mandal},
  {Massaro}, {Meyer-Zhao}, {Mingo}, {Mostert}, {Nair}, {Nakoneczny},
  {Nikiel-Wroczy{\'n}ski}, {Orr{\'u}}, {Pajdosz-{\'S}mierciak}, {Pasini},
  {Prandoni}, {van Piggelen}, {Rajpurohit}, {Retana-Montenegro}, {Riseley},
  {Rowlinson}, {Saxena}, {Schrijvers}, {Sweijen}, {Siewert}, {Timmerman},
  {Vaccari}, {Vink}, {West}, {Wo{\l}owska}, {Zhang}, \& {Zheng}}]{Shimwell2022}
{Shimwell}, T.~W., {Hardcastle}, M.~J., {Tasse}, C., {et~al.} 2022, \aap, 659,
  A1

\bibitem[{{Skalidis} \& {Pelgrims}(2019)}]{Skalidis2019}
{Skalidis}, R. \& {Pelgrims}, V. 2019, \aap, 631, L11

\bibitem[{{Tassis} {et~al.}(2018){Tassis}, {Ramaprakash}, {Readhead}, {Potter},
  {Wehus}, {Panopoulou}, {Blinov}, {Eriksen}, {Hensley}, {Karakci},
  {Kypriotakis}, {Maharana}, {Ntormousi}, {Pavlidou}, {Pearson}, \&
  {Skalidis}}]{Tassis2018}
{Tassis}, K., {Ramaprakash}, A.~N., {Readhead}, A. C.~S., {et~al.} 2018, arXiv
  e-prints, arXiv:1810.05652

\bibitem[{{Tritsis} {et~al.}(2015){Tritsis}, {Panopoulou}, {Mouschovias},
  {Tassis}, \& {Pavlidou}}]{Tritsis2015}
{Tritsis}, A., {Panopoulou}, G.~V., {Mouschovias}, T.~C., {Tassis}, K., \&
  {Pavlidou}, V. 2015, \mnras, 451, 4384

\bibitem[{{Unger} \& {Farrar}(2024)}]{Unger24}
{Unger}, M. \& {Farrar}, G.~R. 2024, \apj, 970, 95

\bibitem[{{Uppal} {et~al.}(2024){Uppal}, {Ganesh}, {Pelgrims}, {Joshi}, \&
  {Sarkar}}]{Uppal2024}
{Uppal}, N., {Ganesh}, S., {Pelgrims}, V., {Joshi}, S., \& {Sarkar}, M. 2024,
  \aap, 690, A49

\bibitem[{{van der Laan}(1962)}]{vanderLaan1962}
{van der Laan}, H. 1962, \mnras, 124, 125

\bibitem[{{Vergely} {et~al.}(2022){Vergely}, {Lallement}, \&
  {Cox}}]{Vergely2022}
{Vergely}, J.~L., {Lallement}, R., \& {Cox}, N.~L.~J. 2022, \aap, 664, A174

\bibitem[{{Vidal} {et~al.}(2015){Vidal}, {Dickinson}, {Davies}, \&
  {Leahy}}]{Vidal2015}
{Vidal}, M., {Dickinson}, C., {Davies}, R.~D., \& {Leahy}, J.~P. 2015, \mnras,
  452, 656

\bibitem[{{West} {et~al.}(2016){West}, {Safi-Harb}, {Jaffe}, {Kothes},
  {Landecker}, \& {Foster}}]{West2016}
{West}, J.~L., {Safi-Harb}, S., {Jaffe}, T., {et~al.} 2016, \aap, 587, A148

\bibitem[{{Wolleben}(2007)}]{Wolleben2007}
{Wolleben}, M. 2007, \apj, 664, 349

\bibitem[{{Yao} {et~al.}(2017){Yao}, {Manchester}, \& {Wang}}]{Yao2017}
{Yao}, J.~M., {Manchester}, R.~N., \& {Wang}, N. 2017, \apj, 835, 29

\end{thebibliography}

\begin{appendix}

\section{Magnetic flux conservation}
\label{sec:Bflux}
In this appendix, we demonstrate that our analytical model for the magnetic field in thick shells of ISM bubbles with the linear displacement field leads to the conservation of the magnetic flux, before and after the explosion.
We use the same notation and convention as in Sect.~\ref{sec:AnalyticModel}.

We start by using the case of a spherical bubble.
The magnetic flux through a plane with unit normal vector $\mathbf{u}$ is given by $\mathfrak{\Phi} = \int {\rm{d}}A \, \mathbf{B}\cdot\mathbf{u}$.
To demonstrate that $\mathfrak{\Phi} = \mathfrak{\Phi}^0$, we consider a plane that contains the explosion center and with a normal vector that makes an angle $\gamma$ with $\mathbf{B}^0$.
To demonstrate that the flux is conserved we need just to integrate $\mathbf{B}\cdot\mathbf{u}$ over the surface delimited by the outer radius of the spherical bubble. The field is unchanged outside it.

Without loss of generality, we consider a spherical coordinate system centered on the explosion center and with the North pole pointing parallel to $\mathbf{u}$. The measure of the flux thus happens in the equatorial plane of this coordinate system, where $\mathbf{u} \cdot \mathbf{e}_\theta = -1$. By construction, we have $B^0_{~\theta} = \mathbf{B}^0 \cdot \mathbf{e}_\theta= - B^0 \cos\gamma$.
The initial flux through the surface is therefore given by $\mathfrak{\Phi}^0 = \pi {R_{\rm{max}}}^2 \, B^0 \cos\gamma$.

To estimate the flux of the present-time magnetic field through the same surface, we need to integrate $\mathbf{B}\cdot\mathbf{u}$. We have
\begin{align}
    \mathfrak{\Phi} &= \int {\rm{d}}A \, \mathbf{B}\cdot\mathbf{u} \\
    &= - \int {\rm{d}}A \, B_\theta \;,
\end{align}
where, for a spherical bubble and with the linear displacement model, $B_\theta(r)$ is given by
\begin{align}
    B_\theta = \left( \frac{R_{\rm{max}}}{R_{\rm{max}} - R_{\rm{min}}}\right)^2 \, \left(\frac{r-R_{\rm{min}}}{r}\right) \, B^0_{~\theta}
\end{align}
in the thick shell, and is zero for $r<R_{\rm{min}}$.
Performing the integral, we have:
\begin{align}
    \mathfrak{\Phi} &= - \int {\rm{d}}A \, B_\theta \\
    &= - \int_0^{2\pi} {\rm{d}}\phi \int_{R_{\rm{min}}}^{R_{\rm{max}}} {\rm{d}}r \, r  \, B_\theta(r) \\
    &= - 2\pi \, B^0_{~\theta} \, \left( \frac{R_{\rm{max}}}{R_{\rm{max}} - R_{\rm{min}}}\right)^2 \int_{R_{\rm{min}}}^{R_{\rm{max}}} \left(r - R_{\rm{min}} \right) \, {\rm{d}}r \\
    &= - \pi \, {R_{\rm{max}}}^2 B^0_{~\theta} \\
    &= \mathfrak{\Phi}^0 \;.
\end{align}
This shows that the flux of the magnetic field before and after the explosion is conserved in any plane containing the explosion center.
This demonstration for the magnetic flux conservation can be generalized to the case of a non-spherical bubble shell. In this case, one needs to take into account the azimuth-angle dependence of the outer and inner radius of the shell in the equatorial plane. In the general case, the flux is given by
\begin{align}
    \mathfrak{\Phi} = \frac{B^0 \cos\gamma}{2} \int_0^{2\pi} {\rm{d}}\phi \, R_{\rm{max}}(\phi) \;,
\end{align}
where $\gamma$ is the angle between the initial magnetic field direction and the normal to the plane in which magnetic flux is measured. This is exactly the expression for $\mathfrak{\Phi}^0$ for such a case.

\section{The thick shell of the Local Bubble}
\label{sec:LBthickShell}
In the last decade, precise measurements of stellar parallaxes from the \textit{Gaia} space mission (\citealt{Gaia2016}) and stellar reddening has enabled to map the dust density distribution in three dimensions (e.g., \citealt{Green2019}; \citealt{Lallement2019}; \citealt{Leike2019}; \citealt{Lallement2022}; \citealt{Vergely2022}; \citealt{Edenhofer2023}). The advent of such 3D maps enables extracting the shell of the Local Bubble as a region of higher dust density and characterizing its geometry directly from observation.
\cite{Pelgrims2020} were the first to extract the shape of the inner surface of the Local Bubble shell based on 3D dust maps. They considered the maps of \cite{Lallement2019} and \cite{Leike2019}, which were the most recent maps at the time and enabled a full-sky study to be carried out.

Drawing sightlines in the data cube outward from the Sun, they constructed radial profiles of the differential of the dust extinction $\delta A_{\rm{v}} \equiv {\rm{d}}A_{\rm{v}}/{\rm{d}}r$ (with $r$ the distance
to the Sun). For each sightline separately, they defined the inner radius of the Local Bubble shell as the distance to the first inflection point where the profile transitions from convex to concave. Having obtained this measure for all sightlines of an HEALPix angular tesselation with $N_{\rm{side}} = 128$, they then derived smoothed versions of the inner surface of the shell using spherical harmonic decompositions of the surface with limiting maximum multipole ($\ell_{\rm{max}}$).
\cite{Pelgrims2020} used these smoothed models of the inner surface of the Local Bubble shell to successfully constrain a model for the smooth component of the magnetic field in the shell, which was assumed to be very thin, through a fit of the dust polarized emission at high-Galactic latitudes. They found that the inner surface extracted from the 3D map of \cite{Lallement2019} leads to more convincing results.

Although there have been new 3D maps of the dust density and also more recent characterization of the shape of the shell of the Local Bubble based on these maps (e.g., \citealt{ONeill2024}), we choose to use the model of \cite{Pelgrims2020} obtained with the maximum multipole $\ell_{\rm{max}} = 6$ for the inner surface of the Local Bubble\footnote{\url{https://doi.org/10.7910/DVN/RHPVNC}}. This choice is motivated by the smooth, closed nature of the surface, which makes it particularly suited to our purposes. We refer to this surface as $r_{\rm{inner}}^{\ell_{\rm{max}}=6}$, both in this Appendix and in the main text. $r_{\rm{inner}}^{\ell_{\rm{max}}=6}$ consists of a sky map of distances from the Sun to the inner edge of the shell of the Local Bubble.

\begin{figure*}
    \centering
    \includegraphics[trim={0.cm .6cm 0.cm 0.cm},clip,width=.48\linewidth]{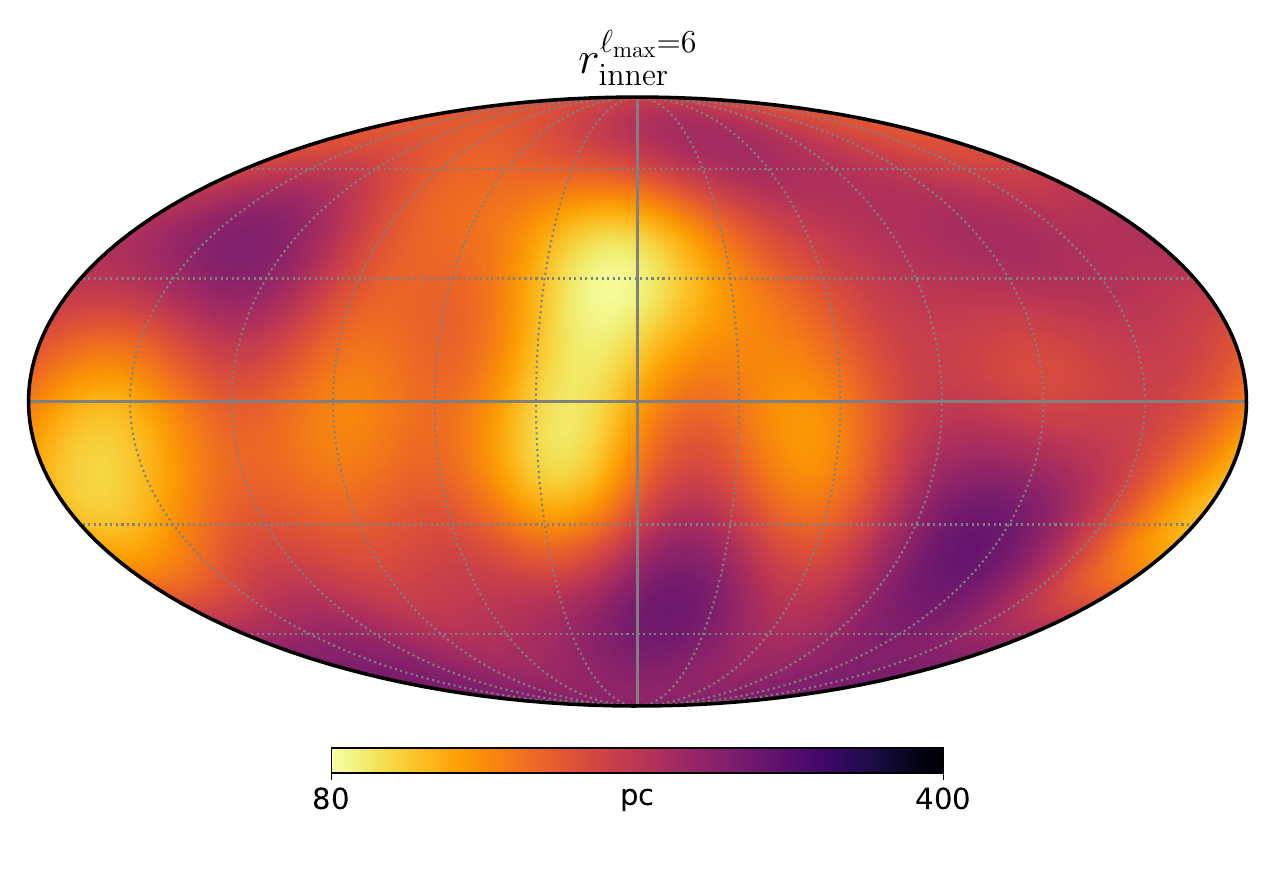}
    \includegraphics[trim={0.cm .6cm 0.cm 0.cm},clip,width=.48\linewidth]{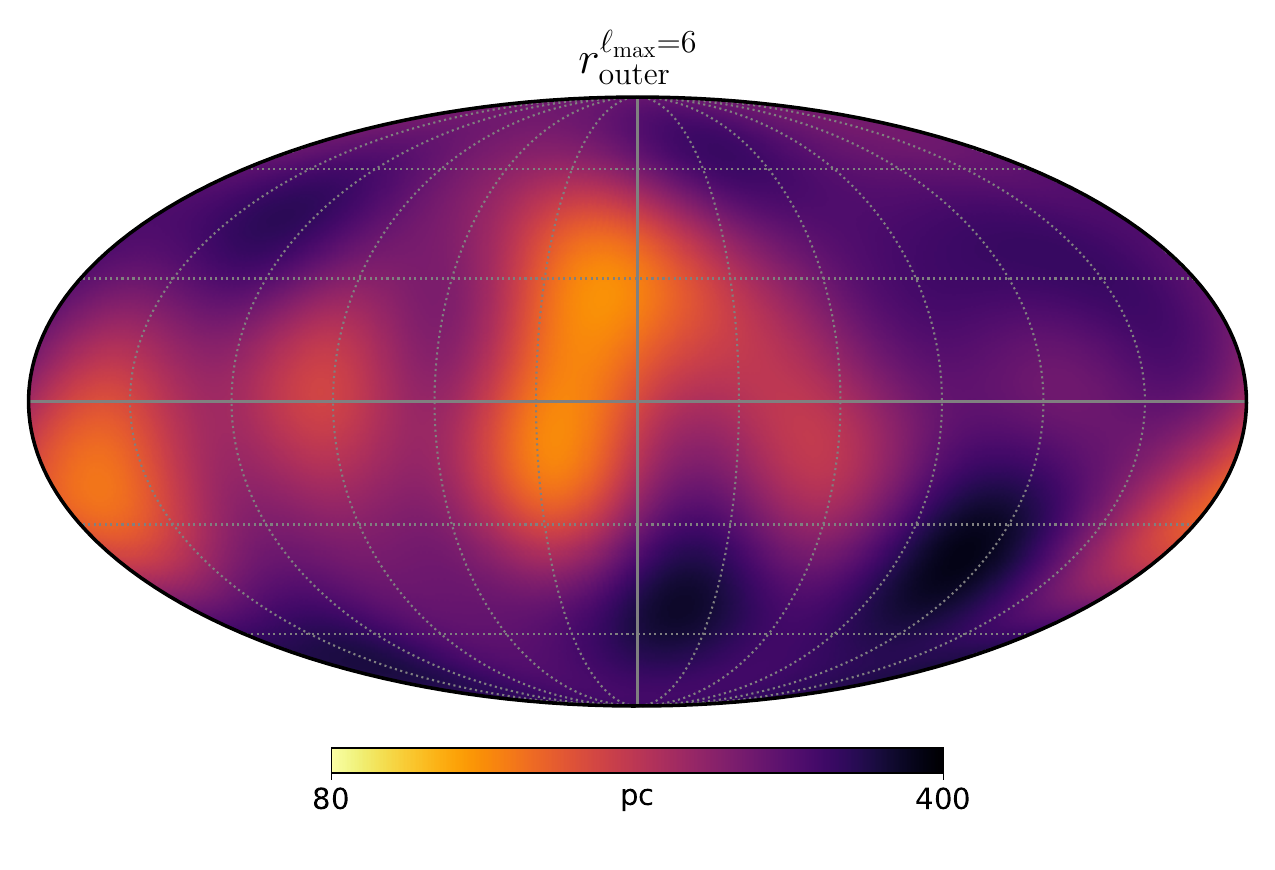}
    \\[-1.5ex]
    \caption{Sky maps of the modeled distances to the inner ($r_{\rm{inner}}^{\ell_{\rm{max}}=6}$, left) and outer ($r_{\rm{outer}}^{\ell_{\rm{max}}=6}$, right) surfaces of the shell of the Local Bubble as measured from the Sun. Projection and conventions are the same as in Fig.~\ref{fig:RMQUmaps_LBscenaros}.}
    \label{fig:r_ell6}
\end{figure*}
\begin{figure*}
    \centering
    \includegraphics[trim={1.cm 0.4cm 2.9cm 0.2cm},clip,height=.29\linewidth]{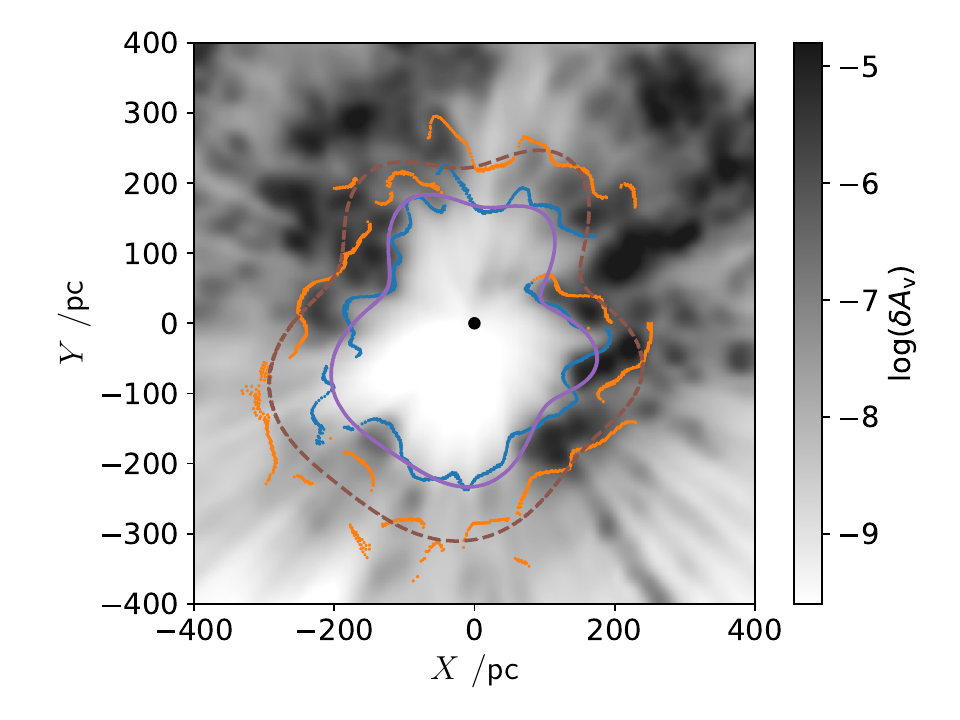}
    \includegraphics[trim={1.cm 0.4cm 2.9cm 0.2cm},clip,height=.29\linewidth]{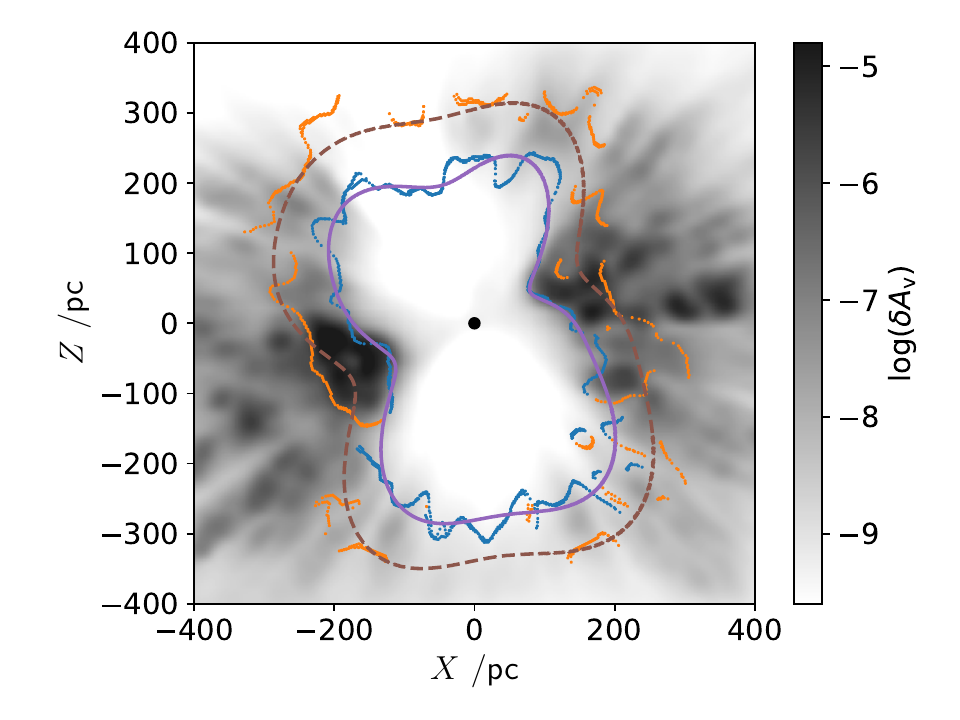}
    \includegraphics[trim={1.cm 0.4cm .6cm 0.2},clip,height=.29\linewidth]{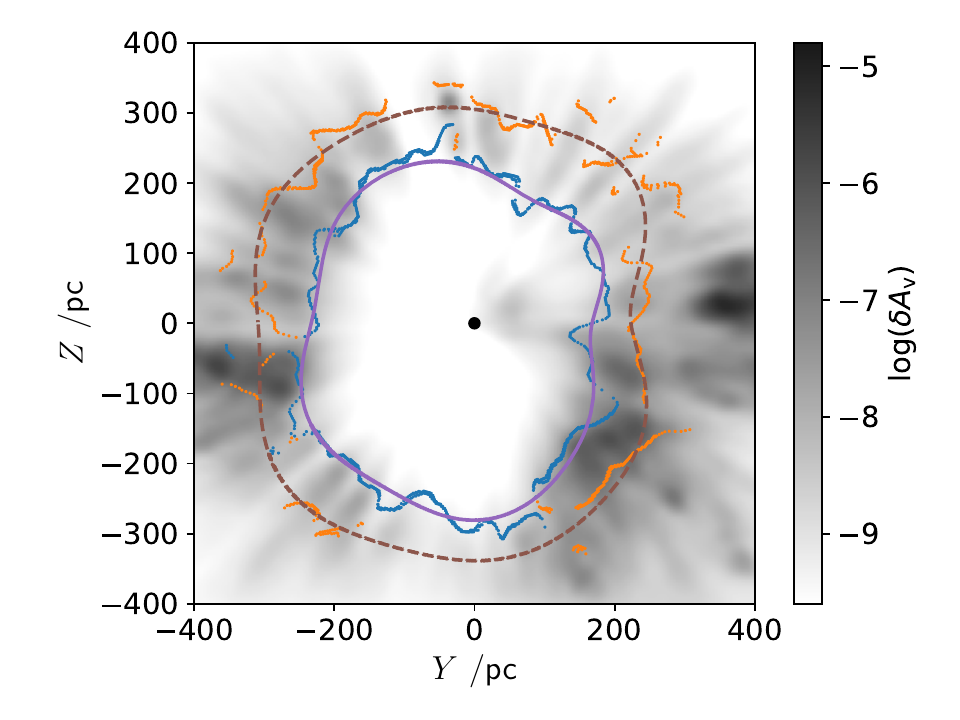}\\[-1.ex]
    \caption{Crosscuts along the planes $XY$, $XZ$, and $YZ$ in the 3D dust extinction map of \cite{Lallement2019}. We use the Heliocentric Galactic coordinates. The (common) gray scale shows $\log(\delta A_{\rm{v}})$, with $\delta A_{\rm{v}}$ the differential of the dust extinction in units of magnitude per parsec. The blue and orange dots mark the inner and outer surfaces of the Local Bubble shell measured from the radial profiles of $\delta A_{\rm{v}}$. The purple and brown continuous lines trace the intersection of our models for $r_{\rm{inner}}^{\ell_{\rm{max}}=6}$ and $r_{\rm{outer}}^{\ell_{\rm{max}}=6}$ with the respective plane, respectively.
    }
    \label{fig:CC_LocalBubbles}
\end{figure*}
\begin{figure*}
    \centering
    \includegraphics[trim={.0cm 0.4cm 0.cm 0.2cm},clip,height=.31\linewidth]{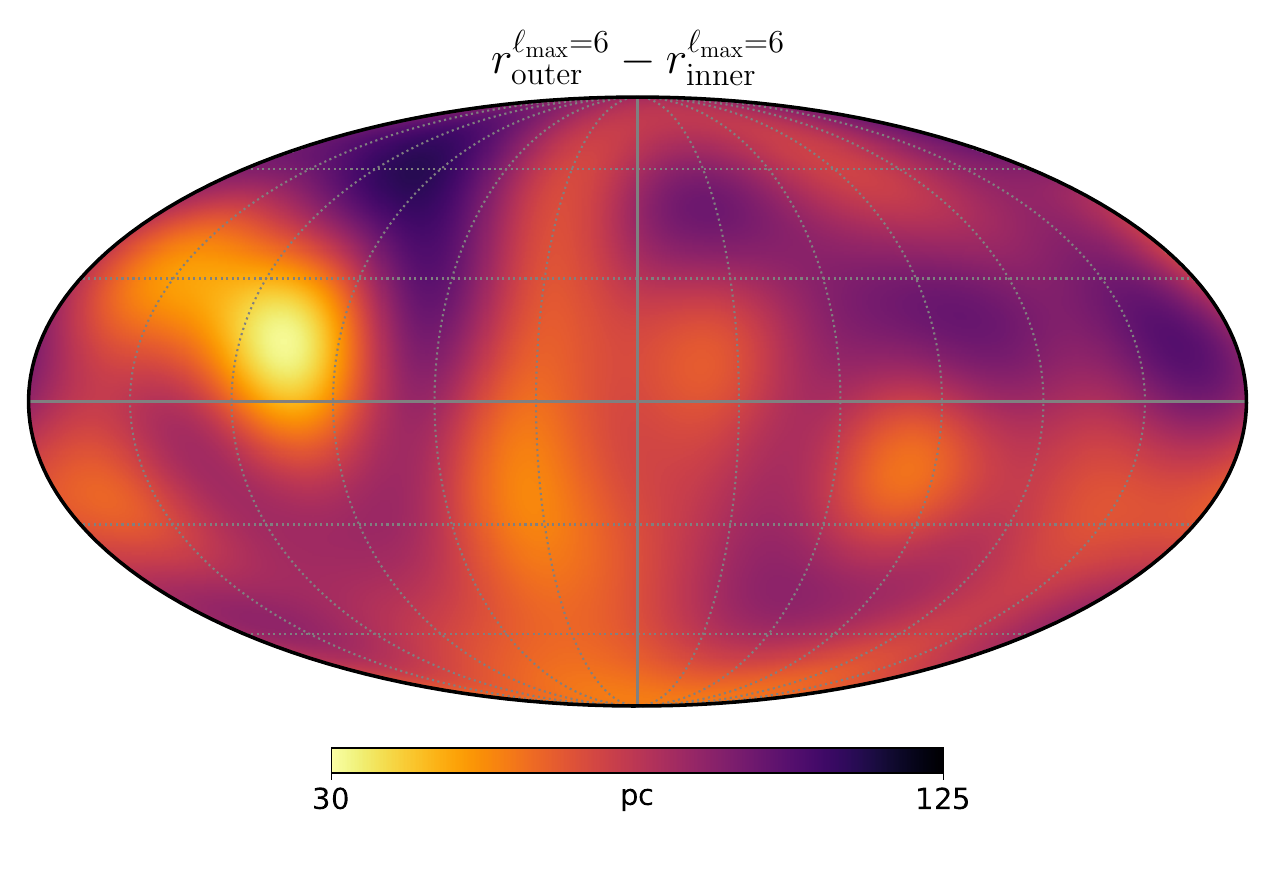}
    \includegraphics[trim={.0cm 0.4cm 0.cm 0.2cm},clip,height=.31\linewidth]{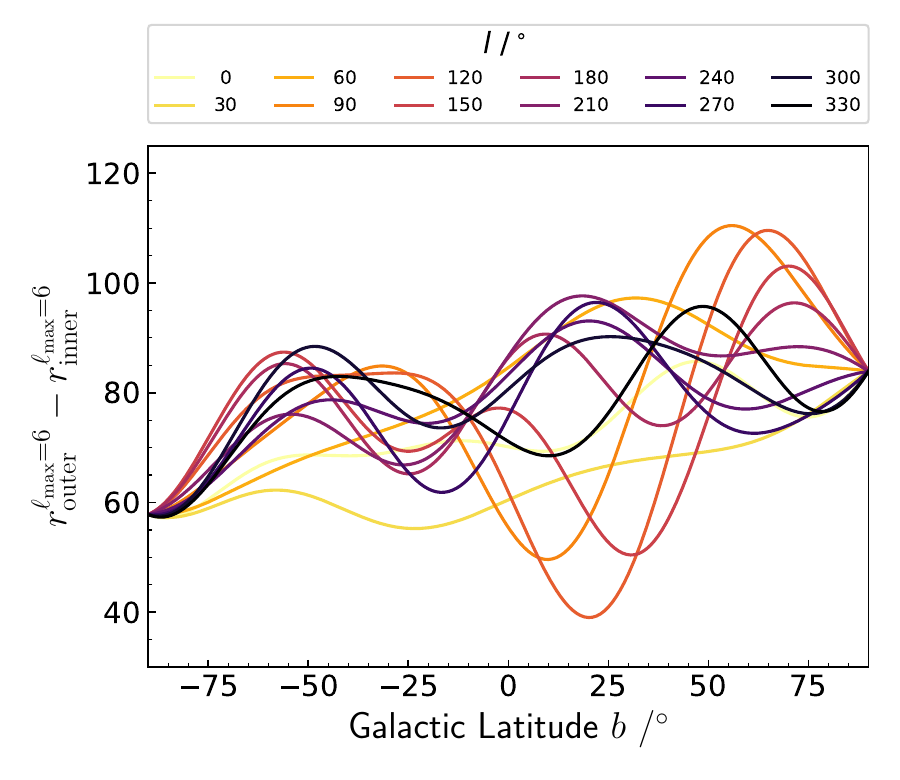}\\[-1.ex]
    \caption{Thickness of the Local Bubble shell as seen from the Sun.
    The full sky map is shown on the left and latitude profiles for longitudes in steps of 30$^\circ$ are shown on the right.
    }
    \label{fig:thickness_LocalBubbles}
\end{figure*}
To describe the outer surface of the Local Bubble shell, we consider two cases. For the first case we assume that the thickness of the shell is constant as seen from a given origin located inside the bubble (see Sect.~\ref{sec:LBscenarios}). Therefore, the outer surface of the shell is fully determined by $r_{\rm{inner}}^{\ell_{\rm{max}}=6}$, the adopted origin, and a constant thickness $\Delta$ which we fix to 35~pc, in agreement with the model of \cite{Yao2017} for the thermal electron density which includes a model for the Local Bubble. A value of 35~pc for the shell thickness gives a ratio of thickness to radius of the order of 10 to 30\% which agrees with what is observed and predicted for ISM super-bubbles (e.g., \citealt{Krause2014}).
However, and to mention only the most obvious, given the stratified and inhomogeneous environment in which the Local Bubble has formed and expanded, it is very unlikely that the thickness of its shell is constant.
So, to consider the case of variable thickness, we use an approach very similar to that used to calculate $r_{\rm{inner}}^{\ell_{\rm{max}}=6}$ in (\citealt{Pelgrims2020}) and derive an estimate of the outer surface of the Local Bubble shell from 3D dust data.
We use the 3D dust map of \cite{Lallement2019} and study the radial profiles of the differential dust extinction. Instead of considering the first inflection point where the profile goes from convex to concave, we consider the second inflection point where, after reaching a local maximum, the profile transitions from concave to convex.
We repeat the process for all sightlines of an HEALPix map with $N_{\rm{side}}=128$.
Because of the limited extent of the 3D map, the second inflection point is either not found before the edge of the map or found very close to it for about 20\% of the sightlines (\citealt{Pelgrims2020}). Therefore, to model the outer surface of the shell, we mask those problematic sightlines and fit the unmasked data by adjusting a surface that is fully described by the spherical harmonic coefficients up to a maximum multipole $\ell_{\rm{max}}$. For consistency with the characterization of the inner surface, we use $\ell_{\rm{max}} = 6$ and define the outer surface of the shell of the Local Bubble, which we denote $r_{\rm{outer}}^{\ell_{\rm{max}}=6}$.
Like $r_{\rm{inner}}^{\ell_{\rm{max}}=6}$, $r_{\rm{outer}}^{\ell_{\rm{max}}=6}$ consists of a sky map of distances from the Sun to the outer edge of the shell of the Local Bubble.
The maps of $r_{\rm{inner}}^{\ell_{\rm{max}}=6}$ and $r_{\rm{outer}}^{\ell_{\rm{max}}=6}$ are shown in the left and right panels of Fig.~\ref{fig:r_ell6}, respectively.
In Fig.~\ref{fig:CC_LocalBubbles}, we show three crosscuts of the Solar neighborhood according to the 3D dust extinction map of \cite{Lallement2019}. This map gives the differential extinction ($\delta A_{\rm{v}}$). These crosscuts show the $XY$, $XZ$, and $YZ$ planes, of the Heliocentric Cartesian coordinate system where the Galactic center is located at $X \approx 8$~kpc.
In this triptych, the blue and orange points show the locations of the first and second inflection points measured on the radial profiles $\delta A_{\rm{v}}(r)$ in the respective plane, when they are well-defined. The continuous lines correspond to the intersection of our models for $r_{\rm{inner}}^{\ell_{\rm{max}}=6}$ and $r_{\rm{outer}}^{\ell_{\rm{max}}=6}$ with the different planes, respectively.

However, we point out that this geometrical construction is likely to overestimate the distance to the outer surface of the Local Bubble shell and, as a result, its thickness. This is primarily due to the limited resolution of the 3D dust map and the post-processing of the radial profiles to extract its inflection points. Their effect is that any density peaks are radially smeared out, leading to lower and higher
values for the inner and outer radius of the shell. In addition, although the geometric approach is in principle suitable for characterizing the geometry of a shell formed in a homogeneous environment, we have little control over the possible effects of initial inhomogeneities on our results. If, for a given line of sight, the shell merges with an initial overdensity, it is possible that the second inflection point does not correspond to the extreme edge of the shell but to that of the overdensity, or to any intermediate value, depending on the relative amplitudes of the density peaks.
In Fig.~\ref{fig:thickness_LocalBubbles}, we show the thickness of the shell of the Local Bubble as measured from the Sun in the form of a sky map and latitude profiles for several longitudes. It is seen that the thickness varies substantially over the sky and is generally larger in the northern hemisphere than in the southern one. We emphasize the fact that this is the thickness of the shell as measured from the Sun and that, due to projection effects, it is not directly related to the amplification of the magnetic field as it is not measured from the explosion center.

\end{appendix}

\end{document}